\documentclass[aps,twocolumn,amsmath,amssymb,showpacs,superscriptaddress,notitlepage,longbibliography]{revtex4-1}
\usepackage[colorlinks=true,linkcolor=blue,anchorcolor=red,citecolor=blue, urlcolor=blue]{hyperref}
\usepackage{bm}
\usepackage{float}
\usepackage{graphicx}
\usepackage{color}
\usepackage{color, colortbl}
\usepackage{soul}
\usepackage[table]{xcolor}
\definecolor{Gray}{gray}{0.9}
\newcolumntype{g}{>{\columncolor{Gray}}c}
\renewcommand{\vec}{\mathbf}

\begin{document}
\title{Numerical study of PbTe-Pb hybrid nanowires for engineering Majorana zero modes}

\author{Zhan Cao}
\affiliation{Beijing Academy of Quantum Information Sciences, Beijing 100193, China}

\author{Dong E. Liu}
\email{dongeliu@mail.tsinghua.edu.cn}
\affiliation{State Key Laboratory of Low Dimensional Quantum Physics, Department of Physics, Tsinghua University, Beijing, 100084, China}
\affiliation{Beijing Academy of Quantum Information Sciences, Beijing 100193, China}
\affiliation{Frontier Science Center for Quantum Information, Beijing 100084, China}

\author{Wan-Xiu He}
\affiliation{Beijing Computational Science Research Center, Beijing 100193, China}

\author{Xin Liu}
\affiliation{School of Physics, Huazhong University of Science and Technology, Wuhan, Hubei 430074, China}

\author{Ke He}
\email{kehe@tsinghua.edu.cn}
\affiliation{State Key Laboratory of Low Dimensional Quantum Physics, Department of Physics, Tsinghua University, Beijing, 100084, China}
\affiliation{Beijing Academy of Quantum Information Sciences, Beijing 100193, China}
\affiliation{Frontier Science Center for Quantum Information, Beijing 100084, China}

\author{Hao Zhang}
\email{hzquantum@mail.tsinghua.edu.cn}
\affiliation{State Key Laboratory of Low Dimensional Quantum Physics, Department of Physics, Tsinghua University, Beijing, 100084, China}
\affiliation{Beijing Academy of Quantum Information Sciences, Beijing 100193, China}
\affiliation{Frontier Science Center for Quantum Information, Beijing 100084, China}

\begin{abstract}
Epitaxial semiconductor-superconductor (SM-SC) hybrid nanowires are potential candidates for implementing Majorana qubits. Recent experimental and theoretical works show that charged impurities in SM remain a major problem in all existing hybrid nanowires, in which the SM is either InAs or InSb while the SC is mainly Al. Here, we theoretically validate the recently proposed PbTe-Pb hybrid nanowire as a potential candidate for Majorana devices. By studying the electrostatic and electronic properties of PbTe nanowires, we demonstrate that the huge dielectric constant of PbTe endows itself a high tolerance of charged impurity, which is a potential advantage over InAs and InSb nanowires. Moreover, we find that the effective axial Land\'{e} $g$ factor and Rashba spin-orbit coupling strength of PbTe nanowires are comparable to those of InAs nanowires. The conceivable merits of using Pb as the SC are (i) Pb has a larger superconducting gap, higher critical temperature, and higher parallel critical magnetic field than those of Al; (ii) a superconducting gap comparable with those of InAs-Al and InSb-Al can be induced in PbTe-Pb even by a weak coupling between Pb and PbTe, which simultaneously relieves the adverse renormalization and induced disorder effects on SM from SC; and (iii) Pb film can be grown on PbTe with a thin buffer CdTe layer in between, solving the lattice mismatch problem as an important source of disorder. In the presence of a parallel magnetic field, we show that the typical BdG energy spectrum and tunneling spectroscopy of PbTe-Pb resemble those of InAs and InSb based hybrid nanowires exposed to a tilting magnetic field, as a result of the highly anisotropic Land\'{e} $g$ factors of PbTe nanowires. The calculated topological phase diagrams of PbTe-Pb indicate that the multivalley character of PbTe makes it easier than InAs and InSb to access topological superconducting phases. Our results could facilitate the experimental realization of PbTe-Pb hybrid nanowires and inspire further theoretical works.
\end{abstract}


\maketitle

\section{Introduction}\label{intr}
In the last decade, Majorana zero modes (MZMs) \cite{read2000paried,kitaev2001unpaired} (see Refs.~\onlinecite{alicea2012new,leijnse2012introduction,beenakker2013search,stanescu2013majorana,aguado2017majorana,prada2020andreev} for review) have attracted tremendous interest due to their exotic non-Abelian braiding statistics \cite{moore1991nonabelions,nayak19962n,nayak2008non} and potential applications to fault-tolerant topological quantum computation \cite{kitaev2003fault,nayak2008non,sarma2015majorana,aasen2016milestones,plugge2017majorana,karzig2017scalble}. Theoretically, MZMs can be engineered in laboratories by contacting a semiconductor (SM) nanowire with strong spin-orbit coupling (SOC) to a conventional $s$-wave superconductor (SC) with the aid of electrostatic gates and a magnetic field \cite{lutchyn2010majorana,oreg2010helical}. Many experiments based on this idea reported zero-bias conductance peaks in tunneling spectroscopy of InAs and InSb nanowires \cite{mourik2012signatures,deng2012anomalous,das2012zero,finck2013anomalous,churchill2013superconductor,deng2016majorana,chen2017experimental,suominen2017zero,nichele2017scaling,
gul2018ballistic,sestoft2018engineering,vaitiekenas2018effective,deng2018nonlocality,de2018electric,bommer2019spin,grivnin2019concomitant,anselmetti2019end,menard2020conductance}, possibly connected with MZMs. Most of the peaks are much lower than the predicted quantized value $2e^2/h$ resulting from MZM mediated resonant Andreev reflections \cite{sengupta2001midgap,law2009majorana,flensberg2010tunneling,wimmer2011quantum}. Recently, large zero-bias conductance peaks approaching to $2e^2/h$ have been observed \cite{nichele2017scaling,zhang2021large,song2021large}, however, perfect MZM quantization with peak height robustly sticking to $2e^2/h$ by varying both magnetic field and all relevant gate voltages is still lacking.

Recently, a theoretical work \cite{pan2020physical} explains that most of the experimental zero-bias conductance peaks in InAs and InSb nanowire devices are likely induced by strong disorder in the chemical potential, proximity-induced superconducting gap, or effective Land\'{e} $g$ factor, according to thorough numerical simulations. The level of disorder (strong versus weak) depends on the variation amplitude of corresponding physical quantity relative to its average.  For the widely explored epitaxially grown SM-SC hybrid nanowires \cite{chang2015hard,krogstrup2015epitaxy}, the interface between the SM and the SC has been confirmed to be pristine by material characterization and the observation of hard induced superconducting gaps at zero magnetic field \cite{takei2013soft}. However, recent numerical simulations \cite{sarma2021disorder} reproduced the tunneling spectroscopy in Ref.~\onlinecite{zhang2021large}, implying that the state-of-the-art InSb-Al hybrid nanowire remains disorder-limited. In realistic devices, disorder may arise from the surface oxidation of the SC, the imperfect substrates and gates, and more seriously, from the unintentional charged impurities \cite{pantelides1978the}, the randomly distributed twin defects \cite{caroff2009controlled}, and the stacking faults \cite{shtrikman2009method} in SM nanowires. Thus far, for a comprehensive understanding of Majorana nanowire systems, various long- and short-range disorder effects \cite{brouwer2011probability,brouwer2011topological,pientka2012enhanced,liu2012zero,kells2012near,pikulin2012zero,rainis2013towards,roy2013topologically,stanescu2013disentangling,hui2015bulk,klinovaja2015fermionic,cole2016proximity,liu2017andreev,liu2018impurity,moore2018quantized,fleckenstein2018decaying,cao2019decay,vuik2019reproducing,kiendl2019proximity,pan2021quantized} have been explored, most of which provide alternative interpretations on experimental results that can be attributed to MZMs. Even though the Majorana filter scheme \cite{liu2013filter,liu2022universal,zhang2022suppressing} could provide a solution to exclude many non-Majorana states due to disorder and enhance the Majorana signals, those disorder contamination can significantly destroy their power for testing non-Abelian braiding statistics and quantum information processing~\cite{nayak2008non}.

The next step towards engineering and detecting MZMs in SM-SC hybrid nanowires has three parallel directions in material aspect. (i) Optimizing the current InAs-Al and InSb-Al hybrid nanowires to reduce possible disorder \cite{lutchyn2018majorana,giustino2020}. (ii) Replacing Al with other SC materials, e.g., Sn \cite{pendharkar2021parity} and Pb \cite{kanne2021epitaxial,jung2021universal}, as possible better candidates. As compared in Table \ref{scpara}, Pb and Sn have larger superconducting gaps, higher critical temperatures, and higher critical parallel magnetic fields than those of Al. (iii) Searching potentially better SM materials, e.g., PbTe, which has been briefly mentioned for the realization of MZMs but not systematically investigated yet \cite{klingshirn2013,kamphuis2021towards,schellingerhout2021growth}. Recently, preliminary experimental efforts on exploring the combination of PbTe and Pb as SM-SC hybrid nanowires are ongoing \cite{schlatmann2021josephson,jiang2021selective}. A few merits of PbTe-Pb was pointed out \cite{kamphuis2021towards}: (i) PbTe is predicted to have a Land\'{e} $g$ factor higher than both InAs and InSb; (ii) PbTe nanowires are expected to show stronger SOC than those in InAs and InSb nanowires; and (iii) Pb shell on PbTe nanowire has less lattice mismatch than InAs-Al and InSb-Al. For the last point, we expect that the lattice mismatch problem could be solved by buffering a thin CdTe layer between PbTe and Pb, as CdTe and PbTe share almost the same lattice constant, while PbTe still maintains the induced superconductivity from Pb.

\begin{table}[t!]
\centering
\caption{Parameters of SCs: Al, Sn, and Pb. The superconducting gap $\Delta$ and critical temperature $T_c$ of bulk superconductors are adapted from Ref.~\onlinecite{kittel1996introduction}, while the critical parallel magnetic field $B_{c,\parallel}$ of superconducting thin shells are adapted from SM-SC hybrid nanowire experiments.}\label{scpara}
\begin{tabular}{cccc}
\hline
\hline
~~~~~~~~~ &~~~~~~~~~Al~~~~~~~~~~ &~~~~~~~~~Sn~~~~~~~~~~ &~~~~~~~~Pb~~~~~~~\\
\hline
$\Delta$ (meV) &0.34 &1.15 &2.73 \\
\hline
$T_c$ (K) &1.14  &3.72 & 7.19 \\
\hline
$B_{c,\parallel}$ (T) &$\sim 2$ \cite{deng2016majorana,vaitiekenas2018effective,zhang2021large} &$\sim 4$ \cite{pendharkar2021parity} &$\sim 8.5$ \cite{kanne2021epitaxial}\\
\hline
\hline
\end{tabular}
\end{table}

In fact, one can anticipate more outstanding merits of PbTe-Pb hybrid nanowires. Firstly, epitaxially grown PbTe has a high tolerance of charged impurity, high electron mobility, and high crystalline quality. Specifically, in submicron constrictions lithographically patterned in PbTe quantum wells, sequential quantized conductance steps were observed despite of a significant concentration of charged defects near the constrictions \cite{grabecki1999quantum,grabecki2005disorder,grabecki2004ballistic,grabecki2006pbte}. Such encouraging results are attributed to that PbTe is a paraelectric material with a huge static dielectric constant being 1350 at 4.2 K, almost two orders of magnitude larger than the value ($\sim 15$) of InAs and InSb. Consequently, charged impurity scattering is effectively screened and very high electron mobilities about $10^6~\textrm{cm}^2/\textrm{Vs}$ can be achieved \cite{springholz1993mbe,ueta1997improved}. Moreover, a high resolution transmission electron microscopy revealed that the epitaxially grown PbTe nanowires are free of stacking fault \cite{dziawa2010defect}. Secondly, by virtue of the sizable parent gap $\Delta$ of Pb (see Table \ref{scpara}), a superconducting gap comparable with those of InAs-Al and InSb-Al can be induced in PbTe-Pb even by a weak coupling between Pb and PbTe, where the effective $g$ factor and SOC in PbTe do not get reduced much by the renormalization effects on SM from SC \cite{stanescu2011majorana,cole2015effects,stanescu2017proximity,reeg2018metallization}. Meanwhile, at weak SM-SC couplings, the induced superconductivity in the SM is immune to any nonmagnetic disorder in the SC \cite{hui2015bulk,cole2016proximity,liu2018impurity}. Last but not least, the heavy element Pb has a strong intrinsic atomic SOC which may render additional SOC \cite{nadj2014observation,li2014topological} in PbTe by proximity effect. A theoretical work \cite{hui2015majorana} suggested that the interplay of the atomic SOC and the $s$-$p$ orbital hybridization of the Cooper pairs in superconductor Pb would give rise to an effective $p$-wave pairing in SMs coupled to Pb.

In this work, we present a quantitative study to examine the main anticipations discussed above on PbTe-Pb hybrid nanowires, and furthermore, to reveal discernible differences of PbTe-Pb from InAs and InSb based hybrid nanowires. Within the framework of $\vec{k}\cdot\vec{p}$ theory, we derive an effective conduction band Hamiltonian of PbTe nanowires with both Zeeman and SOC fields. As PbTe bulk has a highly anisotropic band structure and a fourfold valley degeneracy, we explore PbTe nanowires with nine different orientations that are preferred in experiments. By calculating the electrostatic and electronic properties of PbTe nanowires, we demonstrate that the huge dielectric constant of PbTe endows itself an advantage over InAs and InSb nanowires: a high tolerance of charged impurity as expected above. By contrast to the anticipation, we find that the effective axial Land\'{e} $g$ factor and SOC strength are valley-dependent and anti-correlated, inconsistent with the expected coexistence of a large $g$ factor and a strong SOC. Nevertheless, the relevant parameters of PbTe nanowires are found to be comparable with those of InAs nanowires. Our numerical calculations show that the accessible maximum SOC strength of all valleys increases with decreasing the side length $l$ of squared PbTe nanowires. And they exceed 30 meVnm at $l=30$ nm for valleys with the smallest axial effective electron mass ($0.024~m_e$). This strong SOC would benefit the realization of topological superconductivity. A symmetry analysis and numerical simulations indicate that PbTe nanowires with two of the nine orientations cannot be used for engineering MZMs, as they would never have a nondegenerate valley in any device geometries. In the presence of a parallel magnetic field, we study the topological superconductivity of PbTe-Pb hybrid nanowires in the weak SM-SC coupling regime. We obtain the phase transition conditions associated to topological trivial and nontrivial superconducting phases as well as the metallic phase. We show that the typical BdG energy spectrum and tunneling spectroscopy resemble those of InAs and InSb based hybrid nanowires exposed to a tilting magnetic field, which is a result of the highly anisotropic Land\'{e} $g$ factors of PbTe nanowires. Moreover, the topological phase diagrams of PbTe-Pb in the parameter space of back-gate voltage and parallel magnetic field are calculated. It turns out that the multiple valleys of PbTe make it easier than InAs and InSb to access topological superconducting phases. For comparisons, we summarize in Tables \ref{smpara} and \ref{smscpara} the relevant properties of bare InSb, InAs, PbTe nanowires and their corresponding SM-SC hybrid nanowires. Our results shed valuable light on using PbTe-Pb hybrid nanowires for engineering and detecting MZMs.

\begin{table}[t!]
\centering
\caption{Comparison of the relevant parameters: relative dielectric constant $\varepsilon_r$, zero temperature band-gap $E_g$, axial effective electron mass $m_{z}^\ast$, axial effective Land\'{e} factor $g_{z}^\ast$, and number of valley of bare InSb, InAs, and PbTe nanowires.}\label{smpara}
\begin{tabular}{cccc}
\hline
\hline
~~~~~ &~~~~~InSb \cite{de2018electric}~~~~ &~~~InAs \cite{winkler2019unified}~~~~~~ &~~~~PbTe~~~~~\\
\hline
$\varepsilon_r$ &$\sim 15$  &$\sim 15$ &$\sim 1350$ \\
\hline
$E_g$ (eV) &0.237 &0.418 &0.190 \\
\hline
$m_{z}^\ast/m_e$ &0.0139 &0.023 &0.024$-$0.240\\
\hline
$g_{z}^\ast$ &40.0 &14.9 &15.0$-$59.0\\
\hline
No. of valley &1 &1 &4\\
\hline
\hline
\end{tabular}
\end{table}

\begin{table}[t!]
\centering
\caption{Comparison of the major factors in InSb-Al, InAs-Al, and PbTe-Pb hybrid nanowires for realizing topological superconductivity, such as the degree of lattice mismatch (DOM) $(a_\textrm{SM}-a_\textrm{SC})/a_\textrm{SM}$ with $a_\textrm{SM/SC}$ the lattice constant, the gate induced maximum Rashba SOC strength $\alpha_\textrm{SOC}$ (for nanowires with diameters $\sim 100$ nm), the angle $\theta$ between the SOC and Zeeman fields under a parallel magnetic field, the hole screening on electrostatic gates, and the adverse renormalization effects on SM from SC \cite{stanescu2011majorana,cole2015effects,stanescu2017proximity,reeg2018metallization} when inducing a superconducting gap $\Delta_\textrm{ind}\sim 0.2$ meV in the SM. }\label{smscpara}
\begin{tabular}{cccc}
\hline
\hline
~ &InSb-Al \cite{de2018electric}~~ &~InAs-Al \cite{winkler2019unified}~~ &~PbTe-Pb~~\\
\hline
\textrm{DOM}  &0.38 &0.33 &0.24\\
\hline
$\alpha_\textrm{SOC}$ (meVnm) &$\sim 60$ &$\sim 10$ &$\sim 10$\\
\hline
$\theta$ &$90^\circ$ &$90^\circ$ &$83.1^\circ - 90^\circ$\\
\hline
hole screening on  & & &\\
electrostatic gates &strong &strong &negligible\\
\hline
renormalization  & & &\\
effects on SM for &strong &strong &weak\\
$\Delta_\textrm{ind}\sim 0.2$ meV & & &\\
\hline
\hline
\end{tabular}
\end{table}

The remainder of this paper is organized as follows. In Sec.~\ref{ModelandApp}, we introduce the device, the model, and the approaches. The results and discussion are presented in Sec.~\ref{results}. In Sec.~\ref{dispersions}, we show the gate tunability, dispersions, and impurity sensitivity of PbTe nanowires. In Sec.~\ref{soc}, we study the amplitude and direction of the Rashba SOC field in PbTe nanowires with nine different orientations. In Sec.~\ref{MZMs}, the results pertinent to the topological superconductivity of PbTe-Pb hybrid nanowires are presented. Finally, we give a summary and outlook in Sec.~\ref{Summary}.

\section{Model and approach}\label{ModelandApp}
\begin{figure}[t!]
\centering
\includegraphics[width=\columnwidth]{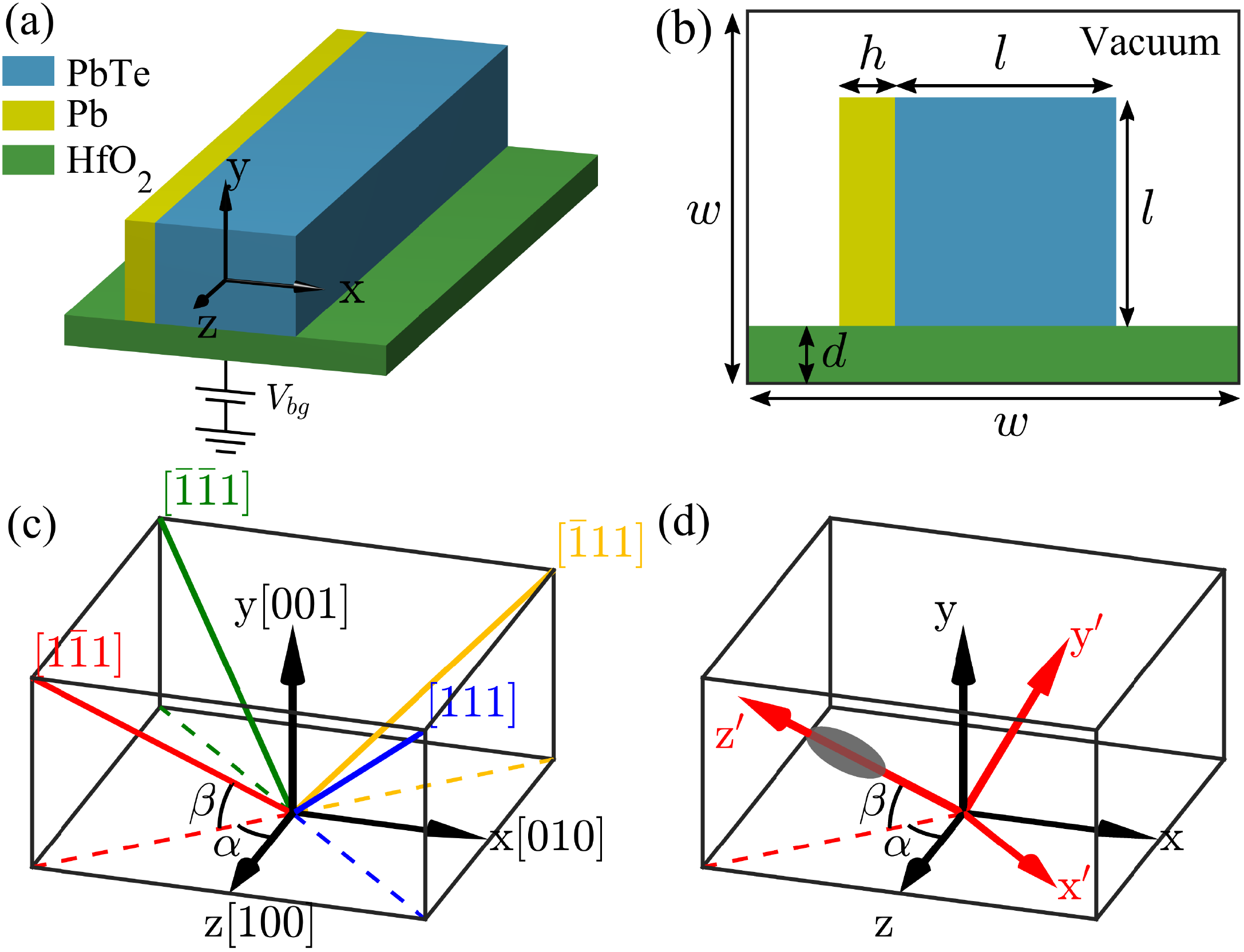}
\caption{(a) Schematic view of a squared PbTe-Pb hybrid nanowire placed on a $\textrm{HfO}_2$ dielectric layer, together with a laboratory coordinate system whose $z$ axis is along the wire's axial direction and the $x$ and $y$ axes are perpendicular to the wire's side facets. The device is gated from below by a back-gate voltage $V_{bg}$. (b) Transverse profile of the device enclosed by a large enough squared box. The relevant length scales are indicated with double arrows. (c) Longitudinal axes (colored solid lines) of the four L valleys of a particular PbTe nanowire grown along $[100]$ direction. The orientation of a valley is defined by azimuth angles $(\alpha,\beta)$. (d) The $\vec{k}\cdot\vec{p}$ Hamiltonian (\ref{Ht_elli})--(\ref{HZ_elli}) for a valley, e.g., $[1\bar{1}1]$, of PbTe is established in a local coordinate system $x'$-$y'$-$z'$ whose $z'$-axis is along the longitudinal axis of the valley. The anisotropic effective electron masses of PbTe manifest as an ellipsoid constant energy surface (in gray) around the $L$ point.}\label{Fig:device}
\end{figure}

We consider a typical device schematically shown in Fig.~\ref{Fig:device}(a), in which an epitaxially grown PbTe-Pb hybrid nanowire with a squared cross section is placed on a $\textrm{HfO}_2$ dielectric layer attached to a back gate with voltage $V_{bg}$. A laboratory coordinate system $x$-$y$-$z$ is indicated at the center of the cross section of the PbTe nanowire. The transverse profile of the device is shown in Fig.~\ref{Fig:device}(b). In Sec.~\ref{Hamiltonian}, we derive an effective conduction band Hamiltonian for each valley of a bare PbTe nanowire in the laboratory coordinate system. In Sec.~\ref{electrostatics}, we outline the approach solving the electrostatics of the device. In Sec.~\ref{effectivemodel}, we present an effective 1D model describing the PbTe-Pb hybrid nanowire and the associated conditions for realizing topological superconducting phases supporting MZMs.

\subsection{Effective conduction band Hamiltonian of PbTe nanowires}\label{Hamiltonian}
PbTe is a IV-VI SM, dramatically different from the III-V SMs such as InAs and InSb. The band extrema of III-V SMs are in the center of the Brillouin zone ($\Gamma$ point), around which the Fermi surface is isotropic. On the contrary, PbTe is a narrow direct band-gap SM with eight equivalent band extrema at the $L$ points of the Brillouin zone, known as the fourfold valley degeneracy \cite{klingshirn2013}. The corresponding Fermi surfaces are highly anisotropic, i.e., four elongated ellipsoids of revolution around the axes $[1\bar{1}1]$, $[\bar{1}\bar{1}1]$, $[\bar{1}11]$, and $[111]$ shown in Fig.~\ref{Fig:device}(c), characterized by longitudinal ($l$) and transverse ($t$) effective masses $m_{l,t}$ and Land\'{e} factors $g_{l,t}$. Considering the four valleys of PbTe being independent of each other, each valley can be modeled by the Dimmock $\vec{k}\cdot\vec{p}$ Hamiltonian \cite{carter1971physics} established in the local coordinate system $x'$-$y'$-$z'$ whose $z'$ axis is along the longitudinal axis of the ellipsoidal valley, as shown in Fig.~\ref{Fig:device}(d). Applying a folding-down procedure to the Dimmock model, an effective conduction band Hamiltonian can be derived as \cite{silva1999optical,hasegawa2003spin,ridolfi2015effective}
\begin{equation}
H^\textrm{elli}=H^\textrm{elli}_\textrm{kin}+H^\textrm{elli}_\textrm{SOC},\label{Ht_elli}
\end{equation}
which consists of the kinetic energy and Rashba SOC
\begin{eqnarray}
H^\textrm{elli}_\textrm{kin}&=&P_{t}^{2}[\hat{k}_{x^\prime}\gamma(\vec{r^\prime})\hat{k}_{x^\prime}+\hat{k}_{y^\prime}\gamma(\vec{r^\prime})\hat{k}_{y^\prime}]\notag\\
&&+P_{l}^{2}\hat{k}_{z^\prime}\gamma(\vec{r^\prime})\hat{k}_{z^\prime}+E_c(\vec{r^\prime}),\label{H1_elli}
\end{eqnarray}
\begin{eqnarray}
H^\textrm{elli}_\textrm{SOC}=&&iP_{t}P_{l}[\hat{k}_{y^\prime}\gamma(\vec{r^\prime})\hat{k}_{z^\prime}-\hat{k}_{z^\prime}\gamma(\vec{r^\prime})\hat{k}_{y^\prime}]\sigma_{x^\prime}\notag\\
&&+iP_{t}P_{l}[\hat{k}_{z^\prime}\gamma(\vec{r^\prime})\hat{k}_{x^\prime}-\hat{k}_{x^\prime}\gamma(\vec{r^\prime})\hat{k}_{z^\prime}]\sigma_{y^\prime}\notag\\
&&+iP_{t}^{2}[\hat{k}_{x^\prime}\gamma(\vec{r^\prime})\hat{k}_{y^\prime}-\hat{k}_{y^\prime}\gamma(\vec{r^\prime})\hat{k}_{x^\prime}]\sigma_{z^\prime},\label{H2_elli}
\end{eqnarray}
where $\hat k_u$ and $\sigma_u$ ($u=x^\prime,y^\prime,z^\prime$) are the momentum and spin-$1/2$ Pauli operators, respectively, $\vec{r^\prime}=(x^\prime,y^\prime,z^\prime)$, $\gamma(\vec{r^\prime})=1/[E-E_v(\vec{r^\prime})]$, $E_{c}(\vec{r^\prime})=-e\phi(\vec{r^\prime})$ and $E_v(\vec{r^\prime})=-E_g-e\phi(\vec{r^\prime})$ are the conduction and valence band edges, respectively, modified by a nonuniform electrostatic potential $\phi(\vec{r^\prime})$, $E_g$ is the SM band-gap, $P_{l(t)}=\hbar(E_g/2m^e_{l(t)})^{1/2}$ is the longitudinal (transverse) conduction-to-valence band coupling with $m^e_{l(t)}$ being the effective electron mass. The conduction band edge in the absence of $\phi(\vec{r^\prime})$ has been chosen as the reference energy. We note that PbTe bulk has a rocksalt crystalline structure which presents inversion symmetry such that the SOC is purely Rashba \cite{hasegawa2003spin,peres2014experimental}. In the presence of an applied magnetic field $\vec{B}$, one should make the fundamental substitution $\vec{k}\rightarrow \vec{\pi}\equiv\vec{k}+e\vec{A}/\hbar$, with $\nabla \times \vec{A}=\vec{B}$, and meanwhile add a Zeeman term \cite{ridolfi2015effective}
\begin{equation}
H^\textrm{elli}_{Z}=\frac{1}{2}\mu_{B}(  g_{t}B_{x^\prime}\sigma_{x^\prime}+g_{t}B_{y^\prime}\sigma_{y^\prime}+g_{l}B_{z^\prime}\sigma_{z^\prime}).\label{HZ_elli}
\end{equation}
where $\mu_B$ is the Bohr magneton. In calculations, we use the low-temperature empirical parameters \cite{silva1999optical,ridolfi2015effective} $g_l=-24.3$, $g_t=-11.4$, $E_g=190$ meV, $m^e_l=0.24~m_e$, $m^e_t=0.024~m_e$, $m^h_l=0.31~m_e$, and $m^h_t=0.022~m_e$, with $m_e$ being the free electron mass.

To facilitate studying low-dimensional multivalley SMs, it is convenient to transform the Hamiltonian of each valley to a common laboratory coordinate system \cite{rahman2005generalized}, i.e., the x-y-z in Fig.~\ref{Fig:device}(d), with the z-axis along the axial direction of the nanowire. In a previous work \cite{aparecida2010ballistic}, such a Hamiltonian of PbTe nanowire is presented but without incorporating the crucial SOC and Zeeman terms. To fill in the gap, we derive an effective conduction band Hamiltonian for each valley of a translational invariant PbTe nanowire exposed to a parallel magnetic field $B_z$ as (see derivation in Appendix \ref{app})
\begin{equation}
H^\textrm{wire}=H^\textrm{wire}_\textrm{kin}+H^\textrm{wire}_\textrm{SOC}+H^\textrm{wire}_\textrm{Z},\label{hwire}
\end{equation}
with
\begin{eqnarray}
H_\textrm{kin}^\textrm{wire}&=&\sum_{u,v=x,y}\hat{\pi}_{u}\frac{M_{uv}}{E_{g}}\hat{\pi}_{v}+\frac{\hbar^{2}k_{z}^{2}}{2m_z^\ast}+E_c(x,y),\label{hwire_kin}\\
H^\textrm{wire}_\textrm{SOC}&=&\sum_u(\Omega_{ux}\hat{\pi}_x+\Omega_{uy}\hat{\pi}_y+\Omega_{uz}k_{z})\sigma_{u},\label{hwire_soc}\\
H^\textrm{wire}_\textrm{Z}&=&\frac{1}{2}\mu_{B}B_{z}\sum_{u}g_{zu}^{\ast}\sigma_{u},\label{hwire_Zeeman}
\end{eqnarray}
where
\begin{eqnarray}
\hspace{-0.6cm}m_{z}^{\ast}&=&\frac{\hbar^{2}E_{g}}{2P_{l}^{2}P_{t}^{2}}[P_{l}^{2}-(P_{l}^{2}-P_{t}^{2})\cos^{2}\alpha\cos^{2}\beta],\label{meff}\\
\hspace{-3.8cm}g_{zu}^{\ast}&=&g_{zu}+\frac{4m_{e}N_{u;xy}}{\hbar^{2}E_{g}},\label{gtensor1}
\end{eqnarray}
\begin{eqnarray}
&&\hspace{-2.2cm}g_{zx}=-(  g_{l}-g_{t})  \sin\alpha\cos\alpha\cos^{2}\beta,\notag\\
&&\hspace{-2.2cm}g_{zy}=(  g_{l}-g_{t})  \cos\alpha\sin\beta\cos\beta,\notag\\
&&\hspace{-2.2cm}g_{zz}=g_{t}+(  g_{l}-g_{t})  \cos^{2}\alpha\cos^{2}\beta,\label{gtensor2}
\end{eqnarray}
\begin{eqnarray}
&&\Omega_{ux}=\frac{eE_{y}(x,y)}{E_{g}^{2}}N_{u;yx},\notag\\
&&\Omega_{uy}=\frac{eE_{x}(x,y)}{E_{g}^{2}}N_{u;xy},\notag\\
&&\Omega_{uz}=\frac{eE_{x}(x,y)}{E_{g}^{2}}\bigg[N_{u;xz}+\frac{M_{xx}M_{zy}-M_{xy}M_{zx}}{M_{xy}^{2}-M_{xx}M_{yy}}N_{u;xy}\bigg]\notag\\
&&\hspace{0.3cm}+\frac{eE_{y}(x,y)}{E_{g}^{2}}\bigg[N_{u;yz}+\frac{M_{yy}M_{zx}-M_{xy}M_{zy}}{M_{xy}^{2}-M_{xx}M_{yy}}N_{u;yx}\bigg],\label{soccoefficient}
\end{eqnarray}
\begin{eqnarray}
&&\hspace{-0.8cm}M_{xx}=P_{t}^{2}+(P_{l}^{2}-P_{t}^{2})\sin^{2}\alpha\cos^{2}\beta,\notag\\
&&\hspace{-0.8cm}M_{yy}=P_{t}^{2}+(P_{l}^{2}-P_{t}^{2})\sin^{2}\beta,\notag\\
&&\hspace{-0.8cm}M_{zz}=P_{t}^{2}+(P_{l}^{2}-P_{t}^{2})\cos^{2}\alpha\cos^{2}\beta,\notag\\
&&\hspace{-0.8cm}M_{xy}=M_{yx}=-(P_{l}^{2}-P_{t}^{2})\sin\alpha\sin\beta\cos\beta,\notag\\
&&\hspace{-0.8cm}M_{xz}=M_{zx}=-(P_{l}^{2}-P_{t}^{2})\sin\alpha\cos\alpha\cos^{2}\beta,\notag\\
&&\hspace{-0.8cm}M_{yz}=M_{zy}=(P_{l}^{2}-P_{t}^{2})\cos\alpha\sin\beta\cos\beta,\label{matrixM}
\end{eqnarray}
\begin{eqnarray}
&&\hspace{-0.6cm}N_{x;yz} =P_{l}P_{t}-P_{t}(  P_{l}-P_{t})  \sin^{2}\alpha\cos^{2}\beta,\notag\\
&&\hspace{-0.6cm}N_{y;zx} =P_{l}P_{t}-P_{t}(  P_{l}-P_{t})  \sin^{2}\beta,\notag\\
&&\hspace{-0.6cm}N_{z;xy}  =P_{l}P_{t}-P_{t}(  P_{l}-P_{t})  \cos^{2}\alpha\cos^{2}\beta,\notag\\
&&\hspace{-0.6cm}N_{x;xy}  =-N_{z;zy}=P_{t}(  P_{l}-P_{t})  \sin\alpha\cos\alpha\cos^{2}\beta,\notag\\
&&\hspace{-0.6cm}N_{y;yz}  =-N_{x;xz}=P_{t}(  P_{l}-P_{t})  \sin\alpha\sin\beta\cos\beta,\notag\\
&&\hspace{-0.6cm}N_{z;zx}=-N_{y;yx}  =-P_{t}(  P_{l}-P_{t})  \cos\alpha\sin\beta\cos\beta. \label{tensorN}
\end{eqnarray}
In Eq.~(\ref{soccoefficient}), $E_{x(y)}(x,y)=-\partial_{x(y)}\phi(x,y)$ is the electric field in the x (y) direction inside the PbTe nanowire. The remaining elements of tensor $N$, not shown in Eq.~(\ref{tensorN}), satisfy the relation $N_{w;uv}=-N_{w;vu}$. This Hamiltonian is directly related to the anisotropic effective masses $m^e_{l,t}$ and Land\'{e} factors $g_{l,t}$ of PbTe. Indeed, the above Hamiltonian would reduce to the one describing InAs and InSb nanowires \cite{de2018electric,winkler2019unified} if $m^e_l=m^e_t$ (i.e., $P_l=P_t$) and $g_l=g_t$. This effective Hamiltonian is also applicable to other IV-VI semiconductors such as PbSe and PbS nanowires with corresponding $\vec{k}\cdot\vec{p}$ parameters. Note that we neglect the inter-valley coupling mediated by the surface of nanowire \cite{boykin2004valley,nestoklon2006spin}, which was predicted to induce appreciable valley splitting for extremely thin nanowires, e.g., for PbSe nanowires with diameters being smaller than 7 nm \cite{avdeev2017valley}. For PbTe-Pb hybrid nanowires, the Pb mediated inter-valley coupling can also be neglected provided the tunnel coupling between PbTe and Pb is weak, as discussed in Sec.~\ref{effectivemodel}.

\begin{figure}[t!]
\centering
\includegraphics[width=\columnwidth]{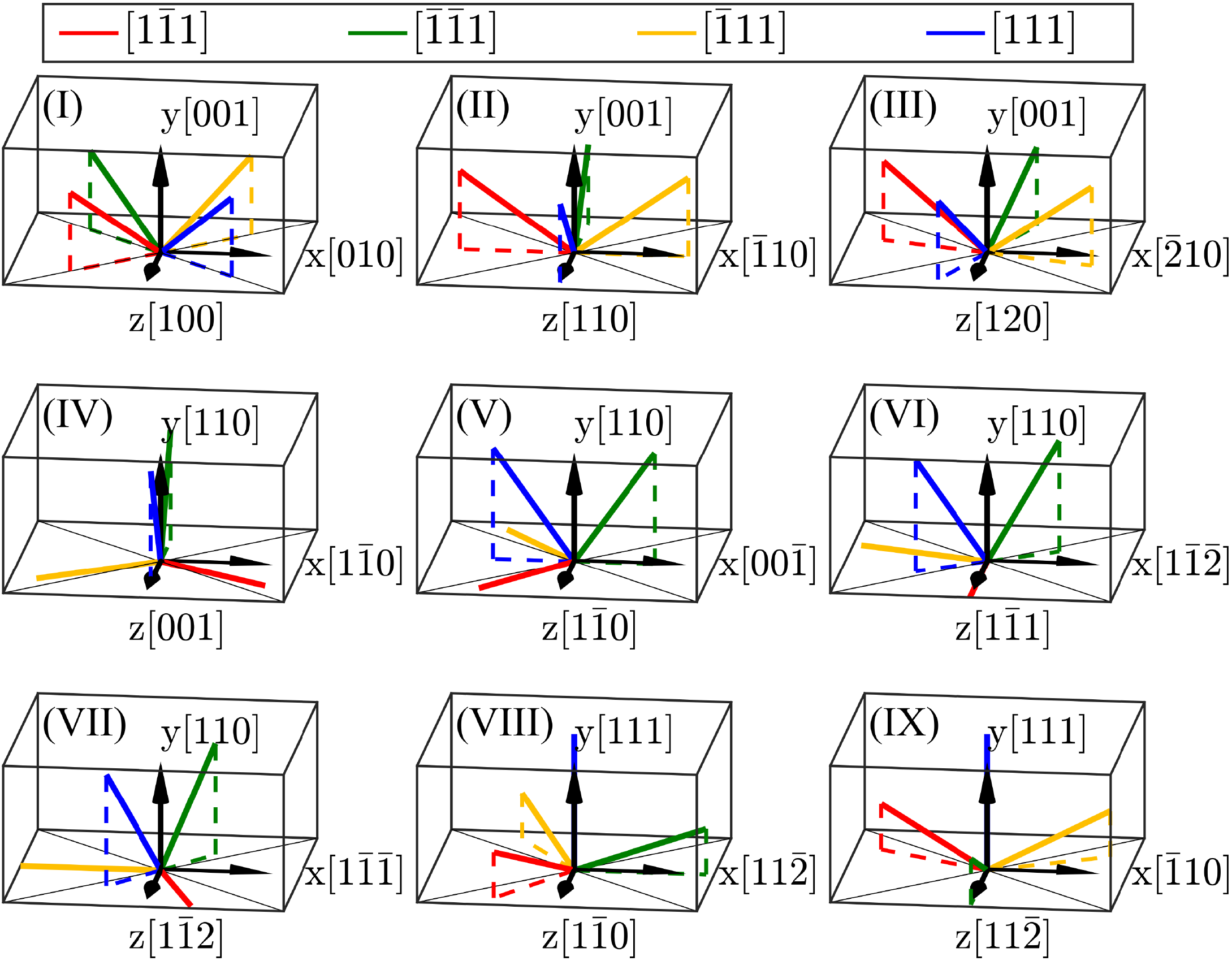}
\caption{View of the longitudinal axes (colored solid lines) of the four valleys in the laboratory coordinate system $x$-$y$-$z$ of PbTe nanowires with nine different growth orientations. The dashed lines are shown to aid the 3D visualization. By molecular beam epitaxy technique, wires (I)-(III) can be grown on CdTe (001) substrate, wires (IV)-(VII) on CdTe (110) substrate, and wires (VIII)-(IX) on CdTe (111) substrate.}\label{Fig:ninewires}
\end{figure}

In experiments, PbTe nanowires along low-index crystal direction are preferred as they can be grown with high crystalline quality on a lattice matched CdTe substrate \cite{jiang2021selective}. In Fig.~\ref{Fig:ninewires}, we show the laboratory coordinate system $x$-$y$-$z$ of nine representative low-index PbTe nanowires together with the longitudinal axes of their four valleys. The primary concern is whether the fourfold valley degeneracy of bulk PbTe can be lifted, at least partly, in these nanowires. This is because that the topological superconducting phase requires odd sub-band occupation \cite{potter2010multichannel,lutchyn2011search,stanescu2011majorana}, therefore, non-degenerate valleys. Our answer is affirmative for most of the nanowire growth directions except for the case I and IV, based on a symmetry analysis and numerical simulations presented in Sec.~\ref{dispersions}. In Table \ref{table}, the model parameters of the nine PbTe nanowires are calculated according to Eqs.~(\ref{meff})-(\ref{gtensor2}), meanwhile, for each nanowire the degenerate valleys are highlighted by the same color for clarity. For all valleys, one can see that the axial Land\'{e} factor $g^\ast_z\equiv(g_{zx}^{\ast 2}+g_{zy}^{\ast 2}+g_{zz}^{\ast 2})^{1/2}$ is positively correlated to the axial effective electron mass $m_z^\ast$, with the maximum $g^\ast_z=59.0$ and minimum $g^\ast_z=15.0$ corresponding to $m^\ast_z=0.240~m_e$ and $m^\ast_z=0.024~m_e$, respectively. Interestingly, this minimum $g_z^\ast$ and $m_z^\ast$ are almost identical to those of InAs nanowires, i.e., $g_z^\ast=14.9$ and $m^\ast_z=0.023~m_e$ \cite{winkler2003spin}. Generally, SM nanowires with a smaller $m_z^\ast$ have a larger carrier mobility \cite{keyes1959correlation}, which facilitates the detection on MZMs by electron transports.

\begin{table*}[htbp]
\centering
\caption{Model parameters in Hamiltonian (\ref{hwire}) for the four valleys of PbTe nanowire with nine different growth directions shown in Fig.~\ref{Fig:ninewires}. For each nanowire with a nonuniform electrostatic potential $\phi(x,y)$, valleys highlighted by the same background color are degenerate at zero magnetic field.}\label{table1}
\begin{tabular}{|c|c|c|c|c|c|c|c|c|c|c|c|c|c|}
\hline
wire &y-axis &x-axis &z-axis &valley &$\sin\alpha$ &$\cos\alpha$ &$\sin\beta$ &$\cos\beta$ &$g^\ast_{zx}$ &$g^\ast_{zy}$ &$g^\ast_{zz}$ &$g^\ast_{z}$ &$m_z^\ast/m_e$\\
\hline
\rowcolor[RGB]{209,209,239}
\cellcolor{white}{} &\cellcolor{white}{} &\cellcolor{white}{} &\cellcolor{white}{} &$[1\bar{1}1]$ &$1/\sqrt{2}$ &$1/\sqrt{2}$ &$1/\sqrt{3}$ &$\sqrt{2}/\sqrt{3}$ &$-14.7$ &14.7 &29.6 &36.2 &0.096\\
\rowcolor[RGB]{209,209,239}
\cellcolor{white}{I} &\cellcolor{white}$[001]$ &\cellcolor{white}$[010]$ &\cellcolor{white}$[100]$ &$[\bar{1}\bar{1}1]$ &$1/\sqrt{2}$ &$-1/\sqrt{2}$ &$1/\sqrt{3}$ &$\sqrt{2}/\sqrt{3}$ &14.7 &$-14.7$ &29.6 &36.2 &0.096\\
\rowcolor[RGB]{186,223,226}
\cellcolor{white}{} &\cellcolor{white}{} &\cellcolor{white}{} &\cellcolor{white}{} &$[\bar{1}11]$ &$-1/\sqrt{2}$ &$-1/\sqrt{2}$ &$1/\sqrt{3}$ &$\sqrt{2}/\sqrt{3}$ &$-14.7$ &$-14.7$ &29.6 &36.2 &0.096\\
\rowcolor[RGB]{186,223,226}
\cellcolor{white}{} &\cellcolor{white}{} &\cellcolor{white}{} &\cellcolor{white}{} &$[111]$ &$-1/\sqrt{2}$ &$1/\sqrt{2}$ &$1/\sqrt{3}$ &$\sqrt{2}/\sqrt{3}$ &14.7 &14.7 &29.6 &36.2 &0.096\\
\hline
\cellcolor{white}{} &\cellcolor{white}{} &\cellcolor{white}{} &\cellcolor{white}{} &$[1\bar{1}1]$ &1 &0 &$1/\sqrt{3}$ &$\sqrt{2}/\sqrt{3}$ &0 &0 &15.0 &15.0 &0.024\\
\rowcolor[RGB]{209,209,239}
\cellcolor{white}{II} &\cellcolor{white}$[001]$ &\cellcolor{white}$[\bar{1}10]$ &\cellcolor{white}$[110]$ &$[\bar{1}\bar{1}1]$ &0 &$-1$ &$1/\sqrt{3}$ &$\sqrt{2}/\sqrt{3}$ &0 &$-20.8$ &44.3 &49.0 &0.168\\
\cellcolor{white}{} &\cellcolor{white}{} &\cellcolor{white}{} &\cellcolor{white}{} &$[\bar{1}11]$ &$-1$ &0 &$1/\sqrt{3}$ &$\sqrt{2}/\sqrt{3}$ &0 &0 &15.0 &15.0 &0.024\\
\rowcolor[RGB]{209,209,239}
\cellcolor{white}{} &\cellcolor{white}{} &\cellcolor{white}{} &\cellcolor{white}{} &$[111]$ &0 &1 &$1/\sqrt{3}$ &$\sqrt{2}/\sqrt{3}$ &0 &20.8 &44.3 &49.0 &0.168\\
\hline
\cellcolor{white}{} &\cellcolor{white}{} &\cellcolor{white}{} &\cellcolor{white}{} &$[1\bar{1}1]$ &$3/\sqrt{10}$ &$-1/\sqrt{10}$ &$1/\sqrt{3}$ &$\sqrt{2}/\sqrt{3}$ &8.8 &$-6.6$ &17.9 &21.0 &0.038\\
\cellcolor{white}{III} &\cellcolor{white}$[001]$ &\cellcolor{white}$[\bar{2}10]$ &\cellcolor{white}$[120]$ &$[\bar{1}\bar{1}1]$ &$-1/\sqrt{10}$ &$-3/\sqrt{10}$ &$1/\sqrt{3}$ &$\sqrt{2}/\sqrt{3}$ &$-8.8$ &$-19.7$ &41.4 &46.7 &0.154\\
\cellcolor{white}{} &\cellcolor{white}{} &\cellcolor{white}{} &\cellcolor{white}{} &$[\bar{1}11]$ &$-3/\sqrt{10}$ &$1/\sqrt{10}$ &$1/\sqrt{3}$ &$\sqrt{2}/\sqrt{3}$ &8.8 &6.6 &17.9 &21.0 &0.038\\
\cellcolor{white}{} &\cellcolor{white}{} &\cellcolor{white}{} &\cellcolor{white}{} &$[111]$ &$1/\sqrt{10}$ &$3/\sqrt{10}$ &$1/\sqrt{3}$ &$\sqrt{2}/\sqrt{3}$ &$-8.8$ &19.7 &41.4 &46.7 &0.154\\
\hline
\rowcolor[RGB]{209,209,239}
\cellcolor{white}{} &\cellcolor{white}{} &\cellcolor{white}{} &\cellcolor{white}{} &$[1\bar{1}1]$ &$-\sqrt{2}/\sqrt{3}$ &$1/\sqrt{3}$ &0 &1 &20.8 &0 &29.6 &36.2 &0.096\\
\rowcolor[RGB]{186,223,226}
\cellcolor{white}{IV} &\cellcolor{white}$[110]$ &\cellcolor{white}$[1\bar{1}0]$ &\cellcolor{white}$[001]$ &$[\bar{1}\bar{1}1]$ &0 &$-1$ &$\sqrt{2}/\sqrt{3}$ &$1/\sqrt{3}$ &0 &$-20.8$ &29.6 &36.2 &0.096\\
\rowcolor[RGB]{209,209,239}
\cellcolor{white}{} &\cellcolor{white}{} &\cellcolor{white}{} &\cellcolor{white}{} &$[\bar{1}11]$ &$\sqrt{2}/\sqrt{3}$ &$1/\sqrt{3}$ &0 &1 &$-20.8$ &0 &29.6 &36.2 &0.096\\
\rowcolor[RGB]{186,223,226}
\cellcolor{white}{} &\cellcolor{white}{} &\cellcolor{white}{} &\cellcolor{white}{} &$[111]$ &0 &1 &$\sqrt{2}/\sqrt{3}$ &$1/\sqrt{3}$ &0 &20.8 &29.6 &36.2 &0.096\\
\hline
\rowcolor[RGB]{209,209,239}
\cellcolor{white}{} &\cellcolor{white}{} &\cellcolor{white}{} &\cellcolor{white}{} &$[1\bar{1}1]$ &$1/\sqrt{3}$ &$\sqrt{2}/\sqrt{3}$ &0 &1 &$-20.8$ &0 &44.3 &49.0 &0.168\\
\cellcolor{white}{V} &\cellcolor{white}$[110]$ &\cellcolor{white}$[00\bar{1}]$ &\cellcolor{white}$[1\bar{1}0]$ &$[\bar{1}\bar{1}1]$ &$-1$ &0 &$\sqrt{2}/\sqrt{3}$ &$1/\sqrt{3}$ &0 &0 &15.0 &15.0 &0.024\\
\rowcolor[RGB]{209,209,239}
\cellcolor{white}{} &\cellcolor{white}{} &\cellcolor{white}{} &\cellcolor{white}{} &$[\bar{1}11]$ &$1/\sqrt{3}$ &$-\sqrt{2}/\sqrt{3}$ &0 &1 &20.8 &0 &44.3 &49.0 &0.168\\
\cellcolor{white}{} &\cellcolor{white}{} &\cellcolor{white}{} &\cellcolor{white}{} &$[111]$ &1 &0 &$\sqrt{2}/\sqrt{3}$ &$1/\sqrt{3}$ &0 &0 &15.0 &15.0 &0.024\\
\hline
\cellcolor{white}{} &\cellcolor{white}{} &\cellcolor{white}{} &\cellcolor{white}{} &$[1\bar{1}1]$ &0 &1 &0 &1 &0 &0 &59.0 &59.0 &0.240\\
\cellcolor{white}{VI} &\cellcolor{white}$[110]$ &\cellcolor{white}$[1\bar{1}\bar{2}]$ &\cellcolor{white}$[1\bar{1}1]$ &$[\bar{1}\bar{1}1]$ &$-\sqrt{2}/\sqrt{3}$ &$-1/\sqrt{3}$ &$\sqrt{2}/\sqrt{3}$ &$1/\sqrt{3}$ &$-6.9$ &$-12.0$ &19.9 &24.2 &0.048\\
\cellcolor{white}{} &\cellcolor{white}{} &\cellcolor{white}{} &\cellcolor{white}{} &$[\bar{1}11]$ &$2\sqrt{2}/3$ &$-1/3$ &0 &1 &13.9 &0 &19.9 &24.2 &0.048\\
\cellcolor{white}{} &\cellcolor{white}{} &\cellcolor{white}{} &\cellcolor{white}{} &$[111]$ &$\sqrt{2}/\sqrt{3}$ &$1/\sqrt{3}$ &$\sqrt{2}/\sqrt{3}$ &$1/\sqrt{3}$ &$-6.9$ &$12.0$ &$19.9$ &$24.2$ &0.048\\
\hline
\cellcolor{white}{} &\cellcolor{white}{} &\cellcolor{white}{} &\cellcolor{white}{} &$[1\bar{1}1]$ &$-1/3$ &$2\sqrt{2}/3$ &0 &1 &13.9 &0 &54.1 &55.9 &0.216\\
\cellcolor{white}{VII} &\cellcolor{white}$[110]$ &\cellcolor{white}$[1\bar{1}\bar{1}]$ &\cellcolor{white}$[1\bar{1}2]$ &$[\bar{1}\bar{1}1]$ &$-1/\sqrt{3}$ &$-\sqrt{2}/\sqrt{3}$ &$\sqrt{2}/\sqrt{3}$ &$1/\sqrt{3}$ &$-6.9$ &$-17.0$ &24.7 &30.8 &0.072\\
\cellcolor{white}{} &\cellcolor{white}{} &\cellcolor{white}{} &\cellcolor{white}{} &$[\bar{1}11]$ &1 &0 &0 &1 &0 &0 &15.0 &15.0 &0.024\\
\cellcolor{white}{} &\cellcolor{white}{} &\cellcolor{white}{} &\cellcolor{white}{} &$[111]$ &$1/\sqrt{3}$ &$\sqrt{2}/\sqrt{3}$ &$\sqrt{2}/\sqrt{3}$ &$1/\sqrt{3}$ &$-6.9$ &$17.0$ &$24.7$ &$30.8$ &$0.072$\\
\hline
\rowcolor[RGB]{209,209,239}
\cellcolor{white}{} &\cellcolor{white}{} &\cellcolor{white}{} &\cellcolor{white}{} &$[1\bar{1}1]$ &1/2 &$\sqrt{3}/2$ &1/3 &$2\sqrt{2}/3$ &$-17.0$ &12.0 &44.3 &49.0 &0.168\\
\cellcolor{white}{VIII} &\cellcolor{white}$[111]$ &\cellcolor{white}$[11\bar{2}]$ &\cellcolor{white}$[1\bar{1}0]$ &$[\bar{1}\bar{1}1]$ &$-1$ &0 &1/3 &$2\sqrt{2}/3$ &0 &0 &15.0 &15.0 &0.024\\
\rowcolor[RGB]{209,209,239}
\cellcolor{white}{} &\cellcolor{white}{} &\cellcolor{white}{} &\cellcolor{white}{} &$[\bar{1}11]$ &1/2 &$-\sqrt{3}/2$ &1/3 &$2\sqrt{2}/3$ &$17.0$ &$-12.0$ &44.3 &49.0 &0.168\\
\cellcolor{white}{} &\cellcolor{white}{} &\cellcolor{white}{} &\cellcolor{white}{} &$[111]$ &0 &1 &1 &0 &0 &0 &15.0 &15.0 &0.024\\
\hline
\cellcolor{white}{} &\cellcolor{white}{} &\cellcolor{white}{} &\cellcolor{white}{} &$[1\bar{1}1]$ &$\sqrt{3}/2$ &$-1/2$ &1/3 &$2\sqrt{2}/3$ &17.0 &$-7.0$ &24.7 &30.8 &0.072\\
\cellcolor{white}{IX} &\cellcolor{white}$[111]$ &\cellcolor{white}$[\bar{1}10]$ &\cellcolor{white}$[11\bar{2}]$ &$[\bar{1}\bar{1}1]$ &0 &1 &1/3 &$2\sqrt{2}/3$ &0 &13.9 &54.1 &55.9 &0.216\\
\cellcolor{white}{} &\cellcolor{white}{} &\cellcolor{white}{} &\cellcolor{white}{} &$[\bar{1}11]$ &$-\sqrt{3}/2$ &$-1/2$ &1/3 &$2\sqrt{2}/3$ &$-17.0$ &$-7.0$ &24.7 &30.8 &0.072\\
\cellcolor{white}{} &\cellcolor{white}{} &\cellcolor{white}{} &\cellcolor{white}{} &$[111]$ &0 &1 &1 &0 &0 &0 &15.0 &15.0 &0.024\\
\hline
\end{tabular}\label{table}
\end{table*}

\subsection{Electrostatic potential}\label{electrostatics}
In Sec.~\ref{Hamiltonian}, both the band edges and Rashba SOC fields are dependent on the electrostatic potential $\phi(x,y)$. For a translational invariant SM-SC hybrid nanowire schematically shown in Fig.~\ref{Fig:device}(a), $\phi(x,y)$ is determined by the 2D Poisson equation
\begin{equation}
\nabla\cdot\left(\varepsilon_0\varepsilon_r(x,y)  \nabla\phi(x,y)\right)=-[\rho_e(x,y)+\rho_h(x,y)+\rho_\textrm{imp}(x,y)],\label{Poisson}
\end{equation}
where $\varepsilon_0$ is the vacuum dielectric constant, $\varepsilon_r(x,y)$ is the relative dielectric constant taking different values inside each material, i.e., $\varepsilon_r^{\textrm{PbTe}}=1350$, $\varepsilon_r^{\textrm{HfO}_2}=7.5$, $\rho_{e(h)}(x,y)$ represents the mobile electron (hole) density in the conduction (valence) band of the nanowire, and $\rho_\textrm{imp}(x,y)$ denotes the charge density carried by the intrinsic impurities in the SM. Besides, positive surface charge \cite{olsson1996charge,degtyarev2017features} has recently been taken into account to study its impacts on the electronic properties of InAs and InSb based hybrid nanowires \cite{wojcik2018tuning,winkler2019unified,escribano2019effects,escribano2020improved,woods2020subband,liu2021electronic}. Here, we neglect possible surface charge in PbTe nanowires as it does not qualitatively affect the results in this work.

As shown in Fig.~\ref{Fig:device}(b), Poisson equation (\ref{Poisson}) is solved within a large enough 2D squared box enclosing the nanowire and its surrounding regions. We set the box size as $w=300$ nm and the thicknesses of the Pb shell and the dielectric layer as $h=7$ nm and $d=20$ nm, respectively. Neumann boundary conditions are imposed on the left, right, and top sides of the squared box \cite{woods2020subband}. Dirichlet boundary condition $\phi=V_{bg}$ is imposed on the bottom of the dielectric layer while $\phi=-W/e$ on the surface of Pb shell, with $W$ being the work function difference between PbTe and Pb. A positive $W$ is crucial for realizing the superconducting proximity effect, as in this case the conduction band edge near the SM-SC interface bends downward to form a potential well confining electrons \cite{mikkelsen2018hybridization,antipov2018effects}. In fact, the exact value of $W$ of a  SM-SC hybrid nanowire is unknown as it depends on the device details. We use a conservative estimation of $W=80$ meV, according to that the electron affinity of bulk PbTe is 4.6 eV \cite{hoang2007theoretical} and the work function of bulk Pb is 4.25 eV \cite{michaelson1977work}.

We use the Thomas-Fermi approximation for a 3D electron gas to determine $\rho_{e}(x,y)$ and $\rho_{h}(x,y)$ inside the PbTe nanowire
\begin{equation}
\rho_{e}(x,y)=\frac{-4e\left\{2m_{d}^{e}[-E_c(x,y)]\Theta[-E_c(x,y)]\right\}^{3/2}}{3\pi^{2}\hbar^3},\label{Thomas-Fermi-electron}
\end{equation}
\begin{equation}
\rho_{h}(x,y)=\frac{4e\left\{2m_{d}^{h}E_v(x,y)\Theta[E_v(x,y)]\right\}^{3/2}}{3\pi^{2}\hbar^3},\label{Thomas-Fermi-hole}
\end{equation}
where $e$ is the modulus of the electron charge, $m_{d}^{e/h}=(m_{l}^{e/h})^{1/3}(m_{t}^{e/h})^{2/3}$ is the density-of-states effective mass of SMs having ellipsoidal energy surfaces \cite{lee2016thermoelectrics}, $E_c(x,y)=-e\phi(x,y)$ and $E_v(x,y)=E_c(x,y)-E_g$ are the conduction and valence band edge, respectively, of the PbTe nanowire, and $\Theta$ denotes the Heaviside step function corresponding to the Fermi-Dirac distribution at zero temperature. The prefactor $4$ takes into account the fourfold valley degeneracy of bulk PbTe. Substituting Eqs.~(\ref{Thomas-Fermi-electron}) and (\ref{Thomas-Fermi-hole}) into Eq.~(\ref{Poisson}) the electrostatic potential $\phi(x,y)$ can be solved using the finite element method.
We note that the $\phi(x,y)$ obtained with the Thomas-Fermi approximation is independent of the nanowire's growth direction. This approximation has been used and justified in studying the electrostatics of InAs and InSb based SM-SC hybrid nanowires \cite{wojcik2018tuning,winkler2019unified,mikkelsen2018hybridization,escribano2019effects,escribano2020improved,liu2021electronic}.

\subsection{1D effective model of PbTe-Pb hybrid nanowires}\label{effectivemodel}
Transport experiments have demonstrated the superconducting proximity effect in SM-SC hybrid structures comprising IV-VI SMs and $s$-wave SCs, such as PbTe-In contact \cite{grabecki2010contact} and Pb$_{0.5}$In$_{0.5}$-PbS-Pb$_{0.5}$In$_{0.5}$ Josephson junction \cite{kim2017strong}. We infer that the proximity induced $s$-wave superconducting pairing in IV-VI SMs occurs in intra-valley rather than inter-valley carriers, as only the former ones possess opposite spin and momentum, e.g., carriers at two opposite L points of the Brillouin zone. Inter-valley couplings mediated by Pb are allowed for high-energy electrons outside the superconducting gap $\Delta_\textrm{Pb}$ of Pb, however, they could be safely neglected when the induced superconducting gap $\Delta_\textrm{ind}$ is far smaller than $\Delta_\textrm{Pb}$, as Majorana physics is associated to the low-energy electrons inside $\Delta_\textrm{ind}$.

Essentially, the magnitude of $\Delta_{\textrm{ind}}$ depends on both the parent gap $\Delta$ of the SC and the SM-SC coupling strength $\Gamma$. Based on the ratio $\Gamma/\Delta$ one can define the strong ($\Gamma/\Delta\gg 1$), intermediate ($\Gamma/\Delta\sim 1$), and weak ($\Gamma/\Delta\ll 1$) coupling regimes. In principle, with the molecular beam epitaxy technique the coupling strength $\Gamma$ can be tailored by growing a buffer layer with a specified thickness between the SM and SC. In InAs-Al and InSb-Al hybrid nanowires, without a thin buffer layer in between, the induced gap $\Delta_{\textrm{ind}}$ at zero magnetic field can reach about $0.2-0.3$ meV, see, e.g., Refs.~\onlinecite{deng2016majorana,vaitiekenas2018effective,zhang2021large}, which is comparable to the parent gap of Al and is considered as a result of a strong SM-SC coupling. However, theories suggest that a strong SM-SC coupling reduces the SOC and the effective Land\'{e} $g$ factors of the SM nanowires by the renormalization effects from the SC \cite{stanescu2011majorana,cole2015effects,stanescu2017proximity,reeg2018metallization}. Moreover, a strong SM-SC coupling can also lead to trivial zero energy modes due to the induced disorder effect in SM by the intrinsic disorder in SC \cite{hui2015bulk,cole2016proximity,liu2018impurity}. This conundrum would be circumvented by using Pb as the SC. Because of the sizable parent gap $\Delta_\textrm{Pb}=2.73$ meV, a weak coupling is sufficient to maintain a $\Delta_{\textrm{ind}}\sim0.2$ meV without the unwanted side effects. For this reason and to neglect the inter-valley couplings mediated by Pb mentioned above, we focus on the weak coupling regime.

Another relevant issue worth mentioning in SM-SC hybrid nanowires is the orbital effects induced by the magnetic field \cite{nijholt2016orbital,winkler2017orbital,winkler2019unified}. It was found that the orbital effects break the chiral symmetry and prevent the appearance of MZMs whenever the magnetic field is not aligned with the wire axis, and moreover, orbital effects suppress the topological gap and increase the coherence length \cite{nijholt2016orbital}. For the topological properties of PbTe-Pb, we focus on thin nanowires with a side length of $l=40$ nm, available in a recent experiment \cite{schellingerhout2021growth}, applied by a large negative $V_{bg}$, which depletes electrons from the gate side, such that the size of the effective cross section is about 10 nm corresponding to the magnetic length of a 6.5 T parallel magnetic field. In this situation, the orbital effects can be neglected and the strong confinement effect makes the energy separations between neighboring subbands, if not occasionally degenerate, much lager than $\Gamma$ such that the decoupled-band approximation \cite{stanescu2013superconducting,stanescu2017proximity} is applicable. As a result, the BdG Hamiltonian describing a PbTe-Pb hybrid nanowire reads $H_\textrm{BdG}=\sum_n H_n$,
with
\begin{eqnarray}
H_n&=&\bigg(\frac{\hbar^{2}k_{z}^{2}}{2m_{z,n}^{\ast}}-\mu_n+\sum_{u}\alpha_{u,n}\sigma_{u}k_{z}\bigg)\tau_z\notag\\
&&+\frac{1}{2}\mu_{B}B_{z}\sum_{u}g^\ast_{zu}\sigma_{u}+\Delta_{\textrm{ind},n}\tau_x,\label{1Dmodel}
\end{eqnarray}
where $H_n$ is the projected 1D effective Hamiltonian associated to the $n$th transverse subband of the PbTe nanowire with \cite{wojcik2018tuning,stanescu2013superconducting}
\begin{eqnarray}
\alpha_{u,n}&=&\int^{l/2}_{-l/2} dx \int^{l/2}_{-l/2} dy \Omega_{uz}(x,y)|\psi_n(x,y)|^2,\label{anu}\\
\Gamma_n&=&\lambda\int^{l/2}_{-l/2} dy |\psi_n(-l/2,y)|^2,\\
\Delta_{\textrm{ind},n}&=&\frac{\Gamma_n\Delta}{\Gamma_n+\Delta}\approx\Gamma_n, \label{induced_gap}
\end{eqnarray}
where $\mu_n=-E_n$, $E_n$ and $\psi_n(x,y)$ are the $n$th eigen energy and eigen wavefunction, respectively, of the 2D Hamiltonian $H^{\textrm{wire}}$ (see Eq.~(\ref{hwire})) with $k_z=0$ and $B_z=0$. $\Omega_{uz}(x,y)$ is defined in Eq.~(\ref{soccoefficient}) and $\lambda=\nu_F|t_\textrm{SM-SC}|^2$ is a constant that depends on the surface density of states $\nu_F$ of the SC in the normal state and the transparency $t_\textrm{SM-SC}$ of the SM-SC interface. In numerical calculations, we set $\lambda=100$ meV such that the $\Delta_\textrm{ind,n}$ of PbTe-Pb hybrid nanowires can reach about 0.3 meV at a sufficiently negative $V_{bg}$.

Whether there exists topological trivial or nontrivial zero-energy solutions to $H_n$ at a given $B_z$ can be answered by analyzing the characteristic polynomial $p_n(k_z,E)=\textrm{det}(H_{n}-E)$ at $E=0$ \cite{lutchyn2010majorana,rex2014tilting},
\begin{eqnarray}
p_n(  k_{z},0)&=&\bigg[  \bigg(  \frac{\hbar^{2}k_{z}^{2}}{2m_{z,n}^{\ast}}-\mu_n\bigg)  ^{2}-\alpha_n^{2}k_{z}^{2}+\Delta_n^{2}-E_{Z}^{2}\bigg]  ^{2}\notag\\
&&+4\alpha_n^{2}(  \Delta_n^{2}-E_{Z}^{2}\cos^{2}\theta_n)  k_{z}^{2}, \label{characteristic}
\end{eqnarray}
where $E_{Z}=g^\ast_z \mu_{B} B_{z}/2$, $\alpha_n=(\alpha_{x,n}^2+\alpha_{y,n}^2+\alpha_{z,n}^2)^{1/2}$, and $\cos\theta_n=(  \alpha_{nx}g^\ast_{zx}+\alpha_{ny}g^\ast_{zy}+\alpha_{nz}g^\ast_{zz})/\alpha_n g^\ast_{z}$, with $\theta_n$ being the angle between the vectors of the SOC and Zeeman fields.
We find that Eq.~(\ref{characteristic}) is identical to the one of the Lutchyn-Oreg model with isotropic Land\'{e} factors and a tilting magnetic field \cite{rex2014tilting}. Indeed, our Zeeman term in Hamiltonian (\ref{1Dmodel}) can be mapped to the latter's form $\frac{1}{2}g^\ast_z\mu_B\sum_u B_u\sigma_u$ with $B_u=B_zg^\ast_{zu}/g^\ast_z$. According to Ref.~\onlinecite{rex2014tilting}, there are two critical magnetic fields for $H_n$ as
\begin{eqnarray}
B_{c1,n}&=&\frac{\sqrt{\Delta_{\textrm{ind},n}^2+\mu_n^2}}{g^\ast_z\mu_B/2},\label{Bc1}\\
B_{c2,n}&=&\frac{\Delta_{\textrm{ind},n}/|\cos\theta_n|}{g^\ast_z\mu_{B}/2}.\label{Bc2}
\end{eqnarray}
Specifically, if $B_{c1,n}<B_{c2,n}$, as $B_z$ increases the bulk superconducting gap closes and reopens at $B_z=B_{c1,n}$ accompanied by the emergence of a pair of MZMs. Upon further increasing $B_z$, the MZMs persist till $B_z=B_{c2,n}$, at which the bulk superconducting gap eventually closes completely, signaling the entrance to the metallic phase. In the opposite case of $B_{c1,n}>B_{c2,n}$, the topological superconducting phase is inaccessible by increasing $B_z$.

For a uniform PbTe-Pb nanowire with a finite length, we discretize Hamiltonian (\ref{1Dmodel}) onto a 1D lattice with a spacing of 0.5 nm that is sufficiently smaller than the relevant Fermi wavelength. Using the Kwant package \cite{groth2014kwant} we can calculate both the BdG energy spectrum and the tunneling spectroscopy as a function of $B_z$ and $V_{bg}$. A delta-like potential barrier with a width of 1 nm and a height of 1000 meV is added to the left end of the hybrid nanowire for calculating tunneling conductance.

Under the decoupled-band approximation, the BdG energy spectrum, tunneling spectroscopy, and phase diagram of a hybrid PbTe-Pb nanowire can be constructed by taking into account the contributions from all independent subbands of the four valleys. Particularly, at a given $V_{bg}$, the hybrid nanowire is in the metallic phase if $B_z>\textrm{min}\{B_{c2,n}\}$ and otherwise is in the superconducting phase. The latter can be divided into topological trivial and nontrivial superconducting phases according to an integer $N$ characterizing the number of MZM pairs. An odd (even) $N$ indicates a topological nontrivial (trivial) superconducting phase \cite{stanescu2011majorana}. With the decoupled-band approximation, the integer $N$ can be obtained by simply counting the number of subbands fulfilling the condition $B_{c1,n}<B_z<B_{c2,n}$, under which a pair of MZMs are hosted by the $n$th subband as mentioned above. Consequently, for a PbTe-Pb hybrid nanowire at $B_z<\textrm{min}\{B_{c2,n}\}$,
\begin{equation}
N(B_z)=\sum_n{} \Theta(B_z-B_{c1,n}).\label{Nindex}
\end{equation}
In numerical calculations, we truncate $n$ by considering only subbands with $\mu_n>-7$ meV, as the other subbands well above the Fermi level are empty and thus irrelevant.

Before closing this section, we remark that Pb element has a strong intrinsic atomic SOC that may induce additional SOC in the SMs coupled to Pb by proximity effect \cite{nadj2014observation,li2014topological}. Moreover, a theoretical work \cite{hui2015majorana} suggested that the interplay of the atomic SOC and the $s$-$p$ orbital hybridization of the Cooper pairs in superconductor Pb would give rise to an effective $p$-wave pairing in the SMs coupled to Pb. For simplicity, these effects are not taken into account in this work.

\section{Numerical results and discussion}\label{results}
\subsection{Gate tunability, dispersions, and impurity sensitivity of PbTe nanowires}\label{dispersions}
\begin{figure}[t!]
\includegraphics[width=\columnwidth]{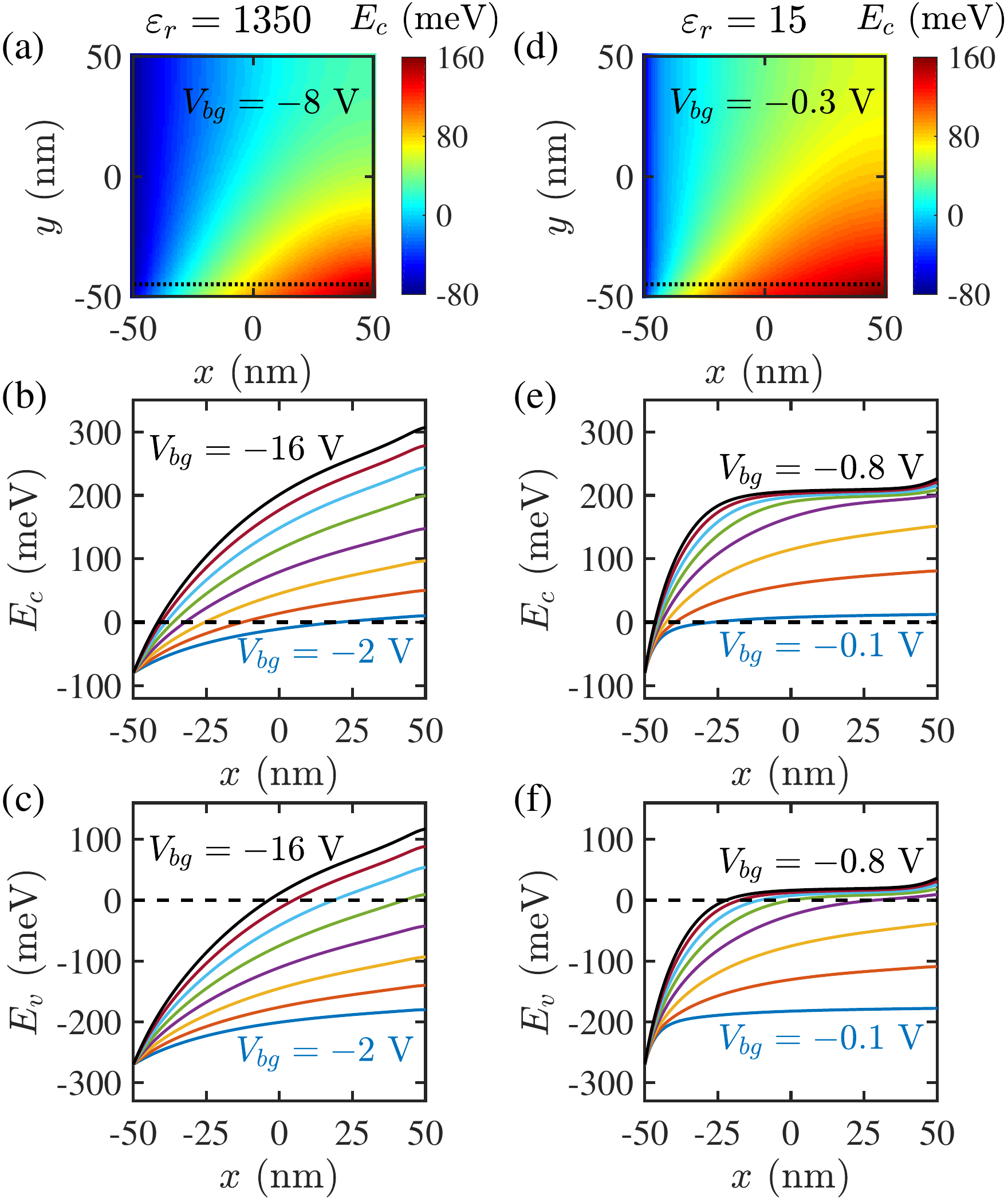}
\caption{(a) Conduction band edge $E_c(x,y)$ of a squared PbTe nanowire with side length $l=100$ nm at back-gate voltage $V_{bg}=-8$ V. [(b) and (c)] Horizontal line cuts of conduction ($E_c$) and valence ($E_v$) band edges, respectively, along the dotted line in (a) at different $V_{bg}$ that are uniformly varied. The dashed lines mark the Fermi level. [(d)--(f)] Results calculated with the same parameters as those of (a)-(c) except that the dielectric constant of the nanowire is changed to 15 that is realistic for InAs and InSb. The associated values of $V_{bg}$ are indicated in each figure.}\label{Fig:bandedges}
\end{figure}

For the device shown in Fig.~\ref{Fig:device}(a), the back gate is used to tune the electrostatic potential $\phi(x,y)$ and hence the conduction and valence band edges $E_c(x,y)$, $E_v(x,y)$ of the nanowire. In Fig.~\ref{Fig:bandedges}, we show the gate tunability of PbTe nanowires (side length $l=100$ nm) with a huge dielectric constant $\varepsilon_r=1350$ (left panel) and with an assumed $\varepsilon_r=15$ (right panel) that is realistic for InAs and InSb nanowires. The striking contrast between the two panels indicates that the band edge profiles and their responses to the back-gate voltage $V_{bg}$ strongly depend on the dielectric constant of the nanowire. Typical contour maps of $E_c(x,y)$ at negative $V_{bg}$ are shown in Figs.~\ref{Fig:bandedges}(a) and \ref{Fig:bandedges}(d). In both figures, at the PbTe-Pb interface ($x=-50$ nm) $E_c(x,y)$ is pined at $-80$ meV due to the work function difference $W=80$ meV between PbTe and Pb. Away from the PbTe-Pb interface $E_c(x,y)$ increases to above the Fermi level due to the negative $V_{bg}$ applied to the dielectric layer. Consequently, near the PbTe-Pb interface a triangle-like potential well is created, which is wider and smoother for $\varepsilon_r=1350$ than for $\varepsilon_r=15$. In Figs.~\ref{Fig:bandedges}(b) and \ref{Fig:bandedges}(e), we show the horizontal line cuts of $E_c(x,y)$ along the dotted lines in Figs.~\ref{Fig:bandedges}(a) and \ref{Fig:bandedges}(d), respectively, at different $V_{bg}$ that are uniformly varied. As $V_{bg}$ decreases, $E_c(x)$ is progressively tuned and the potential well becomes narrower and steeper for $\varepsilon_r=1350$. On the contrary, $E_c(x)$ tends to saturate and so is the potential well for $\varepsilon_r=15$. Such a difference is attributed to the different strengths of screening on the back gate arising from the holes populated in the valence band. To show this effect, the corresponding responses of the valence band edges $E_v(x)$ to $V_{bg}$ are shown in Figs.~\ref{Fig:bandedges}(c) and \ref{Fig:bandedges}(f). They have identical profiles as $E_c(x)$ but differs from the latter by the value of the band-gap $E_g=190$ meV of PbTe. In Fig.~\ref{Fig:bandedges}(f), $V_{bg}$ hardly tunes the $E_v(x)$ once part of it slightly exceeds the Fermi level to populate holes, which screen the gate very effectively due to $\varepsilon_r=15$. Whereas, in Fig.~\ref{Fig:bandedges}(c), the screening from holes is negligible as a result of the huge dielectric constant $\varepsilon_r=1350$ of PbTe.
We note that holes are detrimental as they are localized far away from the PbTe-Pb interface therefore experience no superconducting proximity effect \cite{woods2020subband}. Thus, in PbTe based hybrid nanowire experiments the $V_{bg}$ should not be too negative, otherwise a soft gap can be induced by conductive holes. Comparing Figs.~\ref{Fig:bandedges}(b) and \ref{Fig:bandedges}(e), the conduction band edge in PbTe being smoother than that in InAs and InSb nanowires implies that the electrons in the former are less efficiently depleted than the latter. In order to reach the single or fewer subband occupation regime that is preferred for engineering MZMs \cite{woods2020subband}, one can reduce the diameter of PbTe nanowires to less than 40 nm, which has been reported in a recent experiment \cite{schellingerhout2021growth}.

\begin{figure}[t!]
\centering
\includegraphics[width=\columnwidth]{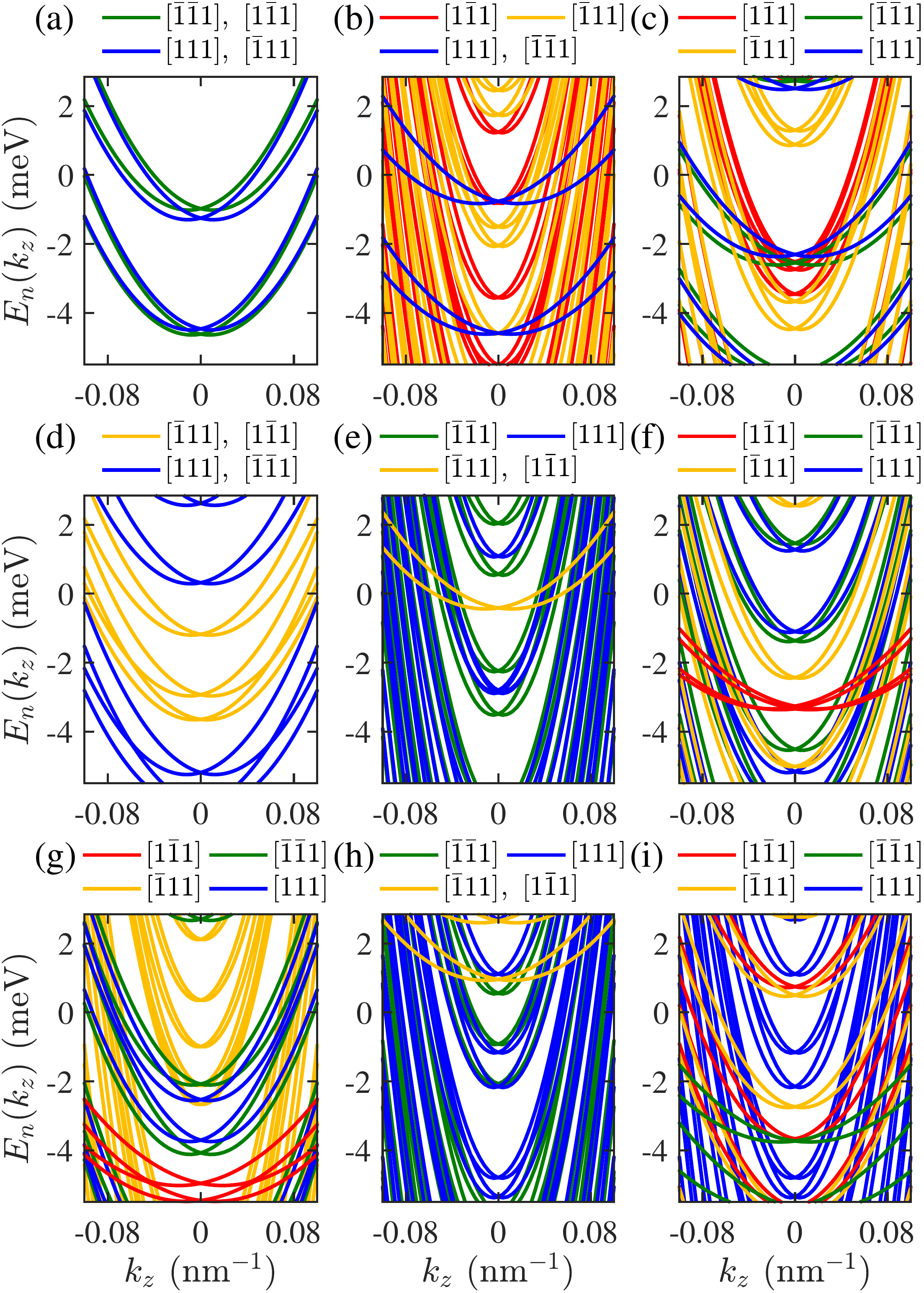}
\caption{Dispersions of the nine PbTe nanowires shown in Fig.~\ref{Fig:ninewires} with side length $l=100$ nm at $V_{bg}=-8$ V and $B_z=0$. Valleys in (a) and (d) are doubly degenerate due to a symmetry protection. Whereas, valley degeneracy is partly lifted in (b), (e), and (h) and is completely lifted in the remaining figures.}\label{Fig:dispersion}
\end{figure}

\begin{figure}[t!]
\centering
\includegraphics[width=\columnwidth]{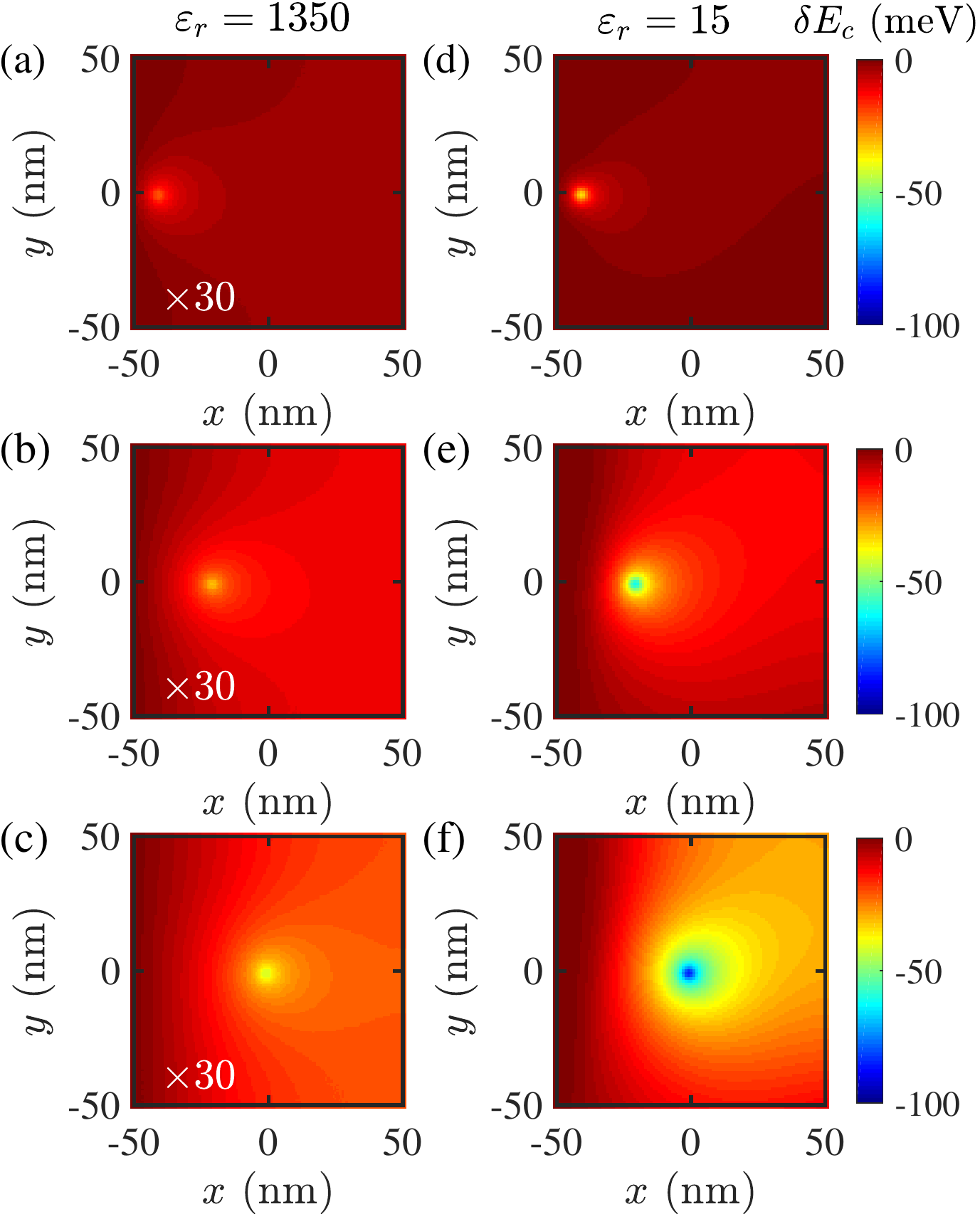}
\caption{The left and right panels show the changes of the conduction band edge $\delta E_c(x,y)$ with respect to Figs.~\ref{Fig:bandedges}(a) and \ref{Fig:bandedges}(d), respectively, induced by a uniform and infinitely long cylinder charged impurity with charge density of $\rho_\textrm{imp}=10^{19} e/cm^3$ and radius of 2 nm centered at different position $(x_{\textrm{imp}},y_{\textrm{imp}})$. [(a) and (d)] $(x_{\textrm{imp}},y_{\textrm{imp}})=(-40,0) ~\textrm{nm}$, [(b) and (e)] $(x_{\textrm{imp}},y_{\textrm{imp}})=(-20,0)~\textrm{nm}$, and [(c) and (f)] $(x_{\textrm{imp}},y_{\textrm{imp}})=(0,0)~\textrm{nm}$. Notably, $\delta E_c$ in the left panel are about 50 times smaller than those in the right, indicating that the huge dielectric constant of PbTe nanowire makes it less sensitive to charged impurity compared to InAs and InSb nanowires.}\label{Fig:diff_phiwithimpurity}
\end{figure}

\begin{figure}[t!]
\centering
\includegraphics[width=\columnwidth]{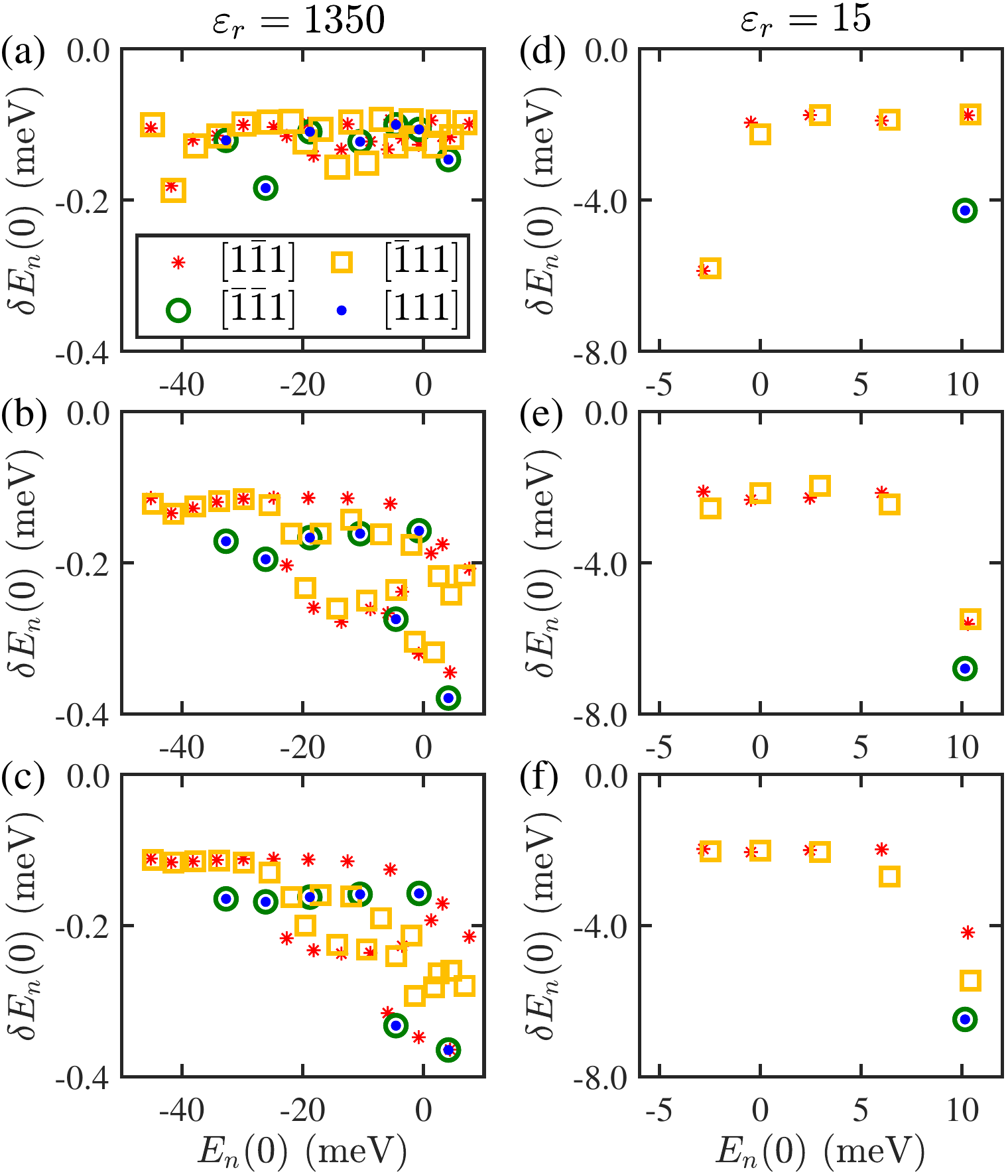}
\caption{Changes of the dispersion at $k_z=0$ of wire case II, which result from the changes of the conduction band edge shown in Fig.~\ref{Fig:diff_phiwithimpurity}. Different symbols denote the subbands belonging to the four valleys indicated in (a). Results of valley $[\bar{1}\bar{1}1]$ and $[111]$ coincide as these two valleys are degenerate.}\label{Fig:dispersionwithimpurity}
\end{figure}

Substituting the $E_c(x,y)$ shown in Fig.~\ref{Fig:bandedges}(a) and the relevant parameters listed in Table \ref{table} into Hamiltonian (\ref{hwire}), the dispersions corresponding to the nine PbTe nanowire orientations shown in Fig.~\ref{Fig:ninewires} are calculated at $B_z=0$, as shown in Fig.~\ref{Fig:dispersion}. Because of the Rashba SOC, the dispersions of all subbands behave as shifted parabolas as those of InAs and InSb nanowires. In each figure, subbands in different colors belong to independent valleys as indicated at the top. For valleys with larger axial effective electron mass $m_z^*$, which are listed in Table \ref{table}, the associated subbands have larger curvature radiuses at the band bottoms and generally larger subband separations. Remarkably, the valleys in Figs.~\ref{Fig:dispersion}(a) and \ref{Fig:dispersion}(d) are doubly degenerate, whereas, the valley degeneracy is partially lifted in Figs.~\ref{Fig:dispersion}(b), \ref{Fig:dispersion}(e), and \ref{Fig:dispersion}(h) and is completely lifted in Figs.~\ref{Fig:dispersion}(c), \ref{Fig:dispersion}(f), \ref{Fig:dispersion}(g), and \ref{Fig:dispersion}(i). The valley degeneracy can be promptly identified by inspecting the valley orientations shown in Fig.~\ref{Fig:ninewires}. To be specific, two valleys are degenerate if the projections of their longitudinal axes onto the $x$-$y$ plane are identical. This can be understood by a symmetry analysis of Hamiltonian (\ref{hwire}). For a general nonuniform $E_c(x,y)$, the Hamiltonian is invariant under the replacement $\alpha\rightarrow \pi-\alpha$, $k_z\rightarrow -k_z$, $\sigma_x\rightarrow -\sigma_x$, $\sigma_y\rightarrow -\sigma_y$, which implies that a pair of valleys that are mirror symmetric about the x-y plane are degenerate. Furthermore, in a special case where $E_c(x,y)$ is uniform or $E_c(x,y)=E_c(-x,y)$, Hamiltonian (\ref{hwire}) is invariant under the replacement $\alpha\rightarrow 2\pi-\alpha$, $x\rightarrow -x$, $\sigma_y\rightarrow -\sigma_y$, $\sigma_z\rightarrow -\sigma_z$, which implies that a pair of valleys that are mirror symmetric about the y-z plane are also degenerate. This symmetry is intentionally broken in the device shown in Fig.~\ref{Fig:device}(a) as $E_c(x,y)\ne E_c(-x,y)$ (see Fig.~\ref{Fig:bandedges}(a)). Basically, all the  nondegenerate valleys of a particular PbTe nanowire can be exploited to engineer MZMs. Therefore, the multivalley character of PbTe probably makes it easier than InAs and InSb based hybrid nanowires to enter into a topological superconducting phase when gating subbands successively across the Fermi level. This conjecture is substantiated by the topological phase diagrams shown in Sec.~\ref{MZMs}.

Figures \ref{Fig:diff_phiwithimpurity} and \ref{Fig:dispersionwithimpurity} validate the anticipation that PbTe nanowires have a higher tolerance of charged impurity than InAs and InSb nanowires. For a translational invariant nanowire, we consider an infinitely long cylinder charged impurity with a radius of 2 nm. The charge density is assumed uniform and set as $\rho_\textrm{imp}=10^{19}~e/cm^3$. This value is of the same order of the electron density near the PbTe-Pb interface, which is estimated to be $-0.5\times10^{19} e/cm^3$ according to Eq.~(\ref{Thomas-Fermi-electron}), comparable to the ones of InAs nanowires \cite{winkler2019unified}. In Fig.~\ref{Fig:diff_phiwithimpurity}, the left and right panels show the changes of the conduction band edge $\delta E_c(x,y)$ with respect to those shown in Figs.~\ref{Fig:bandedges}(a) and \ref{Fig:bandedges}(d), respectively, induced by the charged impurity centered at three different positions $(x_{\textrm{imp}},y_{\textrm{imp}})=(-40,0)~\textrm{nm}$, $(x_{\textrm{imp}},y_{\textrm{imp}})=(-20,0)~\textrm{nm}$, and $(x_{\textrm{imp}},y_{\textrm{imp}})=(0,0)~\textrm{nm}$. Clearly, around the impurity a potential deep is created, such that negatively charged electrons would be populated there to screen the positively charged impurity. Moreover, the impurity closer to the PbTe-Pb interface has a weaker impact on the conduction band edge, since it is screened by its surrounding electrons confined in the potential well. The comparison between the two panels shows that $\delta E_c$ for $\varepsilon_r=1350$ are about 50 times smaller than those for $\varepsilon_r=15$. Note that in Figs.~\ref{Fig:diff_phiwithimpurity}(a)-\ref{Fig:diff_phiwithimpurity}(c), $\delta E_c$ is multiplied by 30 for clarity. Consequently, as shown in Fig.~\ref{Fig:dispersionwithimpurity}, the changes of the dispersion at $k_z=0$ of wire II for $\varepsilon_r=1350$ are on the order of 0.4 meV, which are substantially smaller than those for $\varepsilon_r=15$ by an order of magnitude. By reducing the charge density to lower than $10^{19}~e/cm^3$ the changes of dispersions are expected to be decreased to below 0.1 meV, which is smaller than the proximity induced superconducting gap. For point-like charged impurities randomly distributed in a SM nanowire, the effective chemical potential of a 1D subband fluctuates along the nanowire \cite{woods2021charge}, which can lead to trivial zero-energy bound states mimicking MZMs \cite{pan2020physical,sarma2021disorder}. Due to the high tolerance of charged impurity in PbTe nanowires, we speculate that PbTe nanowires are superior than InAs and InSb nanowires to possess approximately uniform chemical potentials desired for engineering MZMs.

In Fig.~\ref{Fig:dispersionwithimpurity}, subband separations can be inferred from the differences between the abscissas of neighboring scatters in the same color. Evidently, subband separations in the right panel are larger than the left panel, due to that the potential well in Fig.~\ref{Fig:bandedges}(d) is steeper than that in Fig.~\ref{Fig:bandedges}(a). The subbands with wavefunctions having large weights near the impurity are mostly affected. To be specific, the wavefunctions of subbands with low-lying $E_n(0)$ are mostly localized to the PbTe-Pb interface, thus they are dramatically affected by impurities near the interface [Figs.~\ref{Fig:dispersionwithimpurity}(a) and \ref{Fig:dispersionwithimpurity}(d)] but are weakly affected by impurities far away from the interface [Figs.~\ref{Fig:dispersionwithimpurity}(b), \ref{Fig:dispersionwithimpurity}(c), \ref{Fig:dispersionwithimpurity}(e), and \ref{Fig:dispersionwithimpurity}(f)].

\begin{figure}[t!]
\centering
\includegraphics[width=\columnwidth]{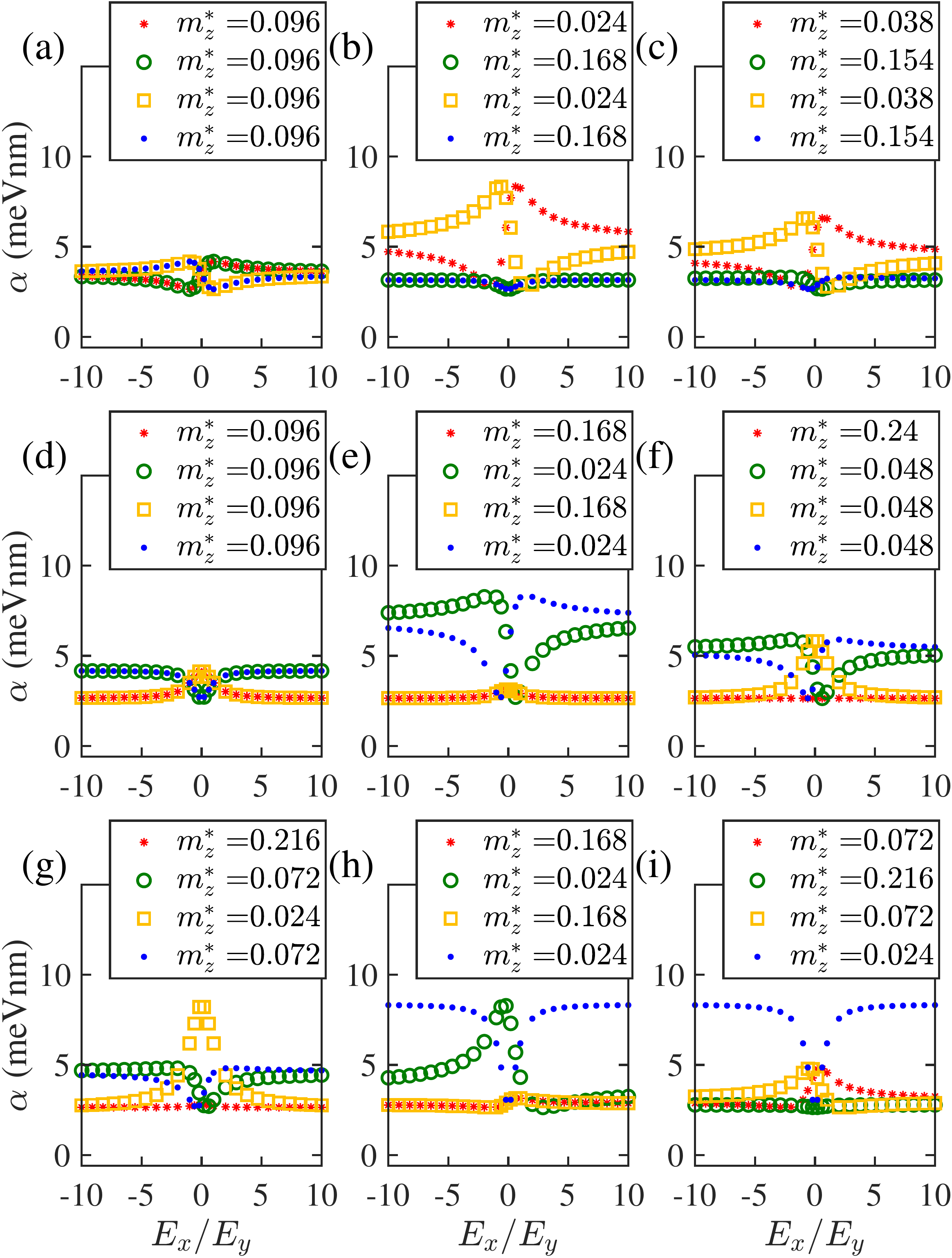}
\caption{Dependence of the SOC strength $\alpha$ of the four valleys of PbTe nanowires grown along the nine different orientations shown in Fig.~\ref{Fig:ninewires} on the ratio of uniform electric fields $E_x$ and $E_y$ with the modulus $E=(E_x^2+E_y^2)^{1/2}$ being fixed at 1 mV/nm. $\alpha=(\alpha_x^2+\alpha_y^2+\alpha_z^2)^{1/2}$ is calculated from Eqs.~(\ref{soccoefficient}) and (\ref{anu}) and it scales linearly with $E$. These results are independent of the device geometry. In each figure, the axial effective electron masses $m_z^\ast$ of valley $[1\bar{1}1]$, $[\bar{1}11]$, $[\bar{1}\bar{1}1]$, $[111]$ are indicated from top to bottom.}\label{Fig:alphavsp}
\end{figure}

\subsection{SOCs of PbTe nanowires}\label{soc}
To realize topological superconductivity in SM-SC hybrid nanowires a strong Rashba SOC field perpendicular to the Zeeman field is desired. Due to the high anisotropy of the band structure of bulk PbTe, the SOCs of the four valleys of the nine wire orientations shown in Fig.~\ref{Fig:ninewires} are presumably quite different. A natural question is which wire orientation and valley has the optimal SOC. As defined in Sec.~\ref{effectivemodel}, Rashba SOC $\alpha_n=(\alpha_{nx}^2+\alpha_{ny}^2+\alpha_{nz}^2)^{1/2}$ and its components $\alpha_{nu}$ ($u=x,y,z$) of the $n$th subband of a PbTe nanowire depend on the electric fields $E_x(x,y)$, $E_y(x,y)$ in the x and y directions. To gain generic insights on the SOC of PbTe nanowires irrelevant to the device geometry, we assume uniform electric fields $E_x$ and $E_y$. In this case, $\alpha_n$ and $\alpha_{nu}$ are independent of the subband index $n$. In Fig.~\ref{Fig:alphavsp}, we show the SOC strength $\alpha$ of the four valleys of the nine PbTe orientations as a function of the ratio $E_x/E_y$ with the modulus $(E_x^2+E_y^2)^{1/2}$ being fixed at 1 mV/nm. Note that $\alpha$ calculated from Eqs.~(\ref{soccoefficient}) and (\ref{anu}) scales linearly with with $E$. In each figure, the colors of the symbols are identical to those of the valleys shown in Fig.~\ref{Fig:ninewires}, and the superposition of symbols with different colors results from valley degeneracy. The axial effective electron mass $m_z^\ast$ of the four valleys are indicated in each figure for clarity. Note that, for all wire orientations, valleys with smaller $m_z^\ast$ generally exhibit stronger SOCs, despite of their diverse dependence on $E_x/E_y$.

\begin{figure}[t!]
\centering
\includegraphics[width=\columnwidth]{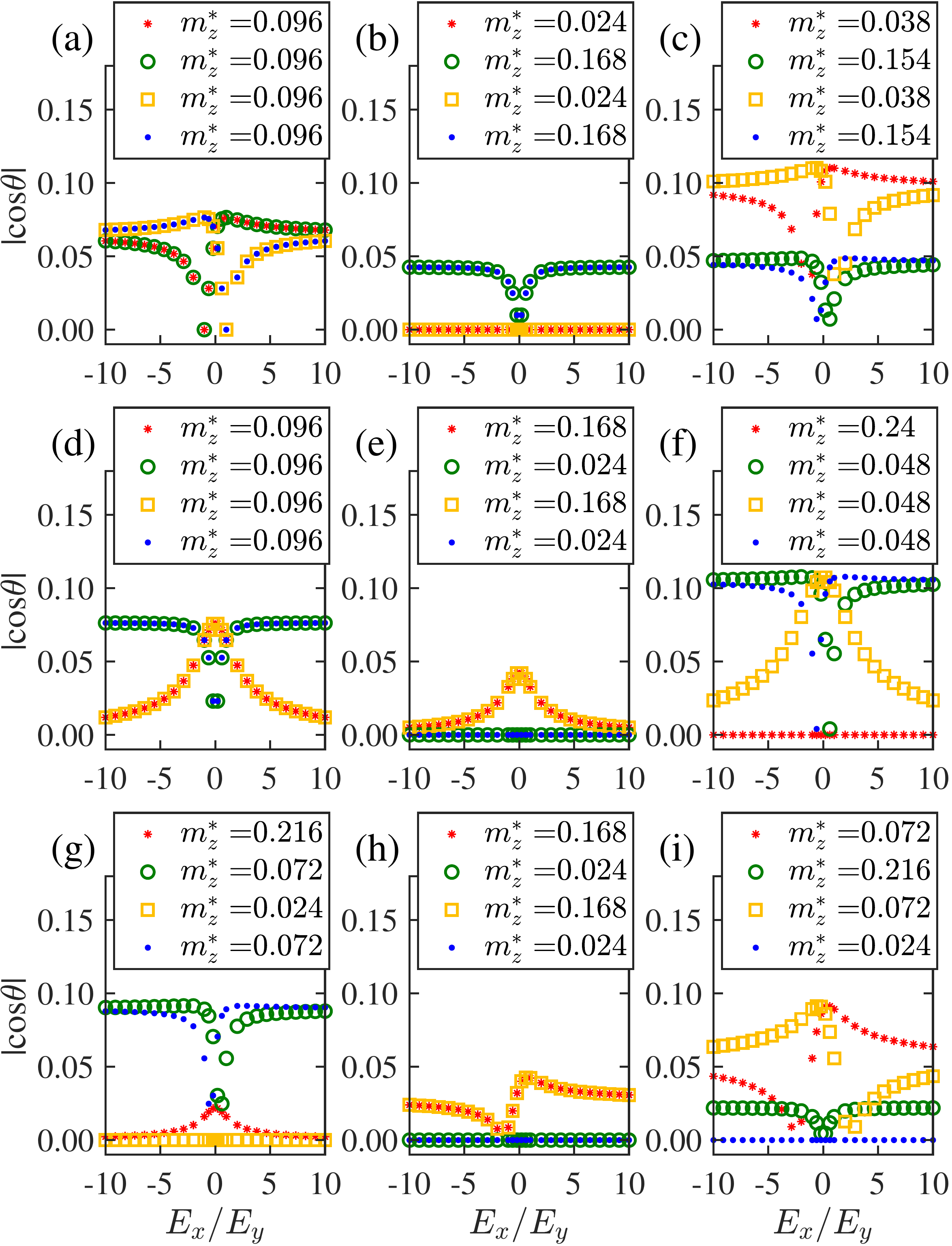}
\caption{Dependence of $\vert\cos\theta\vert$, with $\theta$ being the angle between the vectors of SOC and Zeeman fields, of the four valleys of PbTe nanowires grown along the nine different orientations shown in Fig.~\ref{Fig:ninewires} on the ratio of uniform electric fields $E_x$ and $E_y$. These results are independent of the device geometry. The markers in each figure are the same as those in Fig.~\ref{Fig:alphavsp}.}\label{Fig:costhetavsp}
\end{figure}

Figure \ref{Fig:costhetavsp} shows the corresponding dependence of $\vert\cos\theta\vert$ on $E_x/E_y$, with $\theta$ being the angle between the vectors of the SOC and Zeeman fields, in the presence of a parallel magnetic field. For all valleys, $\vert\cos\theta\vert$ is smaller than 0.12, implying that the SOC and Zeeman fields are nearly orthogonal. Thus a finite $B_z$ may drive nanowires with nondegenerate valleys into the topological superconducting phases. For valleys with $m_z^\ast=0.024~m_e$ and $m_z^\ast=0.24~m_e$ the $\vert\cos\theta\vert$ is pinned exactly at zero despite of the ratio $E_x/E_y$, due to that the associated $g_{zx}^\ast=g_{zy}^\ast=0$ (see Table \ref{table}). The results shown in Figs.~\ref{Fig:alphavsp} and \ref{Fig:costhetavsp} indicate that valleys with $m_z^\ast=0.024$ in wires II, V, VIII, and IX are the optimal candidates for engineering MZMs, as their SOCs are maximum in amplitude and meanwhile are perfectly perpendicular to the Zeeman field. However, the weakness is that their axial effective Land\'{e} factor $g_z^\ast=15$ is the smallest among all valleys. Even so, these values are very close to those of InAs nanowires, i.e., $m^\ast_z=0.023~m_e$ and $g_z^\ast=14.9$ \cite{winkler2019unified}.

\begin{figure}[t!]
\centering
\includegraphics[width=\columnwidth]{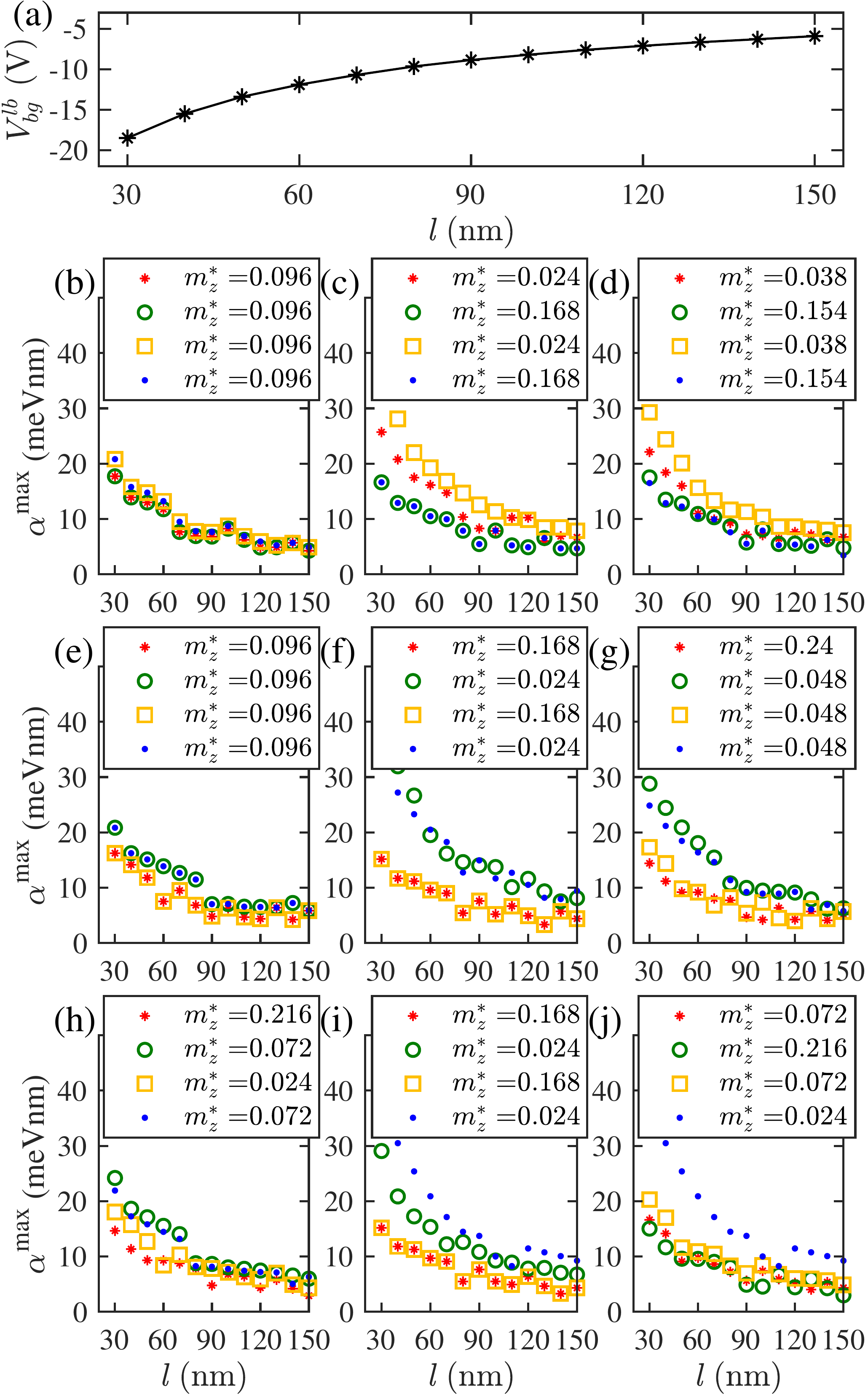}
\caption{(a) Lower bound of the back-gate voltage $V_{bg}^{lb}$ as a function of the side length $l$ of a squared PbTe nanowire. For $V_{bg}$ above $V_{bg}^{lb}$ the entire valence band edge is below the Fermi level such that no hole states are populated. (b)-(j) Side length dependence of the accessible maximum SOC strength of the four valleys of PbTe nanowires grown along the nine different orientations shown in Fig.~\ref{Fig:ninewires}. These results are for the particular device shown in Fig.~\ref{Fig:device}(a) with $V_{bg}=V_{bg}^{lb}$. The markers in each figure are the same as those in Fig.~\ref{Fig:alphavsp}.}\label{Fig:soc}
\end{figure}

Different from the ideal case considered above, in a realistic device model [Fig.~\ref{Fig:device}(a)] the electric fields $E_x(x,y)$ and $E_y(x,y)$ are nonuniform and tuned by the back-gate voltage $V_{bg}$. In this case, the SOC strength $\alpha_n$ depends on the subband index $n$. For a given $V_{bg}$, we focus on the subband of each valley closest to the Fermi level for MZM realization. In Fig.~\ref{Fig:soc} we estimate by numerical simulations the accessible maximum SOC of the nine PbTe wire orientations as a function of the side length $l$ for the particular device geometry shown in Fig.~\ref{Fig:device}(a). Generally speaking, decreasing $V_{bg}$ enhances the electric fields and hence the Rashba SOC inside the PbTe nanowire. However, as noted in Sec.~\ref{dispersions}, detrimental hole states would be populated whenever part of the valence band edge surpasses the Fermi level at too negative $V_{bg}$. As the holes appear at a distance, say $l_h$, away from the PbTe-Pb interface, they can be naturally expelled from the nanowire if $l<l_h$. Indeed, Fig.~\ref{Fig:soc}(a) shows that the lower-bound of the back-gate voltage $V_{bg}^{lb}$ increases monotonously with decreasing $l$. Correspondingly, the accessible maximum SOC strength of each nanowire at $V_{bg}=V_{bg}^{lb}$ becomes stronger for smaller $l$, as shown in Figs.~\ref{Fig:soc}(b)-\ref{Fig:soc}(j). In agreement with Fig.~\ref{Fig:alphavsp}, valleys with smaller axial effective electron mass $m_z^\ast$ generally have larger SOC strengths. For the smallest $m_z^*=0.024$ the SOC strength can reach about $30$ meVnm for $l=40$ nm, see, e.g., in Figs.~\ref{Fig:soc}(f), \ref{Fig:soc}(i), and \ref{Fig:soc}(j). Remarkably, for each wire orientation, though the SOC strengths of the four valleys are quite different the corresponding spin-orbit energies $E_\textrm{SO}=m_z^\ast\alpha^2/2\hbar^2$ are almost comparable, as can be deduced from Fig.~\ref{Fig:soc}.

\subsection{Topological properties of PbTe-Pb hybrid nanowires}\label{MZMs}

\begin{figure}[t!]
\centering
\includegraphics[width=\columnwidth]{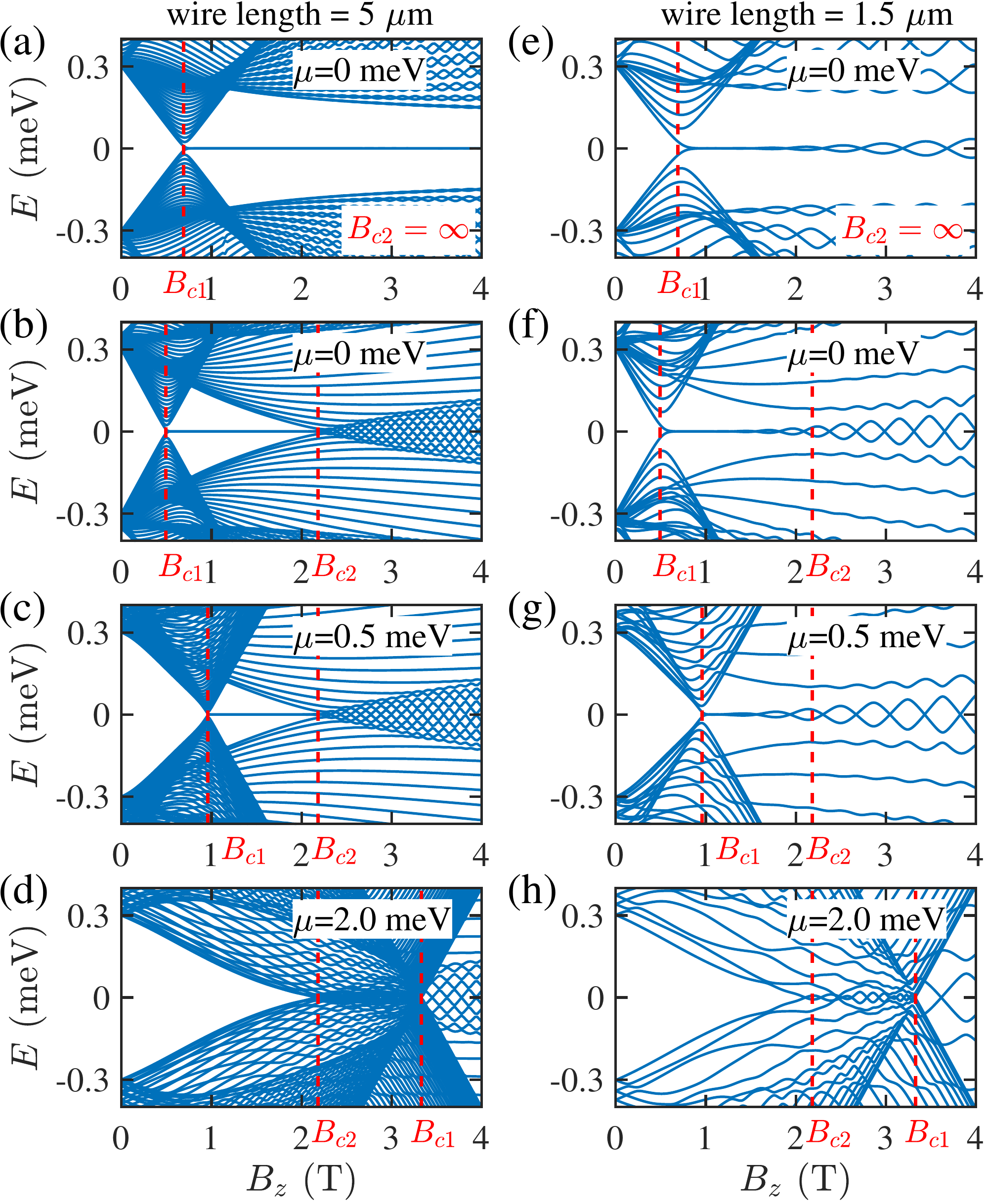}
\caption{Typical BdG energy spectra of the 1D effective Hamiltonian (\ref{1Dmodel}) of PbTe-Pb hybrid nanowire with different parameters. The effective Land\'{e} factors and electron masses are used as those of valley $[1\bar1 1]$ of (a) wire case II and (b)-(d) wire case III as listed in Table \ref{table1}, while the chemical potentials are indicated in each figure. Other parameters used are $\alpha_x=5$ meVnm, $\alpha_y=25$ meVnm, $\alpha_z=0$, and $\Delta_\textrm{ind}=0.3$ meV. $B_{c1}$ and $B_{c2}$ are two critical magnetic fields defined in Eqs.~(\ref{Bc1}) and (\ref{Bc2}), respectively. In (a)-(c), a pair of MZMs exist in the topological superconducting region $B_{c1}<B_z<B_{c2}$. No MZMs exists in (d) because $B_{c2}<B_{c1}$. (e)-(h) are calculated with the same parameters as (a)-(d) except that the wire length is decreased from 5 $\mu m$ to 1.5 $\mu m$. Due to the finite size effects, no gap closing is observed at $B_z=B_{c2}$ in (e)-(g).}\label{Fig:typicalspectrum}
\end{figure}

In Secs.~\ref{dispersions} and \ref{soc}, the Pb shell is only considered to impose a boundary condition of the Poisson equation at the PbTe-Pb interface. We now take into account the superconducting pairing inside PbTe nanowires inherited from the superconducting Pb shell to study the topological properties of PbTe-Pb hybrid nanowires. Figure \ref{Fig:typicalspectrum} shows typical BdG energy spectra of the 1D effective Hamiltonian (\ref{1Dmodel}) with different parameters. The effective Land\'{e} $g$ factors and electron masses are used as those of valley $[1\bar1 1]$ of wire orientation II [Fig.~\ref{Fig:typicalspectrum}(a)] and III [Figs.~\ref{Fig:typicalspectrum}(b)--\ref{Fig:typicalspectrum}(d)], respectively, as listed in Table \ref{table1}, while the chemical potentials $\mu$ are indicated in each figure. Other parameters used are $\alpha_x=5$ meVnm, $\alpha_y=25$ meVnm, $\alpha_z=0$, and $\Delta_\textrm{ind}=0.3$ meV. According to the formulas in Sec.~\ref{effectivemodel}, the angle between the vectors of the SOC and Zeeman fields is calculated as $90^\circ$ for Fig.~\ref{Fig:typicalspectrum}(a) and $76.9^\circ$ for Figs.~\ref{Fig:typicalspectrum}(b)--\ref{Fig:typicalspectrum}(d), and the associated critical magnetic fields ($B_{c1}$, $B_{c2}$) for the four figures are (0.69 T, $\infty$), (0.49 T, 2.18 T), (0.96 T, 2.18 T), and, (3.33 T, 2.18 T), respectively, as marked by the red dashed lines. Clearly, the $B_{c1}$ increases with $\mu$ as shown in Figs.~\ref{Fig:typicalspectrum}(b)--\ref{Fig:typicalspectrum}(d). The spectrum pattern shown in Fig.~\ref{Fig:typicalspectrum}(a) represents the typical one of the Lutchyn-Oreg model \cite{lutchyn2010majorana,oreg2010helical} describing a SM nanowire with an induced $s$-wave pairing and exposed to a parallel magnetic field. As $B_z$ increases, the superconducting bulk gap closes and reopens at the topological phase transition point $B_z=B_{c1}$, beyond which a pair of MZMs persist in the middle of the slowly decayed superconducting bulk gap. In Figs.~\ref{Fig:typicalspectrum}(b)--\ref{Fig:typicalspectrum}(d), as the vectors of the SOC and Zeeman fields are not perpendicular, MZMs survive
only if $B_{c1}<B_z<B_{c2}$. For $B_z$ larger than $B_{c2}$ the superconducting bulk gap closes completely and the metallic phase is arrived. Figures \ref{Fig:typicalspectrum}(b) and \ref{Fig:typicalspectrum}(c) are similar to the spectrum of the Lutchyn-Oreg model with a tilting magnetic field \cite{osca2014effects}, due to that the Zeeman terms of these two models can be mapped exactly to each other as mentioned in Sec.~\ref{effectivemodel}. To reveal the finite-size effects, we decrease the wire length from 5 $\mu m$ to 1.5 $\mu m$ that is realistic in experiments and find that the above energy spectra are modified to Figs.~\ref{Fig:typicalspectrum}(e)--\ref{Fig:typicalspectrum}(h). In Figs.~\ref{Fig:typicalspectrum}(e)--\ref{Fig:typicalspectrum}(g), as $B_z$ increases the MZMs evolve as the Majorana oscillation with an increasing amplitude and period, a result from the increasing overlap between the two Majorana wave functions \cite{sarma2012splitting}. Another distinct finite-size effect in Figs.~\ref{Fig:typicalspectrum}(f) and \ref{Fig:typicalspectrum}(g) is that the second critical magnetic field increases from $B_{c2}=2.18$ T to values larger than 4 T.

\begin{figure}[t!]
\centering
\includegraphics[width=\columnwidth]{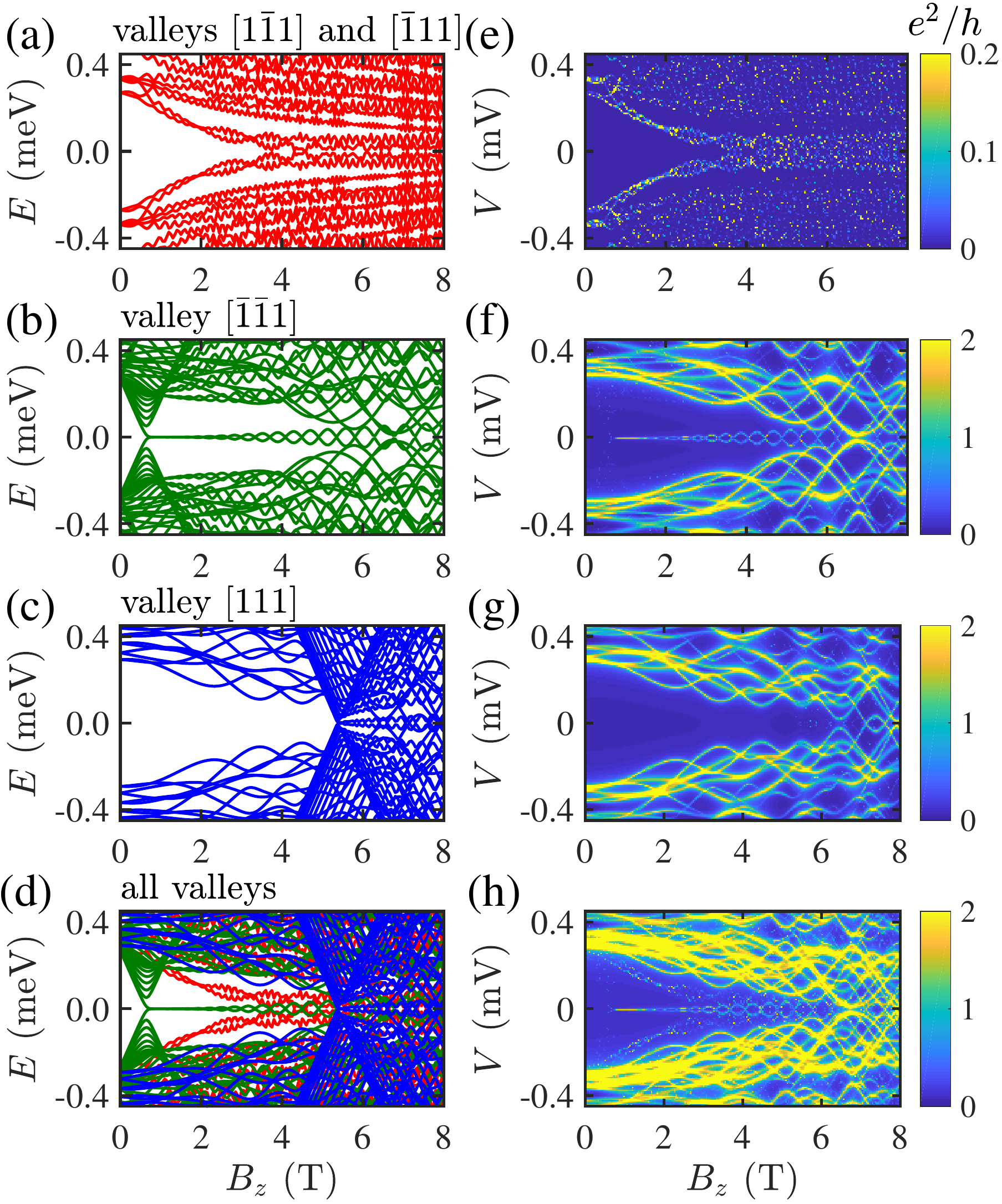}
\caption{BdG energy spectra (left panel) and tunneling conductance maps (right panel) for a PbTe-Pb hybrid nanowire made up of wire case V with a finite length of $2~\mu m$. The results are for the particular device geometry shown in Fig.~\ref{Fig:device}(a) with $l=40$ nm and $V_{bg}=-10.5$ V. (a)-(c) and (e)-(g) show valley-resolved results while (d) and (h) take all valleys into account. Note that valleys $[1\bar{1}1]$ and $[\bar{1}11]$ are degenerate as indicated on top of (a). The MZMs shown in (b) and (d) manifest themselves as sharp quantized zero-bias peaks in (f) and (h), respectively, in a wide scope of $B_z$ ranging from 0.7 to 1.5 T.}\label{Fig:spectrumanddIdV}
\end{figure}

Below, we present the numerical results of the particular device geometry shown in Fig.~\ref{Fig:device}(a) with the side length of PbTe nanowire of $l=40$ nm. Starting from the 1D effective model addressed by Eqs.~(\ref{1Dmodel})-(\ref{induced_gap}), we calculate in Fig.~\ref{Fig:spectrumanddIdV} the BdG energy spectra (left panel) and tunneling spectroscopies (right panel) of the PbTe-Pb hybrid nanowire made up of wire case V with the wire length of 2 $\mu m$. The top three rows are the spectrum of each valley while the last row shows them together. Note that valley $[1\bar{1}1]$ and $[\bar{1}11]$ of wire case V are degenerate. The spectra are associated to all 1D subbands of the nanowire with $\mu_n>-7$ meV. As we can see, the spectra in Fig.~\ref{Fig:spectrumanddIdV}(a) are of the type shown in Fig.~\ref{Fig:typicalspectrum}(d), while those in Figs.~\ref{Fig:spectrumanddIdV}(b) and \ref{Fig:spectrumanddIdV}(c) are of the type shown in Fig.~\ref{Fig:typicalspectrum}(a). At $B_z=0$, while the induced superconducting gaps for the four valleys are slightly different, the smallest one approaches to 0.25 meV that is comparable to the InAs-Al and InSb-Al hybrid nanowires. The back-gate voltage is set as $V_{bg}=-10.5$ V, at which a subband of valley $[\bar{1}\bar{1}1]$ has $\mu_n\approx0$ while the other subbands are well separated from the Fermi level. As a result, as shown in Fig.~\ref{Fig:spectrumanddIdV}(b), a clear phase transition signaled by a gap closing-and-reopening shows up at $B\approx 0.7$ T and a pair of MZMs persist to $B\approx 5$ T protected by a considerable superconducting bulk gap. Because all subbands are independent under the decoupled-band approximation, the total energy and conductance spectrum can be obtained by collecting and summing those of individual subbands, respectively \cite{liu2017andreev}. After superimposing the spectrum of all valleys in Fig.~\ref{Fig:spectrumanddIdV}(d), the topological superconducting region is reduced as the bulk gaps of some topological trivial spectra, especially the red ones, decrease rapidly as $B_z$ increases. Figures \ref{Fig:spectrumanddIdV}(e)-(h) show the corresponding tunneling conductance. Evidently, the MZMs shown in Figs.~\ref{Fig:spectrumanddIdV}(b) and \ref{Fig:spectrumanddIdV}(d) manifest themselves as sharp quantized zero-bias peaks in Figs.~\ref{Fig:spectrumanddIdV}(f) and \ref{Fig:spectrumanddIdV}(h), respectively, for $0.7~\textrm{T}<B_z<1.5~\textrm{T}$. However, the tunneling spectroscopy is a local probe that can not reflect the complete spectra pattern such as the gap closing-and-reopening characteristic, which is associated to spatially extended wave functions in the nanowire \cite{stanescu2012to}.

\begin{figure}[t!]
\centering
\includegraphics[width=\columnwidth]{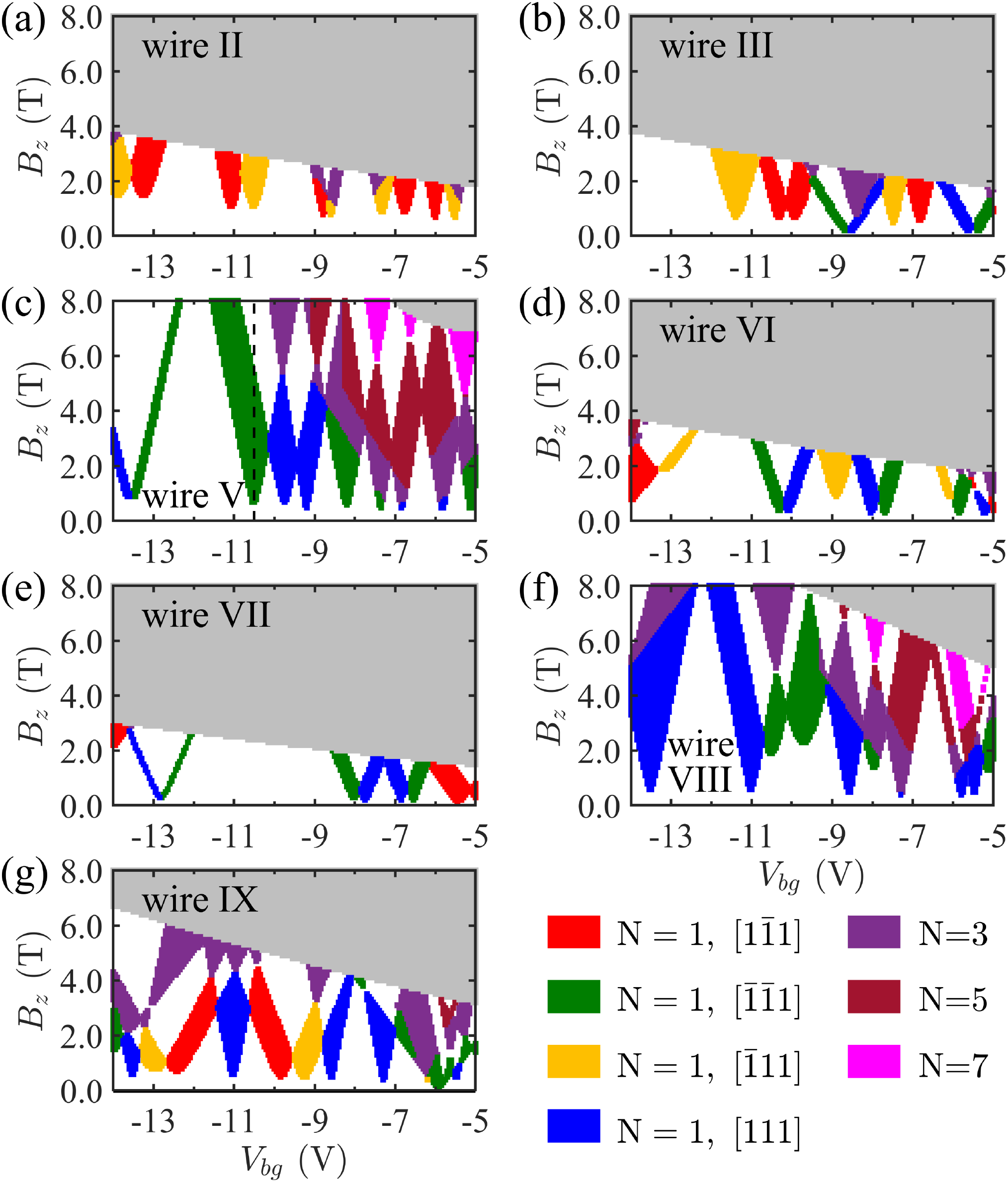}
\caption{Phase diagrams of PbTe-Pb hybrid nanowires made up of PbTe nanowires grown along seven different orientations as indicated. The results are for the particular device geometry shown in Fig.~\ref{Fig:device}(a) with the side length $l=40$ nm. The gray and white regions represent metallic phase and trivial superconducting phase, respectively, while the other colored regions represent topological superconducting phases harboring $N$ pairs of MZMs as indicated in the right bottom. For the $N=1$ phases, the valley from which the MZMs are created is distinguished by different colors. The dashed line in (c) marks the parameter regime explored in Fig.~\ref{Fig:spectrumanddIdV}.}\label{Fig:phasediagramofallwire}
\end{figure}

In Fig.~\ref{Fig:phasediagramofallwire}, we show the topological phase diagrams in the ($V_{bg}$, $B_z$) parameter space of the PbTe-Pb hybrid nanowires made up of PbTe nanowire grown along seven different orientations. The gray regions represent the metallic phase emerging at $B_z$ larger than the minimum of $B_{c2}$ [see Eq.~(\ref{Bc2})] of all subbands. The colored (white) islands represent topological nontrivial (trivial) superconducting phases harboring $N$ pairs of MZMs, with $N$ being an odd (even) number calculated from Eq.~(\ref{Nindex}). The topological superconducting phases of interest are classified by different colors according to the values of $N$ as indicated in the right bottom of the figure. Note that the phases with $N=1$ are further classified by different valley indices to underline from which nondegenerate valley the MZMs are created. In each figure, there are either two or four kinds of $N=1$ phases depending on the valley degeneracy of the associated PbTe nanowire. These phase diagrams explicitly show that the multivalley character of PbTe endows the hybrid nanowires extended topological superconducting phase regions, even for the $N=1$ phases at low $B_z$. The trivial phase diagrams corresponding to wire orientations I and IV are not shown as these two cases would never have a nondegenerate valley in any device geometries due to a symmetry protection, as analyzed in Sec.~\ref{dispersions}, thus are useless for realizing topological superconducting phases. For a special device geometry which has $E_c(x,y)=E_c(-x,y)$, another symmetry analyzed in Sec.~\ref{dispersions} reduces the number of nondegenerate valley from 4 (2) to 2 (0) for wire orientation IX (II and V), therefore the associated topological superconducting phase regions would shrink (disappear). Note that the above phase diagrams would be dramatically modified if the coupling strength between PbTe and Pb is larger than the subband separations such that the decoupled-band assumption we made breaks down. This case is beyond the scope of the present work and thus is not considered here.

\section{Summary and outlook}\label{Summary}
In summary, we have theoretically validated that the recently proposed PbTe-Pb is a promising candidate, potentially better than the existing InAs and InSb based hybrid nanowires for engineering and detecting MZMs. The most attractive property of PbTe is its high tolerance of charged impurity, which is suggested to cause strong disorder leading to trivial zero-energy states in InAs and InSb based hybrid nanowires. Besides, PbTe nanowires can be freely tuned by electrostatic gates without suffering from the screening from holes. These two characteristics of PbTe are due to its huge dielectric constant that is nearly a hundred times larger than InAs and InSb. Different from InAs and InSb which have isotropic Fermi surface, PbTe possesses four highly anisotropic valleys. This multivalley character makes it easier than InAs and InSb to access topological superconducting phases when gating subbands successively across the Fermi level.

We have analyzed PbTe nanowires with nine different growth orientations that are preferred in experiments with high crystalline quality. The effective electron mass, Land\'{e} $g$ factors, and the SOC fields of the four valleys of PbTe nanowires are studied in detail. We found that the axial Land\'{e} $g$ factor and SOC strength of each valley are anti-correlated, which disagrees with previous naive expectation on PbTe nanowires that they can simultaneously have a larger $g$ factor and a stronger SOC than InAs and InSb nanowires. Specifically, for valleys with the smallest (largest) axial effective electron mass $m^\ast_z=0.024~m_e$ ($m^\ast_z=0.24~m_e$), the corresponding $g_z^\ast=15.0$ ($g_z^\ast=59.0$) is the minimum (maximum), while the accessible maximum Rashba SOC strength is $\alpha_\textrm{SOC}^\textrm{max}=10$ meVnm ($\alpha_\textrm{SOC}^\textrm{max}=5$ meVnm) for the particular device model shown in Fig.~\ref{Fig:device}(a) with the side length $l$ being 100 nm. The parameters of the valleys with $m_e^\ast=0.024~m_e$ are comparable to those of an InAs nanowire with a diameter of 100 nm, i.e., $m^\ast_z=0.023~m_e$, $g_z^\ast=14.9$, and $\alpha_\textrm{SOC}^\textrm{max}=10$ meVnm \cite{winkler2019unified}.

The merits of using Pb as the SC have not been investigated comprehensively but are conceivable: (i) Pb has a larger superconducting gap, higher critical temperature, and higher parallel critical magnetic field than those of Al; (ii) Pb can induce a large superconducting gap in PbTe even by a weak coupling between them, which can also relieve the adverse renormalization and induced disorder effects on PbTe from Pb; and (iii) Pb can be grown on PbTe with a clean sharp interface and no interface diffusion. Furthermore, if a thin buffer layer of CdTe is added in between PbTe and Pb, not only the SM-SC coupling strength can be tuned, but also the lattice mismatch between SM-SC can be solved since CdTe is lattice matched with PbTe.

Experimentally, before exploring PbTe-Pb hybrid nanowires, one can perform routinely transport measurements on a bare PbTe nanowires with different growth orientations to study the valley degeneracy, the electron mobility, and the SOC strength with or without an external magnetic field applied in different directions, following previous experiments on bare InAs and InSb nanowires, see, e.g., Refs.~\onlinecite{ford2012observation,van2013quantized,weperen2015spin,kammhuber2016conductance,heedt2016ballistic,kammhuber2017conductance,heedt2017signatures,estrada2018split}. Recently, preliminary electronic transport measurements in single-crystalline PbTe nanowire MOSFET devices have been carried out but still needs further optimization \cite{kamphuis2021towards,schellingerhout2021growth}. Additionally, \textit{ex situ} deposited aluminum superconductor Josephson junctions comprising a PbTe nanowire (Al-PbTe-Al) are also fabricated \cite{schlatmann2021josephson}, in which multiple Andreev reflections and Shapiro steps are explicitly observed. Notably, selective area epitaxy of PbTe-Pb on a lattice matched CdTe substracte has been achieved \cite{jiang2021selective}, holding promise for further transport experiments \cite{zhang2019next}. With these encouraging progresses, we deem that employing PbTe-Pb hybrid nanowires for engineering and detecting MZMs is coming soon.

\section{Acknowledgements}\label{Acknowledgements}
We are grateful to Shuai Yang, Gu Zhang, and Li Chen for helpful discussions. This work was supported by the National Natural Science Foundation of China (Grants No.~12004040, No.~11974198, No.~92065206, and No.~12074133), the National Key Research and Development Program of China (Grant No.~2017YFA0303303), and Tsinghua University Initiative Scientific Research Program.

\appendix
\section{Derivation of Hamiltonian (\ref{hwire})}\label{app}
In this Appendix, we present the derivation of the effective conduction band Hamiltonian (\ref{hwire}) of PbTe nanowires. For PbTe quantum wells, the Hamiltonian in a laboratory coordinate system has been derived from the Dimmock $\vec k \cdot \vec p$ Hamiltonian established in local coordinate system \cite{silva1999optical}. Here, we extend the procedure to the more complicated PbTe nanowire case. As shown in Fig.~\ref{Fig:device}(d), rotating the laboratory coordinate system $x$-$y$-$z$ around the $y$ axis clockwise with an angle $\alpha$ and then around the new $x$ axis clockwise with an angle $\beta$ arrives at the local $x^\prime$-$y^\prime$-$z^\prime$ coordinate system, in which the $\vec k \cdot \vec p$ Hamiltonian $H^\textrm{elli}$ is established. The Hamiltonian in the laboratory coordinate system is obtained through the transformation $H^\textrm{lab}=U^{T}(\alpha,\beta)H^\textrm{elli}U^{\ast}(\alpha,\beta)$ with $U(\alpha,\beta)=e^{i\beta\sigma_{x^\prime}/2}e^{-i\alpha\sigma_{y^\prime}/2}$, and substituting $(\hat{\pi}_{x},\hat{\pi}_{y},\hat{\pi}_{z})^{T}$ and $(  B_{x},B_{y},B_{z})^{T}$ by $R(\alpha,\beta)(\hat{\pi}_{x},\hat{\pi}_{y},\hat{\pi}_{z})^{T}$ and $R(\alpha,\beta)(B_{x},B_{y},B_{z})^{T}$, respectively, with
\begin{eqnarray}
\hspace{-0.5cm}R\left(  \alpha,\beta\right)  =\left(
\begin{array}
[c]{ccc}
\cos\alpha & 0 & \sin\alpha\\
\sin\alpha\sin\beta & \cos\beta & -\cos\alpha\sin\beta\\
-\sin\alpha\cos\beta & \sin\beta & \cos\alpha\cos\beta
\end{array}
\right).
\end{eqnarray}

After some straightforward algebra one obtains
\begin{equation}
H^\textrm{lab}=H^\textrm{lab}_\textrm{kin}+H^\textrm{lab}_\textrm{SOC}+H^\textrm{lab}_\textrm{Z}, \label{Ht_lab}\\
\end{equation}
with
\begin{eqnarray}
H^\textrm{lab}_\textrm{kin}&=&\sum_{u,v}\hat{\pi}_{u}\frac{M_{uv}}{E+E_{g}+e\phi(\vec{r})}\hat{\pi}_{v}-e\phi(\vec{r}),\label{H1_lab}\\
H^\textrm{lab}_\textrm{SOC}&=&\sum_{u,v,w}\bigg(\hat{k}_{u}\frac{iN_{w;uv}}{E+E_{g}+e\phi(\vec{r})  }\bigg)  \hat{\pi}_{v}\sigma_{w},\label{H2_lab}\\
H^\textrm{lab}_\textrm{Z}&=&\frac{\mu_{B}}{2}\sum_{u,w}\bigg[g_{uw}B_{u}+\frac{4m_{e}}{\hbar^{2}}\sum_{v}\frac{iN_{w;uv}(\hat{k}_{u}A_{v})}{E+E_{g}+e\phi(\vec{r})  }\bigg]  \sigma_{w},\label{HZ_lab}\notag\\
\end{eqnarray}
where the elements of $M$ and $N$ are presented in Eqs.~(\ref{matrixM}) and (\ref{tensorN}), respectively, and the anisotropic Land\'{e} $g$ factors are
\begin{eqnarray}
&&g_{xx}=g_{t}+(  g_{l}-g_{t})  \sin^{2}\alpha\cos^{2}\beta,\notag\\
&&g_{yy}=g_{t}+(  g_{l}-g_{t})  \sin^{2}\beta,\notag\\
&&g_{zz}=g_{t}+(  g_{l}-g_{t})  \cos^{2}\alpha\cos^{2}\beta,\notag\\
&&g_{xy}=g_{yx}=-(  g_{l}-g_{t})  \sin\alpha\sin\beta\cos\beta,\notag\\
&&g_{yz}=g_{zy}=(  g_{l}-g_{t})  \cos\alpha\sin\beta\cos\beta,\notag\\
&&g_{zx}=g_{xz}=-(  g_{l}-g_{t})  \sin\alpha\cos\alpha\cos^{2}\beta.
\end{eqnarray}

Considering a nanowire that is translational invariant along the $z$-direction, we replace $\hat{k}_z$ by a constant $k_z$ as it should be a good quantum number and discard the terms $\hat{k}_{z}\phi(\vec{r})$ and $\hat{k}_{z}A_{v}$. Before proceeding, we note that Hamiltonian (\ref{Ht_lab}) depends on $E$, which is itself the eigenvalue of the Hamiltonian and thus requires a self-consistent calculation. This dependency also appears in the $\vec k \cdot \vec p$ Hamiltonians of III-V SMs InAs and InSb. The widely used approximation to remove this dependency is to assume $E_g\gg\vert e\phi(\vec{r}) + E\vert$ and expand the Hamiltonian in series of $1/E_g^n$ and then truncate it to the lowest nonvanishing order \cite{winkler2003spin,wojcik2018tuning,escribano2020improved,darnhofer1993effects}. This approximation is reasonable as the energy scale of Majorana physics is a few meV around the Fermi level, which is far smaller than $E_g$. With this approximation and a gauge $\vec{A}=\left[0,B_{z}(  x-x_{0}),B_{x}(  y-y_{0})  -B_{y}(  x-x_{0})\right]$ the effective conduction band Hamiltonian for an individual valley of a PbTe nanowire is derived as
\begin{equation}
H^\textrm{wire}=H^\textrm{wire}_\textrm{kin}+H^\textrm{wire}_\textrm{SOC}+H^\textrm{wire}_\textrm{Z},\\
\end{equation}
with
\begin{eqnarray}
H^\textrm{wire}_\textrm{kin}&=&\sum_{u,v}\hat{\pi}_{u}\frac{M_{uv}}{E_{g}}\hat{\pi}_{v}-e\phi(x,y),\\
H^\textrm{wire}_\textrm{SOC}&=&\sum_u\bigg[\Omega_{ux}(x,y)\hat{\pi}_x+\Omega_{uy}(x,y)\hat{\pi}_y\notag\\
&&\hspace{0.5cm}+\Omega_{uz}(x,y)(k_{z}+eA_z/\hbar)\bigg]\sigma_{u},\\
H^\textrm{wire}_\textrm{Z}&=&\frac{\mu_{B}}{2}\sum_{u,v}g_{uv}^{\ast}B_{u}\sigma_{v},
\end{eqnarray}
where
\begin{eqnarray}
\hspace{-0.2cm}\Omega_{ux}(x,y)&=&\frac{eE_{y}(x,y)}{E_{g}^{2}}N_{u;yx},\notag\\
\hspace{-0.2cm}\Omega_{uy}(x,y)&=&\frac{eE_{x}(x,y)}{E_{g}^{2}}N_{u;xy},\notag\\
\hspace{-0.2cm}\Omega_{uz}(x,y)&=&\frac{eE_{y}(x,y)}{E_{g}^{2}}N_{u;yz}+\frac{eE_{x}(x,y)}{E_{g}^{2}}N_{u;xz},
\end{eqnarray}
\begin{eqnarray}
\hspace{-0.2cm}g_{xu}^{\ast}&=&g_{xu}+\frac{4m_{e}N_{u;yz}}{\hbar^{2}E_{g}},\notag\\
\hspace{-0.2cm}g_{yu}^{\ast}&=&g_{yu}+\frac{4m_{e}N_{u;zx}}{\hbar^{2}E_{g}},\notag\\
\hspace{-0.2cm}g_{zu}^{\ast}&=&g_{zu}+\frac{4m_{e}N_{u;xy}}{\hbar^{2}E_{g}}.
\end{eqnarray}
$E_{x(y)}(x,y)=-\partial_{x(y)}\phi(x,y)$ is the electric field along the x (y) direction in the nanowire.

For a magnetic field parallel to the nanowire, i.e., $A_z=0$, the linear term in $k_z$ in $H_\textrm{kin}^{\textrm{wire}}$ can be eliminated through a unitary transformation $\widetilde{H}_\textrm{wire}=e^{-S}H_\textrm{wire}e^{S}$ with $S=i(bx+cy)k_z$ and
\begin{equation}
b=\frac{M_{yy}M_{zx}-M_{xy}M_{zy}}{M_{xy}^{2}-M_{xx}M_{yy}},
\end{equation}
\begin{equation}
c=\frac{M_{xx}M_{zy}-M_{xy}M_{zx}}{M_{xy}^{2}-M_{xx}M_{yy}}.
\end{equation}
By noting that $e^{-S}\hat{\pi}_x e^{S}=\hat{\pi}_x+bk_z$ and $e^{-S}\hat{\pi}_y e^{S}=\hat{\pi}_y+ck_z$, one immediately obtains Hamiltonian (\ref{hwire}), where the tilde symbol is neglected for simplicity.


\begin{thebibliography}{146}%
\makeatletter
\providecommand \@ifxundefined [1]{%
 \@ifx{#1\undefined}
}%
\providecommand \@ifnum [1]{%
 \ifnum #1\expandafter \@firstoftwo
 \else \expandafter \@secondoftwo
 \fi
}%
\providecommand \@ifx [1]{%
 \ifx #1\expandafter \@firstoftwo
 \else \expandafter \@secondoftwo
 \fi
}%
\providecommand \natexlab [1]{#1}%
\providecommand \enquote  [1]{``#1''}%
\providecommand \bibnamefont  [1]{#1}%
\providecommand \bibfnamefont [1]{#1}%
\providecommand \citenamefont [1]{#1}%
\providecommand \href@noop [0]{\@secondoftwo}%
\providecommand \href [0]{\begingroup \@sanitize@url \@href}%
\providecommand \@href[1]{\@@startlink{#1}\@@href}%
\providecommand \@@href[1]{\endgroup#1\@@endlink}%
\providecommand \@sanitize@url [0]{\catcode `\\12\catcode `\$12\catcode
  `\&12\catcode `\#12\catcode `\^12\catcode `\_12\catcode `\%12\relax}%
\providecommand \@@startlink[1]{}%
\providecommand \@@endlink[0]{}%
\providecommand \url  [0]{\begingroup\@sanitize@url \@url }%
\providecommand \@url [1]{\endgroup\@href {#1}{\urlprefix }}%
\providecommand \urlprefix  [0]{URL }%
\providecommand \Eprint [0]{\href }%
\providecommand \doibase [0]{http://dx.doi.org/}%
\providecommand \selectlanguage [0]{\@gobble}%
\providecommand \bibinfo  [0]{\@secondoftwo}%
\providecommand \bibfield  [0]{\@secondoftwo}%
\providecommand \translation [1]{[#1]}%
\providecommand \BibitemOpen [0]{}%
\providecommand \bibitemStop [0]{}%
\providecommand \bibitemNoStop [0]{.\EOS\space}%
\providecommand \EOS [0]{\spacefactor3000\relax}%
\providecommand \BibitemShut  [1]{\csname bibitem#1\endcsname}%
\let\auto@bib@innerbib\@empty
\bibitem [{\citenamefont {Read}\ and\ \citenamefont
  {Green}(2000)}]{read2000paried}%
  \BibitemOpen
  \bibfield  {author} {\bibinfo {author} {\bibfnamefont {N.}~\bibnamefont
  {Read}}\ and\ \bibinfo {author} {\bibfnamefont {D.}~\bibnamefont {Green}},\
  }\bibinfo {title} {Paired states of fermions in two dimensions with breaking
  of parity and time-reversal symmetries and the fractional quantum {Hall}
  effect},\ \href {\doibase 10.1103/PhysRevB.61.10267} {\bibfield  {journal}
  {\bibinfo  {journal} {Phys. Rev. B}\ }\textbf {\bibinfo {volume} {61}},\
  \bibinfo {pages} {10267} (\bibinfo {year} {2000})}\BibitemShut {NoStop}%
\bibitem [{\citenamefont {Kitaev}(2001)}]{kitaev2001unpaired}%
  \BibitemOpen
  \bibfield  {author} {\bibinfo {author} {\bibfnamefont {A.~Y.}\ \bibnamefont
  {Kitaev}},\ }\bibinfo {title} {Unpaired {Majorana} fermions in quantum
  wires},\ \href {https://doi.org/10.1070/1063-7869/44/10S/S29} {\bibfield
  {journal} {\bibinfo  {journal} {Phys. Usp.}\ }\textbf {\bibinfo {volume}
  {44}},\ \bibinfo {pages} {131} (\bibinfo {year} {2001})}\BibitemShut
  {NoStop}%
\bibitem [{\citenamefont {Alicea}(2012)}]{alicea2012new}%
  \BibitemOpen
  \bibfield  {author} {\bibinfo {author} {\bibfnamefont {J.}~\bibnamefont
  {Alicea}},\ }\bibinfo {title} {New directions in the pursuit of {Majorana}
  fermions in solid state systems},\ \href
  {https://doi.org/10.1088/0034-4885/75/7/076501} {\bibfield  {journal}
  {\bibinfo  {journal} {Rep. Prog. Phys.}\ }\textbf {\bibinfo {volume} {75}},\
  \bibinfo {pages} {076501} (\bibinfo {year} {2012})}\BibitemShut {NoStop}%
\bibitem [{\citenamefont {Leijnse}\ and\ \citenamefont
  {Flensberg}(2012)}]{leijnse2012introduction}%
  \BibitemOpen
  \bibfield  {author} {\bibinfo {author} {\bibfnamefont {M.}~\bibnamefont
  {Leijnse}}\ and\ \bibinfo {author} {\bibfnamefont {K.}~\bibnamefont
  {Flensberg}},\ }\bibinfo {title} {Introduction to topological
  superconductivity and {Majorana} fermions},\ \href
  {https://doi.org/10.1088/0268-1242/27/12/124003} {\bibfield  {journal}
  {\bibinfo  {journal} {Semicond. Sci. Technol.}\ }\textbf {\bibinfo {volume}
  {27}},\ \bibinfo {pages} {124003} (\bibinfo {year} {2012})}\BibitemShut
  {NoStop}%
\bibitem [{\citenamefont {Beenakker}(2013)}]{beenakker2013search}%
  \BibitemOpen
  \bibfield  {author} {\bibinfo {author} {\bibfnamefont {C.}~\bibnamefont
  {Beenakker}},\ }\bibinfo {title} {Search for {Majorana} fermions in
  superconductors},\ \href
  {https://doi.org/10.1146/annurev-conmatphys-030212-184337} {\bibfield
  {journal} {\bibinfo  {journal} {Annu. Rev. Condens. Matter Phys.}\ }\textbf
  {\bibinfo {volume} {4}},\ \bibinfo {pages} {113} (\bibinfo {year}
  {2013})}\BibitemShut {NoStop}%
\bibitem [{\citenamefont {Stanescu}\ and\ \citenamefont
  {Tewari}(2013{\natexlab{a}})}]{stanescu2013majorana}%
  \BibitemOpen
  \bibfield  {author} {\bibinfo {author} {\bibfnamefont {T.~D.}\ \bibnamefont
  {Stanescu}}\ and\ \bibinfo {author} {\bibfnamefont {S.}~\bibnamefont
  {Tewari}},\ }\bibinfo {title} {{Majorana} fermions in semiconductor
  nanowires: fundamentals, modeling, and experiment},\ \href
  {https://doi.org/10.1088/0953-8984/25/23/233201} {\bibfield  {journal}
  {\bibinfo  {journal} {J. Phys.: Condens. Matter}\ }\textbf {\bibinfo {volume}
  {25}},\ \bibinfo {pages} {233201} (\bibinfo {year}
  {2013}{\natexlab{a}})}\BibitemShut {NoStop}%
\bibitem [{\citenamefont {Aguado}(2017)}]{aguado2017majorana}%
  \BibitemOpen
  \bibfield  {author} {\bibinfo {author} {\bibfnamefont {R.}~\bibnamefont
  {Aguado}},\ }\bibinfo {title} {{Majorana} quasiparticles in condensed
  matter},\ \href {https://doi.org/10.1393/ncr/i2017-10141-9} {\bibfield
  {journal} {\bibinfo  {journal} {Riv. Nuovo Cimento}\ }\textbf {\bibinfo
  {volume} {40}},\ \bibinfo {pages} {523} (\bibinfo {year} {2017})}\BibitemShut
  {NoStop}%
\bibitem [{\citenamefont {Prada}\ \emph {et~al.}(2020)\citenamefont {Prada},
  \citenamefont {San-Jose}, \citenamefont {de~Moor}, \citenamefont {Geresdi},
  \citenamefont {Lee}, \citenamefont {Klinovaja}, \citenamefont {Loss},
  \citenamefont {Nyg{\aa}rd}, \citenamefont {Aguado},\ and\ \citenamefont
  {Kouwenhoven}}]{prada2020andreev}%
  \BibitemOpen
  \bibfield  {author} {\bibinfo {author} {\bibfnamefont {E.}~\bibnamefont
  {Prada}}, \bibinfo {author} {\bibfnamefont {P.}~\bibnamefont {San-Jose}},
  \bibinfo {author} {\bibfnamefont {M.~W.}\ \bibnamefont {de~Moor}}, \bibinfo
  {author} {\bibfnamefont {A.}~\bibnamefont {Geresdi}}, \bibinfo {author}
  {\bibfnamefont {E.~J.}\ \bibnamefont {Lee}}, \bibinfo {author} {\bibfnamefont
  {J.}~\bibnamefont {Klinovaja}}, \bibinfo {author} {\bibfnamefont
  {D.}~\bibnamefont {Loss}}, \bibinfo {author} {\bibfnamefont {J.}~\bibnamefont
  {Nyg{\aa}rd}}, \bibinfo {author} {\bibfnamefont {R.}~\bibnamefont {Aguado}},
  \ and\ \bibinfo {author} {\bibfnamefont {L.~P.}\ \bibnamefont
  {Kouwenhoven}},\ }\bibinfo {title} {From {Andreev} to {Majorana} bound states
  in hybrid superconductor--semiconductor nanowires},\ \href
  {https://doi.org/10.1038/s42254-020-0228-y} {\bibfield  {journal} {\bibinfo
  {journal} {Nat. Rev. Phys.}\ }\textbf {\bibinfo {volume} {2}},\ \bibinfo
  {pages} {575} (\bibinfo {year} {2020})}\BibitemShut {NoStop}%
\bibitem [{\citenamefont {Moore}\ and\ \citenamefont
  {Read}(1991)}]{moore1991nonabelions}%
  \BibitemOpen
  \bibfield  {author} {\bibinfo {author} {\bibfnamefont {G.}~\bibnamefont
  {Moore}}\ and\ \bibinfo {author} {\bibfnamefont {N.}~\bibnamefont {Read}},\
  }\bibinfo {title} {Nonabelions in the fractional quantum {Hall} effect},\
  \href {https://doi.org/10.1016/0550-3213(91)90407-O} {\bibfield  {journal}
  {\bibinfo  {journal} {Nucl. Phys. B}\ }\textbf {\bibinfo {volume} {360}},\
  \bibinfo {pages} {362} (\bibinfo {year} {1991})}\BibitemShut {NoStop}%
\bibitem [{\citenamefont {Nayak}\ and\ \citenamefont
  {Wilczek}(1996)}]{nayak19962n}%
  \BibitemOpen
  \bibfield  {author} {\bibinfo {author} {\bibfnamefont {C.}~\bibnamefont
  {Nayak}}\ and\ \bibinfo {author} {\bibfnamefont {F.}~\bibnamefont
  {Wilczek}},\ }\bibinfo {title} {2n-quasihole states realize
  $2^{n-1}$-dimensional spinor braiding statistics in paired quantum {Hall}
  states},\ \href {https://doi.org/10.1016/0550-3213(96)00430-0} {\bibfield
  {journal} {\bibinfo  {journal} {Nucl. Phys. B}\ }\textbf {\bibinfo {volume}
  {479}},\ \bibinfo {pages} {529} (\bibinfo {year} {1996})}\BibitemShut
  {NoStop}%
\bibitem [{\citenamefont {Nayak}\ \emph {et~al.}(2008)\citenamefont {Nayak},
  \citenamefont {Simon}, \citenamefont {Stern}, \citenamefont {Freedman},\ and\
  \citenamefont {{Das Sarma}}}]{nayak2008non}%
  \BibitemOpen
  \bibfield  {author} {\bibinfo {author} {\bibfnamefont {C.}~\bibnamefont
  {Nayak}}, \bibinfo {author} {\bibfnamefont {S.~H.}\ \bibnamefont {Simon}},
  \bibinfo {author} {\bibfnamefont {A.}~\bibnamefont {Stern}}, \bibinfo
  {author} {\bibfnamefont {M.}~\bibnamefont {Freedman}}, \ and\ \bibinfo
  {author} {\bibfnamefont {S.}~\bibnamefont {{Das Sarma}}},\ }\bibinfo {title}
  {Non-{Abelian} anyons and topological quantum computation},\ \href
  {https://doi.org/10.1103/RevModPhys.80.1083} {\bibfield  {journal} {\bibinfo
  {journal} {Rev. Mod. Phys.}\ }\textbf {\bibinfo {volume} {80}},\ \bibinfo
  {pages} {1083} (\bibinfo {year} {2008})}\BibitemShut {NoStop}%
\bibitem [{\citenamefont {Kitaev}(2003)}]{kitaev2003fault}%
  \BibitemOpen
  \bibfield  {author} {\bibinfo {author} {\bibfnamefont {A.~Y.}\ \bibnamefont
  {Kitaev}},\ }\bibinfo {title} {Fault-tolerant quantum computation by
  anyons},\ \href {https://doi.org/10.1016/S0003-4916(02)00018-0} {\bibfield
  {journal} {\bibinfo  {journal} {Ann. Phys. (Amsterdam)}\ }\textbf {\bibinfo
  {volume} {303}},\ \bibinfo {pages} {2} (\bibinfo {year} {2003})}\BibitemShut
  {NoStop}%
\bibitem [{\citenamefont {{Das Sarma}}\ \emph {et~al.}(2015)\citenamefont {{Das
  Sarma}}, \citenamefont {Freedman},\ and\ \citenamefont
  {Nayak}}]{sarma2015majorana}%
  \BibitemOpen
  \bibfield  {author} {\bibinfo {author} {\bibfnamefont {S.}~\bibnamefont {{Das
  Sarma}}}, \bibinfo {author} {\bibfnamefont {M.}~\bibnamefont {Freedman}}, \
  and\ \bibinfo {author} {\bibfnamefont {C.}~\bibnamefont {Nayak}},\ }\bibinfo
  {title} {{Majorana} zero modes and topological quantum computation},\ \href
  {https://doi.org/10.1038/npjqi.2015.1} {\bibfield  {journal} {\bibinfo
  {journal} {npj Quantum Inf.}\ }\textbf {\bibinfo {volume} {1}},\ \bibinfo
  {pages} {15001} (\bibinfo {year} {2015})}\BibitemShut {NoStop}%
\bibitem [{\citenamefont {Aasen}\ \emph {et~al.}(2016)\citenamefont {Aasen},
  \citenamefont {Hell}, \citenamefont {Mishmash}, \citenamefont {Higginbotham},
  \citenamefont {Danon}, \citenamefont {Leijnse}, \citenamefont {Jespersen},
  \citenamefont {Folk}, \citenamefont {Marcus}, \citenamefont {Flensberg},\
  and\ \citenamefont {Alicea}}]{aasen2016milestones}%
  \BibitemOpen
  \bibfield  {author} {\bibinfo {author} {\bibfnamefont {D.}~\bibnamefont
  {Aasen}}, \bibinfo {author} {\bibfnamefont {M.}~\bibnamefont {Hell}},
  \bibinfo {author} {\bibfnamefont {R.~V.}\ \bibnamefont {Mishmash}}, \bibinfo
  {author} {\bibfnamefont {A.}~\bibnamefont {Higginbotham}}, \bibinfo {author}
  {\bibfnamefont {J.}~\bibnamefont {Danon}}, \bibinfo {author} {\bibfnamefont
  {M.}~\bibnamefont {Leijnse}}, \bibinfo {author} {\bibfnamefont {T.~S.}\
  \bibnamefont {Jespersen}}, \bibinfo {author} {\bibfnamefont {J.~A.}\
  \bibnamefont {Folk}}, \bibinfo {author} {\bibfnamefont {C.~M.}\ \bibnamefont
  {Marcus}}, \bibinfo {author} {\bibfnamefont {K.}~\bibnamefont {Flensberg}}, \
  and\ \bibinfo {author} {\bibfnamefont {J.}~\bibnamefont {Alicea}},\ }\bibinfo
  {title} {Milestones Toward {Majorana}-Based Quantum Computing},\ \href
  {\doibase 10.1103/PhysRevX.6.031016} {\bibfield  {journal} {\bibinfo
  {journal} {Phys. Rev. X}\ }\textbf {\bibinfo {volume} {6}},\ \bibinfo {pages}
  {031016} (\bibinfo {year} {2016})}\BibitemShut {NoStop}%
\bibitem [{\citenamefont {Plugge}\ \emph {et~al.}(2017)\citenamefont {Plugge},
  \citenamefont {Rasmussen}, \citenamefont {Egger},\ and\ \citenamefont
  {Flensberg}}]{plugge2017majorana}%
  \BibitemOpen
  \bibfield  {author} {\bibinfo {author} {\bibfnamefont {S.}~\bibnamefont
  {Plugge}}, \bibinfo {author} {\bibfnamefont {A.}~\bibnamefont {Rasmussen}},
  \bibinfo {author} {\bibfnamefont {R.}~\bibnamefont {Egger}}, \ and\ \bibinfo
  {author} {\bibfnamefont {K.}~\bibnamefont {Flensberg}},\ }\bibinfo {title}
  {{Majorana} box qubits},\ \href {https://doi.org/10.1088/1367-2630/aa54e1}
  {\bibfield  {journal} {\bibinfo  {journal} {New J. Phys.}\ }\textbf {\bibinfo
  {volume} {19}},\ \bibinfo {pages} {012001} (\bibinfo {year}
  {2017})}\BibitemShut {NoStop}%
\bibitem [{\citenamefont {Karzig}\ \emph {et~al.}(2017)\citenamefont {Karzig},
  \citenamefont {Knapp}, \citenamefont {Lutchyn}, \citenamefont {Bonderson},
  \citenamefont {Hastings}, \citenamefont {Nayak}, \citenamefont {Alicea},
  \citenamefont {Flensberg}, \citenamefont {Plugge}, \citenamefont {Oreg},
  \citenamefont {Marcus},\ and\ \citenamefont {Freedman}}]{karzig2017scalble}%
  \BibitemOpen
  \bibfield  {author} {\bibinfo {author} {\bibfnamefont {T.}~\bibnamefont
  {Karzig}}, \bibinfo {author} {\bibfnamefont {C.}~\bibnamefont {Knapp}},
  \bibinfo {author} {\bibfnamefont {R.~M.}\ \bibnamefont {Lutchyn}}, \bibinfo
  {author} {\bibfnamefont {P.}~\bibnamefont {Bonderson}}, \bibinfo {author}
  {\bibfnamefont {M.~B.}\ \bibnamefont {Hastings}}, \bibinfo {author}
  {\bibfnamefont {C.}~\bibnamefont {Nayak}}, \bibinfo {author} {\bibfnamefont
  {J.}~\bibnamefont {Alicea}}, \bibinfo {author} {\bibfnamefont
  {K.}~\bibnamefont {Flensberg}}, \bibinfo {author} {\bibfnamefont
  {S.}~\bibnamefont {Plugge}}, \bibinfo {author} {\bibfnamefont
  {Y.}~\bibnamefont {Oreg}}, \bibinfo {author} {\bibfnamefont {C.~M.}\
  \bibnamefont {Marcus}}, \ and\ \bibinfo {author} {\bibfnamefont {M.~H.}\
  \bibnamefont {Freedman}},\ }\bibinfo {title} {Scalable designs for
  quasiparticle-poisoning-protected topological quantum computation with
  {Majorana} zero modes},\ \href {\doibase 10.1103/PhysRevB.95.235305}
  {\bibfield  {journal} {\bibinfo  {journal} {Phys. Rev. B}\ }\textbf {\bibinfo
  {volume} {95}},\ \bibinfo {pages} {235305} (\bibinfo {year}
  {2017})}\BibitemShut {NoStop}%
\bibitem [{\citenamefont {Lutchyn}\ \emph {et~al.}(2010)\citenamefont
  {Lutchyn}, \citenamefont {Sau},\ and\ \citenamefont {{Das
  Sarma}}}]{lutchyn2010majorana}%
  \BibitemOpen
  \bibfield  {author} {\bibinfo {author} {\bibfnamefont {R.~M.}\ \bibnamefont
  {Lutchyn}}, \bibinfo {author} {\bibfnamefont {J.~D.}\ \bibnamefont {Sau}}, \
  and\ \bibinfo {author} {\bibfnamefont {S.}~\bibnamefont {{Das Sarma}}},\
  }\bibinfo {title} {{Majorana} fermions and a topological phase transition in
  semiconductor-superconductor heterostructures},\ \href
  {https://doi.org/10.1103/PhysRevLett.105.077001} {\bibfield  {journal}
  {\bibinfo  {journal} {Phys. Rev. Lett.}\ }\textbf {\bibinfo {volume} {105}},\
  \bibinfo {pages} {077001} (\bibinfo {year} {2010})}\BibitemShut {NoStop}%
\bibitem [{\citenamefont {Oreg}\ \emph {et~al.}(2010)\citenamefont {Oreg},
  \citenamefont {Refael},\ and\ \citenamefont {von Oppen}}]{oreg2010helical}%
  \BibitemOpen
  \bibfield  {author} {\bibinfo {author} {\bibfnamefont {Y.}~\bibnamefont
  {Oreg}}, \bibinfo {author} {\bibfnamefont {G.}~\bibnamefont {Refael}}, \ and\
  \bibinfo {author} {\bibfnamefont {F.}~\bibnamefont {von Oppen}},\ }\bibinfo
  {title} {Helical liquids and {Majorana} bound states in quantum wires},\
  \href {https://doi.org/10.1103/PhysRevLett.105.177002} {\bibfield  {journal}
  {\bibinfo  {journal} {Phys. Rev. Lett.}\ }\textbf {\bibinfo {volume} {105}},\
  \bibinfo {pages} {177002} (\bibinfo {year} {2010})}\BibitemShut {NoStop}%
\bibitem [{\citenamefont {Mourik}\ \emph {et~al.}(2012)\citenamefont {Mourik},
  \citenamefont {Zuo}, \citenamefont {Frolov}, \citenamefont {Plissard},
  \citenamefont {Bakkers},\ and\ \citenamefont
  {Kouwenhoven}}]{mourik2012signatures}%
  \BibitemOpen
  \bibfield  {author} {\bibinfo {author} {\bibfnamefont {V.}~\bibnamefont
  {Mourik}}, \bibinfo {author} {\bibfnamefont {K.}~\bibnamefont {Zuo}},
  \bibinfo {author} {\bibfnamefont {S.~M.}\ \bibnamefont {Frolov}}, \bibinfo
  {author} {\bibfnamefont {S.}~\bibnamefont {Plissard}}, \bibinfo {author}
  {\bibfnamefont {E.~P.}\ \bibnamefont {Bakkers}}, \ and\ \bibinfo {author}
  {\bibfnamefont {L.~P.}\ \bibnamefont {Kouwenhoven}},\ }\bibinfo {title}
  {Signatures of {Majorana} fermions in hybrid superconductor-semiconductor
  nanowire devices},\ \href {https://doi.org/10.1126/science.1222360}
  {\bibfield  {journal} {\bibinfo  {journal} {Science}\ }\textbf {\bibinfo
  {volume} {336}},\ \bibinfo {pages} {1003} (\bibinfo {year}
  {2012})}\BibitemShut {NoStop}%
\bibitem [{\citenamefont {Deng}\ \emph {et~al.}(2012)\citenamefont {Deng},
  \citenamefont {Yu}, \citenamefont {Huang}, \citenamefont {Larsson},
  \citenamefont {Caroff},\ and\ \citenamefont {Xu}}]{deng2012anomalous}%
  \BibitemOpen
  \bibfield  {author} {\bibinfo {author} {\bibfnamefont {M.~T.}\ \bibnamefont
  {Deng}}, \bibinfo {author} {\bibfnamefont {C.}~\bibnamefont {Yu}}, \bibinfo
  {author} {\bibfnamefont {G.}~\bibnamefont {Huang}}, \bibinfo {author}
  {\bibfnamefont {M.}~\bibnamefont {Larsson}}, \bibinfo {author} {\bibfnamefont
  {P.}~\bibnamefont {Caroff}}, \ and\ \bibinfo {author} {\bibfnamefont
  {H.}~\bibnamefont {Xu}},\ }\bibinfo {title} {Anomalous zero-bias conductance
  peak in a {Nb-InSb nanowire-Nb} hybrid device},\ \href
  {https://doi.org/10.1021/nl303758w} {\bibfield  {journal} {\bibinfo
  {journal} {Nano Lett.}\ }\textbf {\bibinfo {volume} {12}},\ \bibinfo {pages}
  {6414} (\bibinfo {year} {2012})}\BibitemShut {NoStop}%
\bibitem [{\citenamefont {Das}\ \emph {et~al.}(2012)\citenamefont {Das},
  \citenamefont {Ronen}, \citenamefont {Most}, \citenamefont {Oreg},
  \citenamefont {Heiblum},\ and\ \citenamefont {Shtrikman}}]{das2012zero}%
  \BibitemOpen
  \bibfield  {author} {\bibinfo {author} {\bibfnamefont {A.}~\bibnamefont
  {Das}}, \bibinfo {author} {\bibfnamefont {Y.}~\bibnamefont {Ronen}}, \bibinfo
  {author} {\bibfnamefont {Y.}~\bibnamefont {Most}}, \bibinfo {author}
  {\bibfnamefont {Y.}~\bibnamefont {Oreg}}, \bibinfo {author} {\bibfnamefont
  {M.}~\bibnamefont {Heiblum}}, \ and\ \bibinfo {author} {\bibfnamefont
  {H.}~\bibnamefont {Shtrikman}},\ }\bibinfo {title} {Zero-bias peaks and
  splitting in an {Al--InAs} nanowire topological superconductor as a signature
  of {Majorana} fermions},\ \href {https://doi.org/10.1038/nphys2479}
  {\bibfield  {journal} {\bibinfo  {journal} {Nat. Phys.}\ }\textbf {\bibinfo
  {volume} {8}},\ \bibinfo {pages} {887} (\bibinfo {year} {2012})}\BibitemShut
  {NoStop}%
\bibitem [{\citenamefont {Finck}\ \emph {et~al.}(2013)\citenamefont {Finck},
  \citenamefont {Van~Harlingen}, \citenamefont {Mohseni}, \citenamefont
  {Jung},\ and\ \citenamefont {Li}}]{finck2013anomalous}%
  \BibitemOpen
  \bibfield  {author} {\bibinfo {author} {\bibfnamefont {A.~D.~K.}\
  \bibnamefont {Finck}}, \bibinfo {author} {\bibfnamefont {D.~J.}\ \bibnamefont
  {Van~Harlingen}}, \bibinfo {author} {\bibfnamefont {P.~K.}\ \bibnamefont
  {Mohseni}}, \bibinfo {author} {\bibfnamefont {K.}~\bibnamefont {Jung}}, \
  and\ \bibinfo {author} {\bibfnamefont {X.}~\bibnamefont {Li}},\ }\bibinfo
  {title} {Anomalous Modulation of a Zero-Bias Peak in a Hybrid
  Nanowire-Superconductor Device},\ \href {\doibase
  10.1103/PhysRevLett.110.126406} {\bibfield  {journal} {\bibinfo  {journal}
  {Phys. Rev. Lett.}\ }\textbf {\bibinfo {volume} {110}},\ \bibinfo {pages}
  {126406} (\bibinfo {year} {2013})}\BibitemShut {NoStop}%
\bibitem [{\citenamefont {Churchill}\ \emph {et~al.}(2013)\citenamefont
  {Churchill}, \citenamefont {Fatemi}, \citenamefont {Grove-Rasmussen},
  \citenamefont {Deng}, \citenamefont {Caroff}, \citenamefont {Xu},\ and\
  \citenamefont {Marcus}}]{churchill2013superconductor}%
  \BibitemOpen
  \bibfield  {author} {\bibinfo {author} {\bibfnamefont {H.~O.~H.}\
  \bibnamefont {Churchill}}, \bibinfo {author} {\bibfnamefont {V.}~\bibnamefont
  {Fatemi}}, \bibinfo {author} {\bibfnamefont {K.}~\bibnamefont
  {Grove-Rasmussen}}, \bibinfo {author} {\bibfnamefont {M.~T.}\ \bibnamefont
  {Deng}}, \bibinfo {author} {\bibfnamefont {P.}~\bibnamefont {Caroff}},
  \bibinfo {author} {\bibfnamefont {H.~Q.}\ \bibnamefont {Xu}}, \ and\ \bibinfo
  {author} {\bibfnamefont {C.~M.}\ \bibnamefont {Marcus}},\ }\bibinfo {title}
  {Superconductor-nanowire devices from tunneling to the multichannel regime:
  Zero-bias oscillations and magnetoconductance crossover},\ \href {\doibase
  10.1103/PhysRevB.87.241401} {\bibfield  {journal} {\bibinfo  {journal} {Phys.
  Rev. B}\ }\textbf {\bibinfo {volume} {87}},\ \bibinfo {pages} {241401(R)}
  (\bibinfo {year} {2013})}\BibitemShut {NoStop}%
\bibitem [{\citenamefont {Deng}\ \emph {et~al.}(2016)\citenamefont {Deng},
  \citenamefont {Vaitiek{\.e}nas}, \citenamefont {Hansen}, \citenamefont
  {Danon}, \citenamefont {Leijnse}, \citenamefont {Flensberg}, \citenamefont
  {Nyg{\aa}rd}, \citenamefont {Krogstrup},\ and\ \citenamefont
  {Marcus}}]{deng2016majorana}%
  \BibitemOpen
  \bibfield  {author} {\bibinfo {author} {\bibfnamefont {M.~T.}\ \bibnamefont
  {Deng}}, \bibinfo {author} {\bibfnamefont {S.}~\bibnamefont
  {Vaitiek{\.e}nas}}, \bibinfo {author} {\bibfnamefont {E.~B.}\ \bibnamefont
  {Hansen}}, \bibinfo {author} {\bibfnamefont {J.}~\bibnamefont {Danon}},
  \bibinfo {author} {\bibfnamefont {M.}~\bibnamefont {Leijnse}}, \bibinfo
  {author} {\bibfnamefont {K.}~\bibnamefont {Flensberg}}, \bibinfo {author}
  {\bibfnamefont {J.}~\bibnamefont {Nyg{\aa}rd}}, \bibinfo {author}
  {\bibfnamefont {P.}~\bibnamefont {Krogstrup}}, \ and\ \bibinfo {author}
  {\bibfnamefont {C.~M.}\ \bibnamefont {Marcus}},\ }\bibinfo {title}
  {{Majorana} bound state in a coupled quantum-dot hybrid-nanowire system},\
  \href {https://doi.org/10.1126/science.aaf3961} {\bibfield  {journal}
  {\bibinfo  {journal} {Science}\ }\textbf {\bibinfo {volume} {354}},\ \bibinfo
  {pages} {1557} (\bibinfo {year} {2016})}\BibitemShut {NoStop}%
\bibitem [{\citenamefont {Chen}\ \emph {et~al.}(2017)\citenamefont {Chen},
  \citenamefont {Yu}, \citenamefont {Stenger}, \citenamefont {Hocevar},
  \citenamefont {Car}, \citenamefont {Plissard}, \citenamefont {Bakkers},
  \citenamefont {Stanescu},\ and\ \citenamefont
  {Frolov}}]{chen2017experimental}%
  \BibitemOpen
  \bibfield  {author} {\bibinfo {author} {\bibfnamefont {J.}~\bibnamefont
  {Chen}}, \bibinfo {author} {\bibfnamefont {P.}~\bibnamefont {Yu}}, \bibinfo
  {author} {\bibfnamefont {J.}~\bibnamefont {Stenger}}, \bibinfo {author}
  {\bibfnamefont {M.}~\bibnamefont {Hocevar}}, \bibinfo {author} {\bibfnamefont
  {D.}~\bibnamefont {Car}}, \bibinfo {author} {\bibfnamefont {S.~R.}\
  \bibnamefont {Plissard}}, \bibinfo {author} {\bibfnamefont {E.~P.}\
  \bibnamefont {Bakkers}}, \bibinfo {author} {\bibfnamefont {T.~D.}\
  \bibnamefont {Stanescu}}, \ and\ \bibinfo {author} {\bibfnamefont {S.~M.}\
  \bibnamefont {Frolov}},\ }\bibinfo {title} {Experimental phase diagram of
  zero-bias conductance peaks in superconductor/semiconductor nanowire
  devices},\ \href {https://doi.org/10.1126/sciadv.1701476} {\bibfield
  {journal} {\bibinfo  {journal} {Sci. Adv.}\ }\textbf {\bibinfo {volume}
  {3}},\ \bibinfo {pages} {e1701476} (\bibinfo {year} {2017})}\BibitemShut
  {NoStop}%
\bibitem [{\citenamefont {Suominen}\ \emph {et~al.}(2017)\citenamefont
  {Suominen}, \citenamefont {Kjaergaard}, \citenamefont {Hamilton},
  \citenamefont {Shabani}, \citenamefont {Palmstr{\o}m}, \citenamefont
  {Marcus},\ and\ \citenamefont {Nichele}}]{suominen2017zero}%
  \BibitemOpen
  \bibfield  {author} {\bibinfo {author} {\bibfnamefont {H.~J.}\ \bibnamefont
  {Suominen}}, \bibinfo {author} {\bibfnamefont {M.}~\bibnamefont
  {Kjaergaard}}, \bibinfo {author} {\bibfnamefont {A.~R.}\ \bibnamefont
  {Hamilton}}, \bibinfo {author} {\bibfnamefont {J.}~\bibnamefont {Shabani}},
  \bibinfo {author} {\bibfnamefont {C.~J.}\ \bibnamefont {Palmstr{\o}m}},
  \bibinfo {author} {\bibfnamefont {C.~M.}\ \bibnamefont {Marcus}}, \ and\
  \bibinfo {author} {\bibfnamefont {F.}~\bibnamefont {Nichele}},\ }\bibinfo
  {title} {Zero-energy modes from coalescing {Andreev} states in a
  two-dimensional semiconductor-superconductor hybrid platform},\ \href
  {https://doi.org/10.1103/PhysRevLett.119.176805} {\bibfield  {journal}
  {\bibinfo  {journal} {Phys. Rev. Lett.}\ }\textbf {\bibinfo {volume} {119}},\
  \bibinfo {pages} {176805} (\bibinfo {year} {2017})}\BibitemShut {NoStop}%
\bibitem [{\citenamefont {Nichele}\ \emph {et~al.}(2017)\citenamefont
  {Nichele}, \citenamefont {Drachmann}, \citenamefont {Whiticar}, \citenamefont
  {O'Farrell}, \citenamefont {Suominen}, \citenamefont {Fornieri},
  \citenamefont {Wang}, \citenamefont {Gardner}, \citenamefont {Thomas},
  \citenamefont {Hatke}, \citenamefont {Krogstrup}, \citenamefont {Manfra},
  \citenamefont {Flensberg},\ and\ \citenamefont
  {Marcus}}]{nichele2017scaling}%
  \BibitemOpen
  \bibfield  {author} {\bibinfo {author} {\bibfnamefont {F.}~\bibnamefont
  {Nichele}}, \bibinfo {author} {\bibfnamefont {A.~C.~C.}\ \bibnamefont
  {Drachmann}}, \bibinfo {author} {\bibfnamefont {A.~M.}\ \bibnamefont
  {Whiticar}}, \bibinfo {author} {\bibfnamefont {E.~C.~T.}\ \bibnamefont
  {O'Farrell}}, \bibinfo {author} {\bibfnamefont {H.~J.}\ \bibnamefont
  {Suominen}}, \bibinfo {author} {\bibfnamefont {A.}~\bibnamefont {Fornieri}},
  \bibinfo {author} {\bibfnamefont {T.}~\bibnamefont {Wang}}, \bibinfo {author}
  {\bibfnamefont {G.~C.}\ \bibnamefont {Gardner}}, \bibinfo {author}
  {\bibfnamefont {C.}~\bibnamefont {Thomas}}, \bibinfo {author} {\bibfnamefont
  {A.~T.}\ \bibnamefont {Hatke}}, \bibinfo {author} {\bibfnamefont
  {P.}~\bibnamefont {Krogstrup}}, \bibinfo {author} {\bibfnamefont {M.~J.}\
  \bibnamefont {Manfra}}, \bibinfo {author} {\bibfnamefont {K.}~\bibnamefont
  {Flensberg}}, \ and\ \bibinfo {author} {\bibfnamefont {C.~M.}\ \bibnamefont
  {Marcus}},\ }\bibinfo {title} {Scaling of {Majorana} Zero-Bias Conductance
  Peaks},\ \href {\doibase 10.1103/PhysRevLett.119.136803} {\bibfield
  {journal} {\bibinfo  {journal} {Phys. Rev. Lett.}\ }\textbf {\bibinfo
  {volume} {119}},\ \bibinfo {pages} {136803} (\bibinfo {year}
  {2017})}\BibitemShut {NoStop}%
\bibitem [{\citenamefont {G{\"u}l}\ \emph {et~al.}(2018)\citenamefont
  {G{\"u}l}, \citenamefont {Zhang}, \citenamefont {Bommer}, \citenamefont
  {de~Moor}, \citenamefont {Car}, \citenamefont {Plissard}, \citenamefont
  {Bakkers}, \citenamefont {Geresdi}, \citenamefont {Watanabe}, \citenamefont
  {Taniguchi} \emph {et~al.}}]{gul2018ballistic}%
  \BibitemOpen
  \bibfield  {author} {\bibinfo {author} {\bibfnamefont {{\"O}.}~\bibnamefont
  {G{\"u}l}}, \bibinfo {author} {\bibfnamefont {H.}~\bibnamefont {Zhang}},
  \bibinfo {author} {\bibfnamefont {J.~D.}\ \bibnamefont {Bommer}}, \bibinfo
  {author} {\bibfnamefont {M.~W.}\ \bibnamefont {de~Moor}}, \bibinfo {author}
  {\bibfnamefont {D.}~\bibnamefont {Car}}, \bibinfo {author} {\bibfnamefont
  {S.~R.}\ \bibnamefont {Plissard}}, \bibinfo {author} {\bibfnamefont {E.~P.}\
  \bibnamefont {Bakkers}}, \bibinfo {author} {\bibfnamefont {A.}~\bibnamefont
  {Geresdi}}, \bibinfo {author} {\bibfnamefont {K.}~\bibnamefont {Watanabe}},
  \bibinfo {author} {\bibfnamefont {T.}~\bibnamefont {Taniguchi}},  \emph
  {et~al.},\ }\bibinfo {title} {Ballistic {Majorana} nanowire devices},\ \href
  {https://doi.org/10.1038/s41565-017-0032-8} {\bibfield  {journal} {\bibinfo
  {journal} {Nat. Nanotechnol.}\ }\textbf {\bibinfo {volume} {13}},\ \bibinfo
  {pages} {192} (\bibinfo {year} {2018})}\BibitemShut {NoStop}%
\bibitem [{\citenamefont {Sestoft}\ \emph {et~al.}(2018)\citenamefont
  {Sestoft}, \citenamefont {Kanne}, \citenamefont {Gejl}, \citenamefont {von
  Soosten}, \citenamefont {Yodh}, \citenamefont {Sherman}, \citenamefont
  {Tarasinski}, \citenamefont {Wimmer}, \citenamefont {Johnson}, \citenamefont
  {Deng}, \citenamefont {Nyg\aa{}rd}, \citenamefont {Jespersen}, \citenamefont
  {Marcus},\ and\ \citenamefont {Krogstrup}}]{sestoft2018engineering}%
  \BibitemOpen
  \bibfield  {author} {\bibinfo {author} {\bibfnamefont {J.~E.}\ \bibnamefont
  {Sestoft}}, \bibinfo {author} {\bibfnamefont {T.}~\bibnamefont {Kanne}},
  \bibinfo {author} {\bibfnamefont {A.~N.}\ \bibnamefont {Gejl}}, \bibinfo
  {author} {\bibfnamefont {M.}~\bibnamefont {von Soosten}}, \bibinfo {author}
  {\bibfnamefont {J.~S.}\ \bibnamefont {Yodh}}, \bibinfo {author}
  {\bibfnamefont {D.}~\bibnamefont {Sherman}}, \bibinfo {author} {\bibfnamefont
  {B.}~\bibnamefont {Tarasinski}}, \bibinfo {author} {\bibfnamefont
  {M.}~\bibnamefont {Wimmer}}, \bibinfo {author} {\bibfnamefont
  {E.}~\bibnamefont {Johnson}}, \bibinfo {author} {\bibfnamefont
  {M.}~\bibnamefont {Deng}}, \bibinfo {author} {\bibfnamefont {J.}~\bibnamefont
  {Nyg\aa{}rd}}, \bibinfo {author} {\bibfnamefont {T.~S.}\ \bibnamefont
  {Jespersen}}, \bibinfo {author} {\bibfnamefont {C.~M.}\ \bibnamefont
  {Marcus}}, \ and\ \bibinfo {author} {\bibfnamefont {P.}~\bibnamefont
  {Krogstrup}},\ }\bibinfo {title} {Engineering hybrid epitaxial {InAsSb/Al}
  nanowires for stronger topological protection},\ \href {\doibase
  10.1103/PhysRevMaterials.2.044202} {\bibfield  {journal} {\bibinfo  {journal}
  {Phys. Rev. Materials}\ }\textbf {\bibinfo {volume} {2}},\ \bibinfo {pages}
  {044202} (\bibinfo {year} {2018})}\BibitemShut {NoStop}%
\bibitem [{\citenamefont {Vaitiek{\.e}nas}\ \emph {et~al.}(2018)\citenamefont
  {Vaitiek{\.e}nas}, \citenamefont {Deng}, \citenamefont {Nyg\aa{}rd},
  \citenamefont {Krogstrup},\ and\ \citenamefont
  {Marcus}}]{vaitiekenas2018effective}%
  \BibitemOpen
  \bibfield  {author} {\bibinfo {author} {\bibfnamefont {S.}~\bibnamefont
  {Vaitiek{\.e}nas}}, \bibinfo {author} {\bibfnamefont {M.-T.}\ \bibnamefont
  {Deng}}, \bibinfo {author} {\bibfnamefont {J.}~\bibnamefont {Nyg\aa{}rd}},
  \bibinfo {author} {\bibfnamefont {P.}~\bibnamefont {Krogstrup}}, \ and\
  \bibinfo {author} {\bibfnamefont {C.~M.}\ \bibnamefont {Marcus}},\ }\bibinfo
  {title} {Effective $g$ Factor of Subgap States in Hybrid Nanowires},\ \href
  {\doibase 10.1103/PhysRevLett.121.037703} {\bibfield  {journal} {\bibinfo
  {journal} {Phys. Rev. Lett.}\ }\textbf {\bibinfo {volume} {121}},\ \bibinfo
  {pages} {037703} (\bibinfo {year} {2018})}\BibitemShut {NoStop}%
\bibitem [{\citenamefont {Deng}\ \emph {et~al.}(2018)\citenamefont {Deng},
  \citenamefont {Vaitiek{\.e}nas}, \citenamefont {Prada}, \citenamefont
  {San-Jose}, \citenamefont {Nyg\aa{}rd}, \citenamefont {Krogstrup},
  \citenamefont {Aguado},\ and\ \citenamefont {Marcus}}]{deng2018nonlocality}%
  \BibitemOpen
  \bibfield  {author} {\bibinfo {author} {\bibfnamefont {M.-T.}\ \bibnamefont
  {Deng}}, \bibinfo {author} {\bibfnamefont {S.}~\bibnamefont
  {Vaitiek{\.e}nas}}, \bibinfo {author} {\bibfnamefont {E.}~\bibnamefont
  {Prada}}, \bibinfo {author} {\bibfnamefont {P.}~\bibnamefont {San-Jose}},
  \bibinfo {author} {\bibfnamefont {J.}~\bibnamefont {Nyg\aa{}rd}}, \bibinfo
  {author} {\bibfnamefont {P.}~\bibnamefont {Krogstrup}}, \bibinfo {author}
  {\bibfnamefont {R.}~\bibnamefont {Aguado}}, \ and\ \bibinfo {author}
  {\bibfnamefont {C.~M.}\ \bibnamefont {Marcus}},\ }\bibinfo {title}
  {Nonlocality of {Majorana} modes in hybrid nanowires},\ \href {\doibase
  10.1103/PhysRevB.98.085125} {\bibfield  {journal} {\bibinfo  {journal} {Phys.
  Rev. B}\ }\textbf {\bibinfo {volume} {98}},\ \bibinfo {pages} {085125}
  (\bibinfo {year} {2018})}\BibitemShut {NoStop}%
\bibitem [{\citenamefont {de~Moor}\ \emph {et~al.}(2018)\citenamefont
  {de~Moor}, \citenamefont {Bommer}, \citenamefont {Xu}, \citenamefont
  {Winkler}, \citenamefont {Antipov}, \citenamefont {Bargerbos}, \citenamefont
  {Wang}, \citenamefont {van Loo}, \citenamefont {Veld}, \citenamefont
  {Gazibegovic} \emph {et~al.}}]{de2018electric}%
  \BibitemOpen
  \bibfield  {author} {\bibinfo {author} {\bibfnamefont {M.~W.}\ \bibnamefont
  {de~Moor}}, \bibinfo {author} {\bibfnamefont {J.~D.}\ \bibnamefont {Bommer}},
  \bibinfo {author} {\bibfnamefont {D.}~\bibnamefont {Xu}}, \bibinfo {author}
  {\bibfnamefont {G.~W.}\ \bibnamefont {Winkler}}, \bibinfo {author}
  {\bibfnamefont {A.~E.}\ \bibnamefont {Antipov}}, \bibinfo {author}
  {\bibfnamefont {A.}~\bibnamefont {Bargerbos}}, \bibinfo {author}
  {\bibfnamefont {G.}~\bibnamefont {Wang}}, \bibinfo {author} {\bibfnamefont
  {N.}~\bibnamefont {van Loo}}, \bibinfo {author} {\bibfnamefont {R.~L.}\
  \bibnamefont {Veld}}, \bibinfo {author} {\bibfnamefont {S.}~\bibnamefont
  {Gazibegovic}},  \emph {et~al.},\ }\bibinfo {title} {Electric field tunable
  superconductor-semiconductor coupling in {Majorana} nanowires},\ \href
  {https://doi.org/10.1088/1367-2630/aae61d} {\bibfield  {journal} {\bibinfo
  {journal} {New J. Phys.}\ }\textbf {\bibinfo {volume} {20}},\ \bibinfo
  {pages} {103049} (\bibinfo {year} {2018})}\BibitemShut {NoStop}%
\bibitem [{\citenamefont {Bommer}\ \emph {et~al.}(2019)\citenamefont {Bommer},
  \citenamefont {Zhang}, \citenamefont {G\"ul}, \citenamefont {Nijholt},
  \citenamefont {Wimmer}, \citenamefont {Rybakov}, \citenamefont {Garaud},
  \citenamefont {Rodic}, \citenamefont {Babaev}, \citenamefont {Troyer},
  \citenamefont {Car}, \citenamefont {Plissard}, \citenamefont {Bakkers},
  \citenamefont {Watanabe}, \citenamefont {Taniguchi},\ and\ \citenamefont
  {Kouwenhoven}}]{bommer2019spin}%
  \BibitemOpen
  \bibfield  {author} {\bibinfo {author} {\bibfnamefont {J.~D.~S.}\
  \bibnamefont {Bommer}}, \bibinfo {author} {\bibfnamefont {H.}~\bibnamefont
  {Zhang}}, \bibinfo {author} {\bibfnamefont {O.}~\bibnamefont {G\"ul}},
  \bibinfo {author} {\bibfnamefont {B.}~\bibnamefont {Nijholt}}, \bibinfo
  {author} {\bibfnamefont {M.}~\bibnamefont {Wimmer}}, \bibinfo {author}
  {\bibfnamefont {F.~N.}\ \bibnamefont {Rybakov}}, \bibinfo {author}
  {\bibfnamefont {J.}~\bibnamefont {Garaud}}, \bibinfo {author} {\bibfnamefont
  {D.}~\bibnamefont {Rodic}}, \bibinfo {author} {\bibfnamefont
  {E.}~\bibnamefont {Babaev}}, \bibinfo {author} {\bibfnamefont
  {M.}~\bibnamefont {Troyer}}, \bibinfo {author} {\bibfnamefont
  {D.}~\bibnamefont {Car}}, \bibinfo {author} {\bibfnamefont {S.~R.}\
  \bibnamefont {Plissard}}, \bibinfo {author} {\bibfnamefont {E.~P. A.~M.}\
  \bibnamefont {Bakkers}}, \bibinfo {author} {\bibfnamefont {K.}~\bibnamefont
  {Watanabe}}, \bibinfo {author} {\bibfnamefont {T.}~\bibnamefont {Taniguchi}},
  \ and\ \bibinfo {author} {\bibfnamefont {L.~P.}\ \bibnamefont
  {Kouwenhoven}},\ }\bibinfo {title} {Spin-Orbit Protection of Induced
  Superconductivity in {Majorana} Nanowires},\ \href {\doibase
  10.1103/PhysRevLett.122.187702} {\bibfield  {journal} {\bibinfo  {journal}
  {Phys. Rev. Lett.}\ }\textbf {\bibinfo {volume} {122}},\ \bibinfo {pages}
  {187702} (\bibinfo {year} {2019})}\BibitemShut {NoStop}%
\bibitem [{\citenamefont {Grivnin}\ \emph {et~al.}(2019)\citenamefont
  {Grivnin}, \citenamefont {Bor}, \citenamefont {Heiblum}, \citenamefont
  {Oreg},\ and\ \citenamefont {Shtrikman}}]{grivnin2019concomitant}%
  \BibitemOpen
  \bibfield  {author} {\bibinfo {author} {\bibfnamefont {A.}~\bibnamefont
  {Grivnin}}, \bibinfo {author} {\bibfnamefont {E.}~\bibnamefont {Bor}},
  \bibinfo {author} {\bibfnamefont {M.}~\bibnamefont {Heiblum}}, \bibinfo
  {author} {\bibfnamefont {Y.}~\bibnamefont {Oreg}}, \ and\ \bibinfo {author}
  {\bibfnamefont {H.}~\bibnamefont {Shtrikman}},\ }\bibinfo {title}
  {Concomitant opening of a bulk-gap with an emerging possible {Majorana} zero
  mode},\ \href {https://doi.org/10.1038/s41467-019-09771-0} {\bibfield
  {journal} {\bibinfo  {journal} {Nat. Commun.}\ }\textbf {\bibinfo {volume}
  {10}},\ \bibinfo {pages} {1940} (\bibinfo {year} {2019})}\BibitemShut
  {NoStop}%
\bibitem [{\citenamefont {Anselmetti}\ \emph {et~al.}(2019)\citenamefont
  {Anselmetti}, \citenamefont {Martinez}, \citenamefont {M\'enard},
  \citenamefont {Puglia}, \citenamefont {Malinowski}, \citenamefont {Lee},
  \citenamefont {Choi}, \citenamefont {Pendharkar}, \citenamefont
  {Palmstr\o{}m}, \citenamefont {Marcus}, \citenamefont {Casparis},\ and\
  \citenamefont {Higginbotham}}]{anselmetti2019end}%
  \BibitemOpen
  \bibfield  {author} {\bibinfo {author} {\bibfnamefont {G.~L.~R.}\
  \bibnamefont {Anselmetti}}, \bibinfo {author} {\bibfnamefont {E.~A.}\
  \bibnamefont {Martinez}}, \bibinfo {author} {\bibfnamefont {G.~C.}\
  \bibnamefont {M\'enard}}, \bibinfo {author} {\bibfnamefont {D.}~\bibnamefont
  {Puglia}}, \bibinfo {author} {\bibfnamefont {F.~K.}\ \bibnamefont
  {Malinowski}}, \bibinfo {author} {\bibfnamefont {J.~S.}\ \bibnamefont {Lee}},
  \bibinfo {author} {\bibfnamefont {S.}~\bibnamefont {Choi}}, \bibinfo {author}
  {\bibfnamefont {M.}~\bibnamefont {Pendharkar}}, \bibinfo {author}
  {\bibfnamefont {C.~J.}\ \bibnamefont {Palmstr\o{}m}}, \bibinfo {author}
  {\bibfnamefont {C.~M.}\ \bibnamefont {Marcus}}, \bibinfo {author}
  {\bibfnamefont {L.}~\bibnamefont {Casparis}}, \ and\ \bibinfo {author}
  {\bibfnamefont {A.~P.}\ \bibnamefont {Higginbotham}},\ }\bibinfo {title}
  {End-to-end correlated subgap states in hybrid nanowires},\ \href {\doibase
  10.1103/PhysRevB.100.205412} {\bibfield  {journal} {\bibinfo  {journal}
  {Phys. Rev. B}\ }\textbf {\bibinfo {volume} {100}},\ \bibinfo {pages}
  {205412} (\bibinfo {year} {2019})}\BibitemShut {NoStop}%
\bibitem [{\citenamefont {M\'enard}\ \emph {et~al.}(2020)\citenamefont
  {M\'enard}, \citenamefont {Anselmetti}, \citenamefont {Martinez},
  \citenamefont {Puglia}, \citenamefont {Malinowski}, \citenamefont {Lee},
  \citenamefont {Choi}, \citenamefont {Pendharkar}, \citenamefont
  {Palmstr\o{}m}, \citenamefont {Flensberg}, \citenamefont {Marcus},
  \citenamefont {Casparis},\ and\ \citenamefont
  {Higginbotham}}]{menard2020conductance}%
  \BibitemOpen
  \bibfield  {author} {\bibinfo {author} {\bibfnamefont {G.~C.}\ \bibnamefont
  {M\'enard}}, \bibinfo {author} {\bibfnamefont {G.~L.~R.}\ \bibnamefont
  {Anselmetti}}, \bibinfo {author} {\bibfnamefont {E.~A.}\ \bibnamefont
  {Martinez}}, \bibinfo {author} {\bibfnamefont {D.}~\bibnamefont {Puglia}},
  \bibinfo {author} {\bibfnamefont {F.~K.}\ \bibnamefont {Malinowski}},
  \bibinfo {author} {\bibfnamefont {J.~S.}\ \bibnamefont {Lee}}, \bibinfo
  {author} {\bibfnamefont {S.}~\bibnamefont {Choi}}, \bibinfo {author}
  {\bibfnamefont {M.}~\bibnamefont {Pendharkar}}, \bibinfo {author}
  {\bibfnamefont {C.~J.}\ \bibnamefont {Palmstr\o{}m}}, \bibinfo {author}
  {\bibfnamefont {K.}~\bibnamefont {Flensberg}}, \bibinfo {author}
  {\bibfnamefont {C.~M.}\ \bibnamefont {Marcus}}, \bibinfo {author}
  {\bibfnamefont {L.}~\bibnamefont {Casparis}}, \ and\ \bibinfo {author}
  {\bibfnamefont {A.~P.}\ \bibnamefont {Higginbotham}},\ }\bibinfo {title}
  {Conductance-Matrix Symmetries of a Three-Terminal Hybrid Device},\ \href
  {\doibase 10.1103/PhysRevLett.124.036802} {\bibfield  {journal} {\bibinfo
  {journal} {Phys. Rev. Lett.}\ }\textbf {\bibinfo {volume} {124}},\ \bibinfo
  {pages} {036802} (\bibinfo {year} {2020})}\BibitemShut {NoStop}%
\bibitem [{\citenamefont {Sengupta}\ \emph {et~al.}(2001)\citenamefont
  {Sengupta}, \citenamefont {\ifmmode \check{Z}\else
  \v{Z}\fi{}uti\ifmmode~\acute{c}\else \'{c}\fi{}}, \citenamefont {Kwon},
  \citenamefont {Yakovenko},\ and\ \citenamefont
  {Das~Sarma}}]{sengupta2001midgap}%
  \BibitemOpen
  \bibfield  {author} {\bibinfo {author} {\bibfnamefont {K.}~\bibnamefont
  {Sengupta}}, \bibinfo {author} {\bibfnamefont {I.}~\bibnamefont {\ifmmode
  \check{Z}\else \v{Z}\fi{}uti\ifmmode~\acute{c}\else \'{c}\fi{}}}, \bibinfo
  {author} {\bibfnamefont {H.-J.}\ \bibnamefont {Kwon}}, \bibinfo {author}
  {\bibfnamefont {V.~M.}\ \bibnamefont {Yakovenko}}, \ and\ \bibinfo {author}
  {\bibfnamefont {S.}~\bibnamefont {Das~Sarma}},\ }\bibinfo {title} {Midgap
  edge states and pairing symmetry of quasi-one-dimensional organic
  superconductors},\ \href {\doibase 10.1103/PhysRevB.63.144531} {\bibfield
  {journal} {\bibinfo  {journal} {Phys. Rev. B}\ }\textbf {\bibinfo {volume}
  {63}},\ \bibinfo {pages} {144531} (\bibinfo {year} {2001})}\BibitemShut
  {NoStop}%
\bibitem [{\citenamefont {Law}\ \emph {et~al.}(2009)\citenamefont {Law},
  \citenamefont {Lee},\ and\ \citenamefont {Ng}}]{law2009majorana}%
  \BibitemOpen
  \bibfield  {author} {\bibinfo {author} {\bibfnamefont {K.~T.}\ \bibnamefont
  {Law}}, \bibinfo {author} {\bibfnamefont {P.~A.}\ \bibnamefont {Lee}}, \ and\
  \bibinfo {author} {\bibfnamefont {T.~K.}\ \bibnamefont {Ng}},\ }\bibinfo
  {title} {{Majorana} fermion induced resonant {Andreev} reflection},\ \href
  {https://doi.org/10.1103/PhysRevLett.103.237001} {\bibfield  {journal}
  {\bibinfo  {journal} {Phys. Rev. Lett.}\ }\textbf {\bibinfo {volume} {103}},\
  \bibinfo {pages} {237001} (\bibinfo {year} {2009})}\BibitemShut {NoStop}%
\bibitem [{\citenamefont {Flensberg}(2010)}]{flensberg2010tunneling}%
  \BibitemOpen
  \bibfield  {author} {\bibinfo {author} {\bibfnamefont {K.}~\bibnamefont
  {Flensberg}},\ }\bibinfo {title} {Tunneling characteristics of a chain of
  {Majorana} bound states},\ \href {https://doi.org/10.1103/PhysRevB.82.180516}
  {\bibfield  {journal} {\bibinfo  {journal} {Phys. Rev. B}\ }\textbf {\bibinfo
  {volume} {82}},\ \bibinfo {pages} {180516(R)} (\bibinfo {year}
  {2010})}\BibitemShut {NoStop}%
\bibitem [{\citenamefont {Wimmer}\ \emph {et~al.}(2011)\citenamefont {Wimmer},
  \citenamefont {Akhmerov}, \citenamefont {Dahlhaus},\ and\ \citenamefont
  {Beenakker}}]{wimmer2011quantum}%
  \BibitemOpen
  \bibfield  {author} {\bibinfo {author} {\bibfnamefont {M.}~\bibnamefont
  {Wimmer}}, \bibinfo {author} {\bibfnamefont {A.}~\bibnamefont {Akhmerov}},
  \bibinfo {author} {\bibfnamefont {J.}~\bibnamefont {Dahlhaus}}, \ and\
  \bibinfo {author} {\bibfnamefont {C.}~\bibnamefont {Beenakker}},\ }\bibinfo
  {title} {Quantum point contact as a probe of a topological superconductor},\
  \href {https://doi.org/10.1088/1367-2630/13/5/053016} {\bibfield  {journal}
  {\bibinfo  {journal} {New J. Phys.}\ }\textbf {\bibinfo {volume} {13}},\
  \bibinfo {pages} {053016} (\bibinfo {year} {2011})}\BibitemShut {NoStop}%
\bibitem [{\citenamefont {Zhang}\ \emph {et~al.}(2021)\citenamefont {Zhang},
  \citenamefont {de~Moor}, \citenamefont {Bommer}, \citenamefont {Xu},
  \citenamefont {Wang}, \citenamefont {van Loo}, \citenamefont {Liu},
  \citenamefont {Gazibegovic}, \citenamefont {Logan}, \citenamefont {Car} \emph
  {et~al.}}]{zhang2021large}%
  \BibitemOpen
  \bibfield  {author} {\bibinfo {author} {\bibfnamefont {H.}~\bibnamefont
  {Zhang}}, \bibinfo {author} {\bibfnamefont {M.~W.}\ \bibnamefont {de~Moor}},
  \bibinfo {author} {\bibfnamefont {J.~D.}\ \bibnamefont {Bommer}}, \bibinfo
  {author} {\bibfnamefont {D.}~\bibnamefont {Xu}}, \bibinfo {author}
  {\bibfnamefont {G.}~\bibnamefont {Wang}}, \bibinfo {author} {\bibfnamefont
  {N.}~\bibnamefont {van Loo}}, \bibinfo {author} {\bibfnamefont {C.-X.}\
  \bibnamefont {Liu}}, \bibinfo {author} {\bibfnamefont {S.}~\bibnamefont
  {Gazibegovic}}, \bibinfo {author} {\bibfnamefont {J.~A.}\ \bibnamefont
  {Logan}}, \bibinfo {author} {\bibfnamefont {D.}~\bibnamefont {Car}},  \emph
  {et~al.},\ }\bibinfo {title} {Large zero-bias peaks in {InSb-Al} hybrid
  semiconductor-superconductor nanowire devices},\ \href
  {https://arxiv.org/abs/2101.11456} {\bibfield  {journal} {\bibinfo  {journal}
  {arXiv: 2101.11456}\ } (\bibinfo {year} {2021})}\BibitemShut {NoStop}%
\bibitem [{\citenamefont {Song}\ \emph {et~al.}(2021)\citenamefont {Song},
  \citenamefont {Zhang}, \citenamefont {Pan}, \citenamefont {Liu},
  \citenamefont {Wang}, \citenamefont {Cao}, \citenamefont {Liu}, \citenamefont
  {Wen}, \citenamefont {Liao}, \citenamefont {Zhuo} \emph
  {et~al.}}]{song2021large}%
  \BibitemOpen
  \bibfield  {author} {\bibinfo {author} {\bibfnamefont {H.}~\bibnamefont
  {Song}}, \bibinfo {author} {\bibfnamefont {Z.}~\bibnamefont {Zhang}},
  \bibinfo {author} {\bibfnamefont {D.}~\bibnamefont {Pan}}, \bibinfo {author}
  {\bibfnamefont {D.}~\bibnamefont {Liu}}, \bibinfo {author} {\bibfnamefont
  {Z.}~\bibnamefont {Wang}}, \bibinfo {author} {\bibfnamefont {Z.}~\bibnamefont
  {Cao}}, \bibinfo {author} {\bibfnamefont {L.}~\bibnamefont {Liu}}, \bibinfo
  {author} {\bibfnamefont {L.}~\bibnamefont {Wen}}, \bibinfo {author}
  {\bibfnamefont {D.}~\bibnamefont {Liao}}, \bibinfo {author} {\bibfnamefont
  {R.}~\bibnamefont {Zhuo}},  \emph {et~al.},\ }\bibinfo {title} {Large zero
  bias peaks and dips in a four-terminal thin {InAs-Al} nanowire device},\
  \href {https://arxiv.org/abs/2107.08282} {\bibfield  {journal} {\bibinfo
  {journal} {arXiv: 2107.08282}\ } (\bibinfo {year} {2021})}\BibitemShut
  {NoStop}%
\bibitem [{\citenamefont {Pan}\ and\ \citenamefont
  {Das~Sarma}(2020)}]{pan2020physical}%
  \BibitemOpen
  \bibfield  {author} {\bibinfo {author} {\bibfnamefont {H.}~\bibnamefont
  {Pan}}\ and\ \bibinfo {author} {\bibfnamefont {S.}~\bibnamefont
  {Das~Sarma}},\ }\bibinfo {title} {Physical mechanisms for zero-bias
  conductance peaks in {Majorana} nanowires},\ \href {\doibase
  10.1103/PhysRevResearch.2.013377} {\bibfield  {journal} {\bibinfo  {journal}
  {Phys. Rev. Research}\ }\textbf {\bibinfo {volume} {2}},\ \bibinfo {pages}
  {013377} (\bibinfo {year} {2020})}\BibitemShut {NoStop}%
\bibitem [{\citenamefont {Chang}\ \emph {et~al.}(2015)\citenamefont {Chang},
  \citenamefont {Albrecht}, \citenamefont {Jespersen}, \citenamefont
  {Kuemmeth}, \citenamefont {Krogstrup}, \citenamefont {Nyg{\aa}rd},\ and\
  \citenamefont {Marcus}}]{chang2015hard}%
  \BibitemOpen
  \bibfield  {author} {\bibinfo {author} {\bibfnamefont {W.}~\bibnamefont
  {Chang}}, \bibinfo {author} {\bibfnamefont {S.}~\bibnamefont {Albrecht}},
  \bibinfo {author} {\bibfnamefont {T.}~\bibnamefont {Jespersen}}, \bibinfo
  {author} {\bibfnamefont {F.}~\bibnamefont {Kuemmeth}}, \bibinfo {author}
  {\bibfnamefont {P.}~\bibnamefont {Krogstrup}}, \bibinfo {author}
  {\bibfnamefont {J.}~\bibnamefont {Nyg{\aa}rd}}, \ and\ \bibinfo {author}
  {\bibfnamefont {C.~M.}\ \bibnamefont {Marcus}},\ }\bibinfo {title} {Hard gap
  in epitaxial semiconductor--superconductor nanowires},\ \href
  {https://doi.org/10.1038/nnano.2014.306} {\bibfield  {journal} {\bibinfo
  {journal} {Nat. Nanotechnol.}\ }\textbf {\bibinfo {volume} {10}},\ \bibinfo
  {pages} {232} (\bibinfo {year} {2015})}\BibitemShut {NoStop}%
\bibitem [{\citenamefont {Krogstrup}\ \emph {et~al.}(2015)\citenamefont
  {Krogstrup}, \citenamefont {Ziino}, \citenamefont {Chang}, \citenamefont
  {Albrecht}, \citenamefont {Madsen}, \citenamefont {Johnson}, \citenamefont
  {Nyg{\aa}rd}, \citenamefont {Marcus},\ and\ \citenamefont
  {Jespersen}}]{krogstrup2015epitaxy}%
  \BibitemOpen
  \bibfield  {author} {\bibinfo {author} {\bibfnamefont {P.}~\bibnamefont
  {Krogstrup}}, \bibinfo {author} {\bibfnamefont {N.}~\bibnamefont {Ziino}},
  \bibinfo {author} {\bibfnamefont {W.}~\bibnamefont {Chang}}, \bibinfo
  {author} {\bibfnamefont {S.}~\bibnamefont {Albrecht}}, \bibinfo {author}
  {\bibfnamefont {M.}~\bibnamefont {Madsen}}, \bibinfo {author} {\bibfnamefont
  {E.}~\bibnamefont {Johnson}}, \bibinfo {author} {\bibfnamefont
  {J.}~\bibnamefont {Nyg{\aa}rd}}, \bibinfo {author} {\bibfnamefont
  {C.}~\bibnamefont {Marcus}}, \ and\ \bibinfo {author} {\bibfnamefont
  {T.}~\bibnamefont {Jespersen}},\ }\bibinfo {title} {Epitaxy of
  semiconductor-superconductor nanowires},\ \href
  {https://doi.org/10.1038/nmat4176} {\bibfield  {journal} {\bibinfo  {journal}
  {Nat. Mater.}\ }\textbf {\bibinfo {volume} {14}},\ \bibinfo {pages} {400}
  (\bibinfo {year} {2015})}\BibitemShut {NoStop}%
\bibitem [{\citenamefont {Takei}\ \emph {et~al.}(2013)\citenamefont {Takei},
  \citenamefont {Fregoso}, \citenamefont {Hui}, \citenamefont {Lobos},\ and\
  \citenamefont {Das~Sarma}}]{takei2013soft}%
  \BibitemOpen
  \bibfield  {author} {\bibinfo {author} {\bibfnamefont {S.}~\bibnamefont
  {Takei}}, \bibinfo {author} {\bibfnamefont {B.~M.}\ \bibnamefont {Fregoso}},
  \bibinfo {author} {\bibfnamefont {H.-Y.}\ \bibnamefont {Hui}}, \bibinfo
  {author} {\bibfnamefont {A.~M.}\ \bibnamefont {Lobos}}, \ and\ \bibinfo
  {author} {\bibfnamefont {S.}~\bibnamefont {Das~Sarma}},\ }\bibinfo {title}
  {Soft Superconducting Gap in Semiconductor {Majorana} Nanowires},\ \href
  {\doibase 10.1103/PhysRevLett.110.186803} {\bibfield  {journal} {\bibinfo
  {journal} {Phys. Rev. Lett.}\ }\textbf {\bibinfo {volume} {110}},\ \bibinfo
  {pages} {186803} (\bibinfo {year} {2013})}\BibitemShut {NoStop}%
\bibitem [{\citenamefont {Das~Sarma}\ and\ \citenamefont
  {Pan}(2021)}]{sarma2021disorder}%
  \BibitemOpen
  \bibfield  {author} {\bibinfo {author} {\bibfnamefont {S.}~\bibnamefont
  {Das~Sarma}}\ and\ \bibinfo {author} {\bibfnamefont {H.}~\bibnamefont
  {Pan}},\ }\bibinfo {title} {Disorder-induced zero-bias peaks in {Majorana}
  nanowires},\ \href {\doibase 10.1103/PhysRevB.103.195158} {\bibfield
  {journal} {\bibinfo  {journal} {Phys. Rev. B}\ }\textbf {\bibinfo {volume}
  {103}},\ \bibinfo {pages} {195158} (\bibinfo {year} {2021})}\BibitemShut
  {NoStop}%
\bibitem [{\citenamefont {Pantelides}(1978)}]{pantelides1978the}%
  \BibitemOpen
  \bibfield  {author} {\bibinfo {author} {\bibfnamefont {S.~T.}\ \bibnamefont
  {Pantelides}},\ }\bibinfo {title} {The electronic structure of impurities and
  other point defects in semiconductors},\ \href {\doibase
  10.1103/RevModPhys.50.797} {\bibfield  {journal} {\bibinfo  {journal} {Rev.
  Mod. Phys.}\ }\textbf {\bibinfo {volume} {50}},\ \bibinfo {pages} {797}
  (\bibinfo {year} {1978})}\BibitemShut {NoStop}%
\bibitem [{\citenamefont {Caroff}\ \emph {et~al.}(2009)\citenamefont {Caroff},
  \citenamefont {Dick}, \citenamefont {Johansson}, \citenamefont {Messing},
  \citenamefont {Deppert},\ and\ \citenamefont
  {Samuelson}}]{caroff2009controlled}%
  \BibitemOpen
  \bibfield  {author} {\bibinfo {author} {\bibfnamefont {P.}~\bibnamefont
  {Caroff}}, \bibinfo {author} {\bibfnamefont {K.~A.}\ \bibnamefont {Dick}},
  \bibinfo {author} {\bibfnamefont {J.}~\bibnamefont {Johansson}}, \bibinfo
  {author} {\bibfnamefont {M.~E.}\ \bibnamefont {Messing}}, \bibinfo {author}
  {\bibfnamefont {K.}~\bibnamefont {Deppert}}, \ and\ \bibinfo {author}
  {\bibfnamefont {L.}~\bibnamefont {Samuelson}},\ }\bibinfo {title} {Controlled
  polytypic and twin-plane superlattices in {III-V} nanowires},\ \href
  {https://doi.org/10.1038/nnano.2008.359} {\bibfield  {journal} {\bibinfo
  {journal} {Nat. Nanotechnol.}\ }\textbf {\bibinfo {volume} {4}},\ \bibinfo
  {pages} {50} (\bibinfo {year} {2009})}\BibitemShut {NoStop}%
\bibitem [{\citenamefont {Shtrikman}\ \emph {et~al.}(2009)\citenamefont
  {Shtrikman}, \citenamefont {Popovitz-Biro}, \citenamefont {Kretinin},
  \citenamefont {Houben}, \citenamefont {Heiblum}, \citenamefont {Buka{\l}a},
  \citenamefont {Galicka}, \citenamefont {Buczko},\ and\ \citenamefont
  {Kacman}}]{shtrikman2009method}%
  \BibitemOpen
  \bibfield  {author} {\bibinfo {author} {\bibfnamefont {H.}~\bibnamefont
  {Shtrikman}}, \bibinfo {author} {\bibfnamefont {R.}~\bibnamefont
  {Popovitz-Biro}}, \bibinfo {author} {\bibfnamefont {A.}~\bibnamefont
  {Kretinin}}, \bibinfo {author} {\bibfnamefont {L.}~\bibnamefont {Houben}},
  \bibinfo {author} {\bibfnamefont {M.}~\bibnamefont {Heiblum}}, \bibinfo
  {author} {\bibfnamefont {M.}~\bibnamefont {Buka{\l}a}}, \bibinfo {author}
  {\bibfnamefont {M.}~\bibnamefont {Galicka}}, \bibinfo {author} {\bibfnamefont
  {R.}~\bibnamefont {Buczko}}, \ and\ \bibinfo {author} {\bibfnamefont
  {P.}~\bibnamefont {Kacman}},\ }\bibinfo {title} {Method for suppression of
  stacking faults in wurtzite {III- V} nanowires},\ \href
  {https://doi.org/10.1021/nl803524s} {\bibfield  {journal} {\bibinfo
  {journal} {Nano Lett.}\ }\textbf {\bibinfo {volume} {9}},\ \bibinfo {pages}
  {1506} (\bibinfo {year} {2009})}\BibitemShut {NoStop}%
\bibitem [{\citenamefont {Brouwer}\ \emph
  {et~al.}(2011{\natexlab{a}})\citenamefont {Brouwer}, \citenamefont
  {Duckheim}, \citenamefont {Romito},\ and\ \citenamefont {von
  Oppen}}]{brouwer2011probability}%
  \BibitemOpen
  \bibfield  {author} {\bibinfo {author} {\bibfnamefont {P.~W.}\ \bibnamefont
  {Brouwer}}, \bibinfo {author} {\bibfnamefont {M.}~\bibnamefont {Duckheim}},
  \bibinfo {author} {\bibfnamefont {A.}~\bibnamefont {Romito}}, \ and\ \bibinfo
  {author} {\bibfnamefont {F.}~\bibnamefont {von Oppen}},\ }\bibinfo {title}
  {Probability Distribution of {Majorana} End-State Energies in Disordered
  Wires},\ \href {\doibase 10.1103/PhysRevLett.107.196804} {\bibfield
  {journal} {\bibinfo  {journal} {Phys. Rev. Lett.}\ }\textbf {\bibinfo
  {volume} {107}},\ \bibinfo {pages} {196804} (\bibinfo {year}
  {2011}{\natexlab{a}})}\BibitemShut {NoStop}%
\bibitem [{\citenamefont {Brouwer}\ \emph
  {et~al.}(2011{\natexlab{b}})\citenamefont {Brouwer}, \citenamefont
  {Duckheim}, \citenamefont {Romito},\ and\ \citenamefont {von
  Oppen}}]{brouwer2011topological}%
  \BibitemOpen
  \bibfield  {author} {\bibinfo {author} {\bibfnamefont {P.~W.}\ \bibnamefont
  {Brouwer}}, \bibinfo {author} {\bibfnamefont {M.}~\bibnamefont {Duckheim}},
  \bibinfo {author} {\bibfnamefont {A.}~\bibnamefont {Romito}}, \ and\ \bibinfo
  {author} {\bibfnamefont {F.}~\bibnamefont {von Oppen}},\ }\bibinfo {title}
  {Topological superconducting phases in disordered quantum wires with strong
  spin-orbit coupling},\ \href {\doibase 10.1103/PhysRevB.84.144526} {\bibfield
   {journal} {\bibinfo  {journal} {Phys. Rev. B}\ }\textbf {\bibinfo {volume}
  {84}},\ \bibinfo {pages} {144526} (\bibinfo {year}
  {2011}{\natexlab{b}})}\BibitemShut {NoStop}%
\bibitem [{\citenamefont {Pientka}\ \emph {et~al.}(2012)\citenamefont
  {Pientka}, \citenamefont {Kells}, \citenamefont {Romito}, \citenamefont
  {Brouwer},\ and\ \citenamefont {von Oppen}}]{pientka2012enhanced}%
  \BibitemOpen
  \bibfield  {author} {\bibinfo {author} {\bibfnamefont {F.}~\bibnamefont
  {Pientka}}, \bibinfo {author} {\bibfnamefont {G.}~\bibnamefont {Kells}},
  \bibinfo {author} {\bibfnamefont {A.}~\bibnamefont {Romito}}, \bibinfo
  {author} {\bibfnamefont {P.~W.}\ \bibnamefont {Brouwer}}, \ and\ \bibinfo
  {author} {\bibfnamefont {F.}~\bibnamefont {von Oppen}},\ }\bibinfo {title}
  {Enhanced Zero-Bias {Majorana} Peak in the Differential Tunneling Conductance
  of Disordered Multisubband Quantum-Wire/Superconductor Junctions},\ \href
  {\doibase 10.1103/PhysRevLett.109.227006} {\bibfield  {journal} {\bibinfo
  {journal} {Phys. Rev. Lett.}\ }\textbf {\bibinfo {volume} {109}},\ \bibinfo
  {pages} {227006} (\bibinfo {year} {2012})}\BibitemShut {NoStop}%
\bibitem [{\citenamefont {Liu}\ \emph {et~al.}(2012)\citenamefont {Liu},
  \citenamefont {Potter}, \citenamefont {Law},\ and\ \citenamefont
  {Lee}}]{liu2012zero}%
  \BibitemOpen
  \bibfield  {author} {\bibinfo {author} {\bibfnamefont {J.}~\bibnamefont
  {Liu}}, \bibinfo {author} {\bibfnamefont {A.~C.}\ \bibnamefont {Potter}},
  \bibinfo {author} {\bibfnamefont {K.~T.}\ \bibnamefont {Law}}, \ and\
  \bibinfo {author} {\bibfnamefont {P.~A.}\ \bibnamefont {Lee}},\ }\bibinfo
  {title} {Zero-Bias Peaks in the Tunneling Conductance of Spin-Orbit-Coupled
  Superconducting Wires with and without {Majorana} End-States},\ \href
  {\doibase 10.1103/PhysRevLett.109.267002} {\bibfield  {journal} {\bibinfo
  {journal} {Phys. Rev. Lett.}\ }\textbf {\bibinfo {volume} {109}},\ \bibinfo
  {pages} {267002} (\bibinfo {year} {2012})}\BibitemShut {NoStop}%
\bibitem [{\citenamefont {Kells}\ \emph {et~al.}(2012)\citenamefont {Kells},
  \citenamefont {Meidan},\ and\ \citenamefont {Brouwer}}]{kells2012near}%
  \BibitemOpen
  \bibfield  {author} {\bibinfo {author} {\bibfnamefont {G.}~\bibnamefont
  {Kells}}, \bibinfo {author} {\bibfnamefont {D.}~\bibnamefont {Meidan}}, \
  and\ \bibinfo {author} {\bibfnamefont {P.~W.}\ \bibnamefont {Brouwer}},\
  }\bibinfo {title} {Near-zero-energy end states in topologically trivial
  spin-orbit coupled superconducting nanowires with a smooth confinement},\
  \href {\doibase 10.1103/PhysRevB.86.100503} {\bibfield  {journal} {\bibinfo
  {journal} {Phys. Rev. B}\ }\textbf {\bibinfo {volume} {86}},\ \bibinfo
  {pages} {100503(R)} (\bibinfo {year} {2012})}\BibitemShut {NoStop}%
\bibitem [{\citenamefont {Pikulin}\ \emph {et~al.}(2012)\citenamefont
  {Pikulin}, \citenamefont {Dahlhaus}, \citenamefont {Wimmer}, \citenamefont
  {Schomerus},\ and\ \citenamefont {Beenakker}}]{pikulin2012zero}%
  \BibitemOpen
  \bibfield  {author} {\bibinfo {author} {\bibfnamefont {D.~I.}\ \bibnamefont
  {Pikulin}}, \bibinfo {author} {\bibfnamefont {J.}~\bibnamefont {Dahlhaus}},
  \bibinfo {author} {\bibfnamefont {M.}~\bibnamefont {Wimmer}}, \bibinfo
  {author} {\bibfnamefont {H.}~\bibnamefont {Schomerus}}, \ and\ \bibinfo
  {author} {\bibfnamefont {C.}~\bibnamefont {Beenakker}},\ }\bibinfo {title} {A
  zero-voltage conductance peak from weak antilocalization in a {Majorana}
  nanowire},\ \href {https://doi.org/10.1088/1367-2630/14/12/125011} {\bibfield
   {journal} {\bibinfo  {journal} {New J. Phys.}\ }\textbf {\bibinfo {volume}
  {14}},\ \bibinfo {pages} {125011} (\bibinfo {year} {2012})}\BibitemShut
  {NoStop}%
\bibitem [{\citenamefont {Rainis}\ \emph {et~al.}(2013)\citenamefont {Rainis},
  \citenamefont {Trifunovic}, \citenamefont {Klinovaja},\ and\ \citenamefont
  {Loss}}]{rainis2013towards}%
  \BibitemOpen
  \bibfield  {author} {\bibinfo {author} {\bibfnamefont {D.}~\bibnamefont
  {Rainis}}, \bibinfo {author} {\bibfnamefont {L.}~\bibnamefont {Trifunovic}},
  \bibinfo {author} {\bibfnamefont {J.}~\bibnamefont {Klinovaja}}, \ and\
  \bibinfo {author} {\bibfnamefont {D.}~\bibnamefont {Loss}},\ }\bibinfo
  {title} {Towards a realistic transport modeling in a superconducting nanowire
  with {Majorana} fermions},\ \href
  {https://doi.org/10.1103/PhysRevB.87.024515} {\bibfield  {journal} {\bibinfo
  {journal} {Phys. Rev. B}\ }\textbf {\bibinfo {volume} {87}},\ \bibinfo
  {pages} {024515} (\bibinfo {year} {2013})}\BibitemShut {NoStop}%
\bibitem [{\citenamefont {Roy}\ \emph {et~al.}(2013)\citenamefont {Roy},
  \citenamefont {Bondyopadhaya},\ and\ \citenamefont
  {Tewari}}]{roy2013topologically}%
  \BibitemOpen
  \bibfield  {author} {\bibinfo {author} {\bibfnamefont {D.}~\bibnamefont
  {Roy}}, \bibinfo {author} {\bibfnamefont {N.}~\bibnamefont {Bondyopadhaya}},
  \ and\ \bibinfo {author} {\bibfnamefont {S.}~\bibnamefont {Tewari}},\
  }\bibinfo {title} {Topologically trivial zero-bias conductance peak in
  semiconductor {Majorana} wires from boundary effects},\ \href {\doibase
  10.1103/PhysRevB.88.020502} {\bibfield  {journal} {\bibinfo  {journal} {Phys.
  Rev. B}\ }\textbf {\bibinfo {volume} {88}},\ \bibinfo {pages} {020502(R)}
  (\bibinfo {year} {2013})}\BibitemShut {NoStop}%
\bibitem [{\citenamefont {Stanescu}\ and\ \citenamefont
  {Tewari}(2013{\natexlab{b}})}]{stanescu2013disentangling}%
  \BibitemOpen
  \bibfield  {author} {\bibinfo {author} {\bibfnamefont {T.~D.}\ \bibnamefont
  {Stanescu}}\ and\ \bibinfo {author} {\bibfnamefont {S.}~\bibnamefont
  {Tewari}},\ }\bibinfo {title} {Disentangling {Majorana} fermions from
  topologically trivial low-energy states in semiconductor {Majorana} wires},\
  \href {\doibase 10.1103/PhysRevB.87.140504} {\bibfield  {journal} {\bibinfo
  {journal} {Phys. Rev. B}\ }\textbf {\bibinfo {volume} {87}},\ \bibinfo
  {pages} {140504(R)} (\bibinfo {year} {2013}{\natexlab{b}})}\BibitemShut
  {NoStop}%
\bibitem [{\citenamefont {Hui}\ \emph {et~al.}(2015{\natexlab{a}})\citenamefont
  {Hui}, \citenamefont {Sau},\ and\ \citenamefont {Das~Sarma}}]{hui2015bulk}%
  \BibitemOpen
  \bibfield  {author} {\bibinfo {author} {\bibfnamefont {H.-Y.}\ \bibnamefont
  {Hui}}, \bibinfo {author} {\bibfnamefont {J.~D.}\ \bibnamefont {Sau}}, \ and\
  \bibinfo {author} {\bibfnamefont {S.}~\bibnamefont {Das~Sarma}},\ }\bibinfo
  {title} {Bulk disorder in the superconductor affects proximity-induced
  topological superconductivity},\ \href {\doibase 10.1103/PhysRevB.92.174512}
  {\bibfield  {journal} {\bibinfo  {journal} {Phys. Rev. B}\ }\textbf {\bibinfo
  {volume} {92}},\ \bibinfo {pages} {174512} (\bibinfo {year}
  {2015}{\natexlab{a}})}\BibitemShut {NoStop}%
\bibitem [{\citenamefont {Klinovaja}\ and\ \citenamefont
  {Loss}(2015)}]{klinovaja2015fermionic}%
  \BibitemOpen
  \bibfield  {author} {\bibinfo {author} {\bibfnamefont {J.}~\bibnamefont
  {Klinovaja}}\ and\ \bibinfo {author} {\bibfnamefont {D.}~\bibnamefont
  {Loss}},\ }\bibinfo {title} {Fermionic and {Majorana} bound states in hybrid
  nanowires with non-uniform spin-orbit interaction},\ \href
  {https://doi.org/10.1140/epjb/e2015-50882-2} {\bibfield  {journal} {\bibinfo
  {journal} {Eur. Phys. J. B}\ }\textbf {\bibinfo {volume} {88}},\ \bibinfo
  {pages} {62} (\bibinfo {year} {2015})}\BibitemShut {NoStop}%
\bibitem [{\citenamefont {Cole}\ \emph {et~al.}(2016)\citenamefont {Cole},
  \citenamefont {Sau},\ and\ \citenamefont {Das~Sarma}}]{cole2016proximity}%
  \BibitemOpen
  \bibfield  {author} {\bibinfo {author} {\bibfnamefont {W.~S.}\ \bibnamefont
  {Cole}}, \bibinfo {author} {\bibfnamefont {J.~D.}\ \bibnamefont {Sau}}, \
  and\ \bibinfo {author} {\bibfnamefont {S.}~\bibnamefont {Das~Sarma}},\
  }\bibinfo {title} {Proximity effect and {Majorana} bound states in clean
  semiconductor nanowires coupled to disordered superconductors},\ \href
  {\doibase 10.1103/PhysRevB.94.140505} {\bibfield  {journal} {\bibinfo
  {journal} {Phys. Rev. B}\ }\textbf {\bibinfo {volume} {94}},\ \bibinfo
  {pages} {140505(R)} (\bibinfo {year} {2016})}\BibitemShut {NoStop}%
\bibitem [{\citenamefont {Liu}\ \emph {et~al.}(2017)\citenamefont {Liu},
  \citenamefont {Sau}, \citenamefont {Stanescu},\ and\ \citenamefont {{Das
  Sarma}}}]{liu2017andreev}%
  \BibitemOpen
  \bibfield  {author} {\bibinfo {author} {\bibfnamefont {C.-X.}\ \bibnamefont
  {Liu}}, \bibinfo {author} {\bibfnamefont {J.~D.}\ \bibnamefont {Sau}},
  \bibinfo {author} {\bibfnamefont {T.~D.}\ \bibnamefont {Stanescu}}, \ and\
  \bibinfo {author} {\bibfnamefont {S.}~\bibnamefont {{Das Sarma}}},\ }\bibinfo
  {title} {{Andreev} bound states versus {Majorana} bound states in quantum
  dot-nanowire-superconductor hybrid structures: Trivial versus topological
  zero-bias conductance peaks},\ \href
  {https://doi.org/10.1103/PhysRevB.96.075161} {\bibfield  {journal} {\bibinfo
  {journal} {Phys. Rev. B}\ }\textbf {\bibinfo {volume} {96}},\ \bibinfo
  {pages} {075161} (\bibinfo {year} {2017})}\BibitemShut {NoStop}%
\bibitem [{\citenamefont {Liu}\ \emph {et~al.}(2018)\citenamefont {Liu},
  \citenamefont {Rossi},\ and\ \citenamefont {Lutchyn}}]{liu2018impurity}%
  \BibitemOpen
  \bibfield  {author} {\bibinfo {author} {\bibfnamefont {D.~E.}\ \bibnamefont
  {Liu}}, \bibinfo {author} {\bibfnamefont {E.}~\bibnamefont {Rossi}}, \ and\
  \bibinfo {author} {\bibfnamefont {R.~M.}\ \bibnamefont {Lutchyn}},\ }\bibinfo
  {title} {Impurity-induced states in superconducting heterostructures},\ \href
  {\doibase 10.1103/PhysRevB.97.161408} {\bibfield  {journal} {\bibinfo
  {journal} {Phys. Rev. B}\ }\textbf {\bibinfo {volume} {97}},\ \bibinfo
  {pages} {161408(R)} (\bibinfo {year} {2018})}\BibitemShut {NoStop}%
\bibitem [{\citenamefont {Moore}\ \emph {et~al.}(2018)\citenamefont {Moore},
  \citenamefont {Zeng}, \citenamefont {Stanescu},\ and\ \citenamefont
  {Tewari}}]{moore2018quantized}%
  \BibitemOpen
  \bibfield  {author} {\bibinfo {author} {\bibfnamefont {C.}~\bibnamefont
  {Moore}}, \bibinfo {author} {\bibfnamefont {C.}~\bibnamefont {Zeng}},
  \bibinfo {author} {\bibfnamefont {T.~D.}\ \bibnamefont {Stanescu}}, \ and\
  \bibinfo {author} {\bibfnamefont {S.}~\bibnamefont {Tewari}},\ }\bibinfo
  {title} {Quantized zero-bias conductance plateau in
  semiconductor-superconductor heterostructures without topological {Majorana}
  zero modes},\ \href {\doibase 10.1103/PhysRevB.98.155314} {\bibfield
  {journal} {\bibinfo  {journal} {Phys. Rev. B}\ }\textbf {\bibinfo {volume}
  {98}},\ \bibinfo {pages} {155314} (\bibinfo {year} {2018})}\BibitemShut
  {NoStop}%
\bibitem [{\citenamefont {Fleckenstein}\ \emph {et~al.}(2018)\citenamefont
  {Fleckenstein}, \citenamefont {Dom\'{\i}nguez}, \citenamefont
  {Traverso~Ziani},\ and\ \citenamefont
  {Trauzettel}}]{fleckenstein2018decaying}%
  \BibitemOpen
  \bibfield  {author} {\bibinfo {author} {\bibfnamefont {C.}~\bibnamefont
  {Fleckenstein}}, \bibinfo {author} {\bibfnamefont {F.}~\bibnamefont
  {Dom\'{\i}nguez}}, \bibinfo {author} {\bibfnamefont {N.}~\bibnamefont
  {Traverso~Ziani}}, \ and\ \bibinfo {author} {\bibfnamefont {B.}~\bibnamefont
  {Trauzettel}},\ }\bibinfo {title} {Decaying spectral oscillations in a
  {Majorana} wire with finite coherence length},\ \href {\doibase
  10.1103/PhysRevB.97.155425} {\bibfield  {journal} {\bibinfo  {journal} {Phys.
  Rev. B}\ }\textbf {\bibinfo {volume} {97}},\ \bibinfo {pages} {155425}
  (\bibinfo {year} {2018})}\BibitemShut {NoStop}%
\bibitem [{\citenamefont {Cao}\ \emph {et~al.}(2019)\citenamefont {Cao},
  \citenamefont {Zhang}, \citenamefont {L\"u}, \citenamefont {He},
  \citenamefont {Lu},\ and\ \citenamefont {Xie}}]{cao2019decay}%
  \BibitemOpen
  \bibfield  {author} {\bibinfo {author} {\bibfnamefont {Z.}~\bibnamefont
  {Cao}}, \bibinfo {author} {\bibfnamefont {H.}~\bibnamefont {Zhang}}, \bibinfo
  {author} {\bibfnamefont {H.-F.}\ \bibnamefont {L\"u}}, \bibinfo {author}
  {\bibfnamefont {W.-X.}\ \bibnamefont {He}}, \bibinfo {author} {\bibfnamefont
  {H.-Z.}\ \bibnamefont {Lu}}, \ and\ \bibinfo {author} {\bibfnamefont {X.~C.}\
  \bibnamefont {Xie}},\ }\bibinfo {title} {Decays of {Majorana} or {Andreev}
  Oscillations Induced by Steplike Spin-Orbit Coupling},\ \href {\doibase
  10.1103/PhysRevLett.122.147701} {\bibfield  {journal} {\bibinfo  {journal}
  {Phys. Rev. Lett.}\ }\textbf {\bibinfo {volume} {122}},\ \bibinfo {pages}
  {147701} (\bibinfo {year} {2019})}\BibitemShut {NoStop}%
\bibitem [{\citenamefont {Vuik}\ \emph {et~al.}(2019)\citenamefont {Vuik},
  \citenamefont {Nijholt}, \citenamefont {Akhmerov},\ and\ \citenamefont
  {Wimmer}}]{vuik2019reproducing}%
  \BibitemOpen
  \bibfield  {author} {\bibinfo {author} {\bibfnamefont {A.}~\bibnamefont
  {Vuik}}, \bibinfo {author} {\bibfnamefont {B.}~\bibnamefont {Nijholt}},
  \bibinfo {author} {\bibfnamefont {A.}~\bibnamefont {Akhmerov}}, \ and\
  \bibinfo {author} {\bibfnamefont {M.}~\bibnamefont {Wimmer}},\ }\bibinfo
  {title} {Reproducing topological properties with quasi-Majorana states},\
  \href {https://scipost.org/10.21468/SciPostPhys.7.5.061} {\bibfield
  {journal} {\bibinfo  {journal} {SciPost Phys.}\ }\textbf {\bibinfo {volume}
  {7}},\ \bibinfo {pages} {061} (\bibinfo {year} {2019})}\BibitemShut {NoStop}%
\bibitem [{\citenamefont {Kiendl}\ \emph {et~al.}(2019)\citenamefont {Kiendl},
  \citenamefont {von Oppen},\ and\ \citenamefont
  {Brouwer}}]{kiendl2019proximity}%
  \BibitemOpen
  \bibfield  {author} {\bibinfo {author} {\bibfnamefont {T.}~\bibnamefont
  {Kiendl}}, \bibinfo {author} {\bibfnamefont {F.}~\bibnamefont {von Oppen}}, \
  and\ \bibinfo {author} {\bibfnamefont {P.~W.}\ \bibnamefont {Brouwer}},\
  }\bibinfo {title} {Proximity-induced gap in nanowires with a thin
  superconducting shell},\ \href {\doibase 10.1103/PhysRevB.100.035426}
  {\bibfield  {journal} {\bibinfo  {journal} {Phys. Rev. B}\ }\textbf {\bibinfo
  {volume} {100}},\ \bibinfo {pages} {035426} (\bibinfo {year}
  {2019})}\BibitemShut {NoStop}%
\bibitem [{\citenamefont {Pan}\ \emph {et~al.}(2021)\citenamefont {Pan},
  \citenamefont {Liu}, \citenamefont {Wimmer},\ and\ \citenamefont
  {Das~Sarma}}]{pan2021quantized}%
  \BibitemOpen
  \bibfield  {author} {\bibinfo {author} {\bibfnamefont {H.}~\bibnamefont
  {Pan}}, \bibinfo {author} {\bibfnamefont {C.-X.}\ \bibnamefont {Liu}},
  \bibinfo {author} {\bibfnamefont {M.}~\bibnamefont {Wimmer}}, \ and\ \bibinfo
  {author} {\bibfnamefont {S.}~\bibnamefont {Das~Sarma}},\ }\bibinfo {title}
  {Quantized and unquantized zero-bias tunneling conductance peaks in
  {Majorana} nanowires: Conductance below and above $2{e}^{2}/h$},\ \href
  {\doibase 10.1103/PhysRevB.103.214502} {\bibfield  {journal} {\bibinfo
  {journal} {Phys. Rev. B}\ }\textbf {\bibinfo {volume} {103}},\ \bibinfo
  {pages} {214502} (\bibinfo {year} {2021})}\BibitemShut {NoStop}%
\bibitem [{\citenamefont {Liu}(2013)}]{liu2013filter}%
  \BibitemOpen
  \bibfield  {author} {\bibinfo {author} {\bibfnamefont {D.~E.}\ \bibnamefont
  {Liu}},\ }\bibinfo {title} {Proposed Method for Tunneling Spectroscopy with
  Ohmic Dissipation Using Resistive Electrodes: A Possible {Majorana} Filter},\
  \href {\doibase 10.1103/PhysRevLett.111.207003} {\bibfield  {journal}
  {\bibinfo  {journal} {Phys. Rev. Lett.}\ }\textbf {\bibinfo {volume} {111}},\
  \bibinfo {pages} {207003} (\bibinfo {year} {2013})}\BibitemShut {NoStop}%
\bibitem [{\citenamefont {Liu}\ \emph {et~al.}(2022)\citenamefont {Liu},
  \citenamefont {Zhang}, \citenamefont {Cao}, \citenamefont {Zhang},\ and\
  \citenamefont {Liu}}]{liu2022universal}%
  \BibitemOpen
  \bibfield  {author} {\bibinfo {author} {\bibfnamefont {D.}~\bibnamefont
  {Liu}}, \bibinfo {author} {\bibfnamefont {G.}~\bibnamefont {Zhang}}, \bibinfo
  {author} {\bibfnamefont {Z.}~\bibnamefont {Cao}}, \bibinfo {author}
  {\bibfnamefont {H.}~\bibnamefont {Zhang}}, \ and\ \bibinfo {author}
  {\bibfnamefont {D.~E.}\ \bibnamefont {Liu}},\ }\bibinfo {title} {Universal
  Conductance Scaling of {Andreev} Reflections Using a Dissipative Probe},\
  \href {\doibase 10.1103/PhysRevLett.128.076802} {\bibfield  {journal}
  {\bibinfo  {journal} {Phys. Rev. Lett.}\ }\textbf {\bibinfo {volume} {128}},\
  \bibinfo {pages} {076802} (\bibinfo {year} {2022})}\BibitemShut {NoStop}%
\bibitem [{\citenamefont {Zhang}\ \emph {et~al.}(2022)\citenamefont {Zhang},
  \citenamefont {Wang}, \citenamefont {Pan}, \citenamefont {Li}, \citenamefont
  {Lu}, \citenamefont {Li}, \citenamefont {Zhang}, \citenamefont {Liu},
  \citenamefont {Cao}, \citenamefont {Liu}, \citenamefont {Wen}, \citenamefont
  {Liao}, \citenamefont {Zhuo}, \citenamefont {Shang}, \citenamefont {Liu},
  \citenamefont {Zhao},\ and\ \citenamefont {Zhang}}]{zhang2022suppressing}%
  \BibitemOpen
  \bibfield  {author} {\bibinfo {author} {\bibfnamefont {S.}~\bibnamefont
  {Zhang}}, \bibinfo {author} {\bibfnamefont {Z.}~\bibnamefont {Wang}},
  \bibinfo {author} {\bibfnamefont {D.}~\bibnamefont {Pan}}, \bibinfo {author}
  {\bibfnamefont {H.}~\bibnamefont {Li}}, \bibinfo {author} {\bibfnamefont
  {S.}~\bibnamefont {Lu}}, \bibinfo {author} {\bibfnamefont {Z.}~\bibnamefont
  {Li}}, \bibinfo {author} {\bibfnamefont {G.}~\bibnamefont {Zhang}}, \bibinfo
  {author} {\bibfnamefont {D.}~\bibnamefont {Liu}}, \bibinfo {author}
  {\bibfnamefont {Z.}~\bibnamefont {Cao}}, \bibinfo {author} {\bibfnamefont
  {L.}~\bibnamefont {Liu}}, \bibinfo {author} {\bibfnamefont {L.}~\bibnamefont
  {Wen}}, \bibinfo {author} {\bibfnamefont {D.}~\bibnamefont {Liao}}, \bibinfo
  {author} {\bibfnamefont {R.}~\bibnamefont {Zhuo}}, \bibinfo {author}
  {\bibfnamefont {R.}~\bibnamefont {Shang}}, \bibinfo {author} {\bibfnamefont
  {D.~E.}\ \bibnamefont {Liu}}, \bibinfo {author} {\bibfnamefont
  {J.}~\bibnamefont {Zhao}}, \ and\ \bibinfo {author} {\bibfnamefont
  {H.}~\bibnamefont {Zhang}},\ }\bibinfo {title} {Suppressing {Andreev} Bound
  State Zero Bias Peaks Using a Strongly Dissipative Lead},\ \href {\doibase
  10.1103/PhysRevLett.128.076803} {\bibfield  {journal} {\bibinfo  {journal}
  {Phys. Rev. Lett.}\ }\textbf {\bibinfo {volume} {128}},\ \bibinfo {pages}
  {076803} (\bibinfo {year} {2022})}\BibitemShut {NoStop}%
\bibitem [{\citenamefont {Lutchyn}\ \emph {et~al.}(2018)\citenamefont
  {Lutchyn}, \citenamefont {Bakkers}, \citenamefont {Kouwenhoven},
  \citenamefont {Krogstrup}, \citenamefont {Marcus},\ and\ \citenamefont
  {Oreg}}]{lutchyn2018majorana}%
  \BibitemOpen
  \bibfield  {author} {\bibinfo {author} {\bibfnamefont {R.}~\bibnamefont
  {Lutchyn}}, \bibinfo {author} {\bibfnamefont {E.}~\bibnamefont {Bakkers}},
  \bibinfo {author} {\bibfnamefont {L.}~\bibnamefont {Kouwenhoven}}, \bibinfo
  {author} {\bibfnamefont {P.}~\bibnamefont {Krogstrup}}, \bibinfo {author}
  {\bibfnamefont {C.}~\bibnamefont {Marcus}}, \ and\ \bibinfo {author}
  {\bibfnamefont {Y.}~\bibnamefont {Oreg}},\ }\bibinfo {title} {{Majorana} zero
  modes in superconductor-semiconductor heterostructures},\ \href
  {https://doi.org/10.1038/s41578-018-0003-1} {\bibfield  {journal} {\bibinfo
  {journal} {Nat. Rev. Mater.}\ }\textbf {\bibinfo {volume} {3}},\ \bibinfo
  {pages} {52} (\bibinfo {year} {2018})}\BibitemShut {NoStop}%
\bibitem [{\citenamefont {Giustino}\ \emph {et~al.}(2020)\citenamefont
  {Giustino}, \citenamefont {Lee}, \citenamefont {Trier}, \citenamefont
  {Bibes}, \citenamefont {Winter}, \citenamefont {Valent{\'\i}}, \citenamefont
  {Son}, \citenamefont {Taillefer}, \citenamefont {Heil}, \citenamefont
  {Figueroa} \emph {et~al.}}]{giustino2020}%
  \BibitemOpen
  \bibfield  {author} {\bibinfo {author} {\bibfnamefont {F.}~\bibnamefont
  {Giustino}}, \bibinfo {author} {\bibfnamefont {J.~H.}\ \bibnamefont {Lee}},
  \bibinfo {author} {\bibfnamefont {F.}~\bibnamefont {Trier}}, \bibinfo
  {author} {\bibfnamefont {M.}~\bibnamefont {Bibes}}, \bibinfo {author}
  {\bibfnamefont {S.~M.}\ \bibnamefont {Winter}}, \bibinfo {author}
  {\bibfnamefont {R.}~\bibnamefont {Valent{\'\i}}}, \bibinfo {author}
  {\bibfnamefont {Y.-W.}\ \bibnamefont {Son}}, \bibinfo {author} {\bibfnamefont
  {L.}~\bibnamefont {Taillefer}}, \bibinfo {author} {\bibfnamefont
  {C.}~\bibnamefont {Heil}}, \bibinfo {author} {\bibfnamefont {A.~I.}\
  \bibnamefont {Figueroa}},  \emph {et~al.},\ }\bibinfo {title} {The 2021
  quantum materials roadmap},\ \href {https://doi.org/10.1088/2515-7639/abb74e}
  {\bibfield  {journal} {\bibinfo  {journal} {J. Phys. Mater.}\ }\textbf
  {\bibinfo {volume} {3}},\ \bibinfo {pages} {042006} (\bibinfo {year}
  {2020})}\BibitemShut {NoStop}%
\bibitem [{\citenamefont {Pendharkar}\ \emph {et~al.}(2021)\citenamefont
  {Pendharkar}, \citenamefont {Zhang}, \citenamefont {Wu}, \citenamefont
  {Zarassi}, \citenamefont {Zhang}, \citenamefont {Dempsey}, \citenamefont
  {Lee}, \citenamefont {Harrington}, \citenamefont {Badawy}, \citenamefont
  {Gazibegovic} \emph {et~al.}}]{pendharkar2021parity}%
  \BibitemOpen
  \bibfield  {author} {\bibinfo {author} {\bibfnamefont {M.}~\bibnamefont
  {Pendharkar}}, \bibinfo {author} {\bibfnamefont {B.}~\bibnamefont {Zhang}},
  \bibinfo {author} {\bibfnamefont {H.}~\bibnamefont {Wu}}, \bibinfo {author}
  {\bibfnamefont {A.}~\bibnamefont {Zarassi}}, \bibinfo {author} {\bibfnamefont
  {P.}~\bibnamefont {Zhang}}, \bibinfo {author} {\bibfnamefont
  {C.}~\bibnamefont {Dempsey}}, \bibinfo {author} {\bibfnamefont
  {J.}~\bibnamefont {Lee}}, \bibinfo {author} {\bibfnamefont {S.}~\bibnamefont
  {Harrington}}, \bibinfo {author} {\bibfnamefont {G.}~\bibnamefont {Badawy}},
  \bibinfo {author} {\bibfnamefont {S.}~\bibnamefont {Gazibegovic}},  \emph
  {et~al.},\ }\bibinfo {title} {Parity-preserving and magnetic field--resilient
  superconductivity in {InSb} nanowires with {Sn} shells},\ \href
  {https://doi.org/10.1126/science.aba5211} {\bibfield  {journal} {\bibinfo
  {journal} {Science}\ }\textbf {\bibinfo {volume} {372}},\ \bibinfo {pages}
  {508} (\bibinfo {year} {2021})}\BibitemShut {NoStop}%
\bibitem [{\citenamefont {Kanne}\ \emph {et~al.}(2021)\citenamefont {Kanne},
  \citenamefont {Marnauza}, \citenamefont {Olsteins}, \citenamefont {Carrad},
  \citenamefont {Sestoft}, \citenamefont {de~Bruijckere}, \citenamefont {Zeng},
  \citenamefont {Johnson}, \citenamefont {Olsson}, \citenamefont
  {Grove-Rasmussen} \emph {et~al.}}]{kanne2021epitaxial}%
  \BibitemOpen
  \bibfield  {author} {\bibinfo {author} {\bibfnamefont {T.}~\bibnamefont
  {Kanne}}, \bibinfo {author} {\bibfnamefont {M.}~\bibnamefont {Marnauza}},
  \bibinfo {author} {\bibfnamefont {D.}~\bibnamefont {Olsteins}}, \bibinfo
  {author} {\bibfnamefont {D.~J.}\ \bibnamefont {Carrad}}, \bibinfo {author}
  {\bibfnamefont {J.~E.}\ \bibnamefont {Sestoft}}, \bibinfo {author}
  {\bibfnamefont {J.}~\bibnamefont {de~Bruijckere}}, \bibinfo {author}
  {\bibfnamefont {L.}~\bibnamefont {Zeng}}, \bibinfo {author} {\bibfnamefont
  {E.}~\bibnamefont {Johnson}}, \bibinfo {author} {\bibfnamefont
  {E.}~\bibnamefont {Olsson}}, \bibinfo {author} {\bibfnamefont
  {K.}~\bibnamefont {Grove-Rasmussen}},  \emph {et~al.},\ }\bibinfo {title}
  {Epitaxial {Pb} on {InAs} nanowires for quantum devices},\ \href
  {https://doi.org/10.1038/s41565-021-00900-9} {\bibfield  {journal} {\bibinfo
  {journal} {Nat. Nanotechnol.}\ }\textbf {\bibinfo {volume} {16}},\ \bibinfo
  {pages} {776} (\bibinfo {year} {2021})}\BibitemShut {NoStop}%
\bibitem [{\citenamefont {Jung}\ \emph {et~al.}(2021)\citenamefont {Jung},
  \citenamefont {Op~het Veld}, \citenamefont {Benoist}, \citenamefont {van~der
  Molen}, \citenamefont {Manders}, \citenamefont {Verheijen},\ and\
  \citenamefont {Bakkers}}]{jung2021universal}%
  \BibitemOpen
  \bibfield  {author} {\bibinfo {author} {\bibfnamefont {J.}~\bibnamefont
  {Jung}}, \bibinfo {author} {\bibfnamefont {R.~L.}\ \bibnamefont {Op~het
  Veld}}, \bibinfo {author} {\bibfnamefont {R.}~\bibnamefont {Benoist}},
  \bibinfo {author} {\bibfnamefont {O.~A.}\ \bibnamefont {van~der Molen}},
  \bibinfo {author} {\bibfnamefont {C.}~\bibnamefont {Manders}}, \bibinfo
  {author} {\bibfnamefont {M.~A.}\ \bibnamefont {Verheijen}}, \ and\ \bibinfo
  {author} {\bibfnamefont {E.~P.}\ \bibnamefont {Bakkers}},\ }\bibinfo {title}
  {Universal Platform for Scalable Semiconductor-Superconductor Nanowire
  Networks},\ \href {https://doi.org/10.1002/adfm.202103062} {\bibfield
  {journal} {\bibinfo  {journal} {Adv. Funct. Mater.}\ }\textbf {\bibinfo
  {volume} {2021}},\ \bibinfo {pages} {2103062} (\bibinfo {year}
  {2021})}\BibitemShut {NoStop}%
\bibitem [{\citenamefont {Springholz}\ and\ \citenamefont
  {Bauer}(2013)}]{klingshirn2013}%
  \BibitemOpen
  \bibfield  {author} {\bibinfo {author} {\bibfnamefont {G.}~\bibnamefont
  {Springholz}}\ and\ \bibinfo {author} {\bibfnamefont {G.}~\bibnamefont
  {Bauer}},\ }\href@noop {} {\emph {\bibinfo {title} {Semiconductor Quantum
  Structures--Growth and Structuring}}},\ edited by\ \bibinfo {editor}
  {\bibfnamefont {C.}~\bibnamefont {Klingshirn}}\ (\bibinfo  {publisher}
  {Springer},\ \bibinfo {year} {2013})\BibitemShut {NoStop}%
\bibitem [{\citenamefont {Kamphuis}(2021)}]{kamphuis2021towards}%
  \BibitemOpen
  \bibfield  {author} {\bibinfo {author} {\bibfnamefont {M.~J.~G.}\
  \bibnamefont {Kamphuis}},\ }\href@noop {} {\emph {\bibinfo {title} {Towards
  quantum transport in single-crystalline {PbTe} nanowire MOSFET devices}}}\
  (\bibinfo  {publisher} {Bachelor thesis, Eindhoven University of
  Technology},\ \bibinfo {year} {2021})\BibitemShut {NoStop}%
\bibitem [{\citenamefont {Schellingerhout}\ \emph {et~al.}(2021)\citenamefont
  {Schellingerhout}, \citenamefont {de~Jong}, \citenamefont {Gomanko},
  \citenamefont {Guan}, \citenamefont {Jiang}, \citenamefont {Hoskam},
  \citenamefont {Koelling}, \citenamefont {Moutanabbir}, \citenamefont
  {Verheijen}, \citenamefont {Frolov} \emph
  {et~al.}}]{schellingerhout2021growth}%
  \BibitemOpen
  \bibfield  {author} {\bibinfo {author} {\bibfnamefont {S.~G.}\ \bibnamefont
  {Schellingerhout}}, \bibinfo {author} {\bibfnamefont {E.~J.}\ \bibnamefont
  {de~Jong}}, \bibinfo {author} {\bibfnamefont {M.}~\bibnamefont {Gomanko}},
  \bibinfo {author} {\bibfnamefont {X.}~\bibnamefont {Guan}}, \bibinfo {author}
  {\bibfnamefont {Y.}~\bibnamefont {Jiang}}, \bibinfo {author} {\bibfnamefont
  {M.~S.~M.}\ \bibnamefont {Hoskam}}, \bibinfo {author} {\bibfnamefont
  {S.}~\bibnamefont {Koelling}}, \bibinfo {author} {\bibfnamefont
  {O.}~\bibnamefont {Moutanabbir}}, \bibinfo {author} {\bibfnamefont {M.~A.}\
  \bibnamefont {Verheijen}}, \bibinfo {author} {\bibfnamefont {S.~M.}\
  \bibnamefont {Frolov}},  \emph {et~al.},\ }\bibinfo {title} {Growth of {PbTe}
  nanowires by Molecular Beam Epitaxy},\ \href
  {https://arxiv.org/abs/2110.12789} {\bibfield  {journal} {\bibinfo  {journal}
  {arXiv: 2110.12789}\ } (\bibinfo {year} {2021})}\BibitemShut {NoStop}%
\bibitem [{\citenamefont {Schlatmann}(2021)}]{schlatmann2021josephson}%
  \BibitemOpen
  \bibfield  {author} {\bibinfo {author} {\bibfnamefont {J.}~\bibnamefont
  {Schlatmann}},\ }\href@noop {} {\emph {\bibinfo {title} {Josephson junction
  characterization of {PbTe} nanowires--Towards {Majorana} devices in
  semiconductor-superconductor hybrid nanowires}}}\ (\bibinfo  {publisher}
  {Master thesis, Eindhoven University of Technology},\ \bibinfo {year}
  {2021})\BibitemShut {NoStop}%
\bibitem [{\citenamefont {Jiang}\ \emph {et~al.}(2021)\citenamefont {Jiang},
  \citenamefont {Yang}, \citenamefont {Li}, \citenamefont {Song}, \citenamefont
  {Miao}, \citenamefont {Tong}, \citenamefont {Geng}, \citenamefont {Gao},
  \citenamefont {Li}, \citenamefont {Zhang} \emph
  {et~al.}}]{jiang2021selective}%
  \BibitemOpen
  \bibfield  {author} {\bibinfo {author} {\bibfnamefont {Y.}~\bibnamefont
  {Jiang}}, \bibinfo {author} {\bibfnamefont {S.}~\bibnamefont {Yang}},
  \bibinfo {author} {\bibfnamefont {L.}~\bibnamefont {Li}}, \bibinfo {author}
  {\bibfnamefont {W.}~\bibnamefont {Song}}, \bibinfo {author} {\bibfnamefont
  {W.}~\bibnamefont {Miao}}, \bibinfo {author} {\bibfnamefont {B.}~\bibnamefont
  {Tong}}, \bibinfo {author} {\bibfnamefont {Z.}~\bibnamefont {Geng}}, \bibinfo
  {author} {\bibfnamefont {Y.}~\bibnamefont {Gao}}, \bibinfo {author}
  {\bibfnamefont {R.}~\bibnamefont {Li}}, \bibinfo {author} {\bibfnamefont
  {Q.}~\bibnamefont {Zhang}},  \emph {et~al.},\ }\bibinfo {title} {Selective
  area epitaxy of {PbTe-Pb} hybrid nanowires on a lattice-matched substrate},\
  \href {https://arxiv.org/abs/2110.13642} {\bibfield  {journal} {\bibinfo
  {journal} {arXiv: 2110.13642}\ } (\bibinfo {year} {2021})}\BibitemShut
  {NoStop}%
\bibitem [{\citenamefont {Kittel}(2005)}]{kittel1996introduction}%
  \BibitemOpen
  \bibfield  {author} {\bibinfo {author} {\bibfnamefont {C.}~\bibnamefont
  {Kittel}},\ }\href@noop {} {\emph {\bibinfo {title} {Introduction to Solid
  State Physics}}}\ (\bibinfo  {publisher} {Wiley, New York},\ \bibinfo {year}
  {2005})\BibitemShut {NoStop}%
\bibitem [{\citenamefont {Grabecki}\ \emph {et~al.}(1999)\citenamefont
  {Grabecki}, \citenamefont {Wr\'obel}, \citenamefont {Dietl}, \citenamefont
  {Byczuk}, \citenamefont {Papis}, \citenamefont {Kami\ifmmode~\acute{n}\else
  \'{n}\fi{}ska}, \citenamefont {Piotrowska}, \citenamefont {Springholz},
  \citenamefont {Pinczolits},\ and\ \citenamefont
  {Bauer}}]{grabecki1999quantum}%
  \BibitemOpen
  \bibfield  {author} {\bibinfo {author} {\bibfnamefont {G.}~\bibnamefont
  {Grabecki}}, \bibinfo {author} {\bibfnamefont {J.}~\bibnamefont {Wr\'obel}},
  \bibinfo {author} {\bibfnamefont {T.}~\bibnamefont {Dietl}}, \bibinfo
  {author} {\bibfnamefont {K.}~\bibnamefont {Byczuk}}, \bibinfo {author}
  {\bibfnamefont {E.}~\bibnamefont {Papis}}, \bibinfo {author} {\bibfnamefont
  {E.}~\bibnamefont {Kami\ifmmode~\acute{n}\else \'{n}\fi{}ska}}, \bibinfo
  {author} {\bibfnamefont {A.}~\bibnamefont {Piotrowska}}, \bibinfo {author}
  {\bibfnamefont {G.}~\bibnamefont {Springholz}}, \bibinfo {author}
  {\bibfnamefont {M.}~\bibnamefont {Pinczolits}}, \ and\ \bibinfo {author}
  {\bibfnamefont {G.}~\bibnamefont {Bauer}},\ }\bibinfo {title} {Quantum
  ballistic transport in constrictions of n-{PbTe}},\ \href {\doibase
  10.1103/PhysRevB.60.R5133(R)} {\bibfield  {journal} {\bibinfo  {journal}
  {Phys. Rev. B}\ }\textbf {\bibinfo {volume} {60}},\ \bibinfo {pages} {R5133}
  (\bibinfo {year} {1999})}\BibitemShut {NoStop}%
\bibitem [{\citenamefont {Grabecki}\ \emph {et~al.}(2005)\citenamefont
  {Grabecki}, \citenamefont {Wr\'obel}, \citenamefont {Dietl}, \citenamefont
  {Janik}, \citenamefont {Aleszkiewicz}, \citenamefont {Papis}, \citenamefont
  {Kami\ifmmode~\acute{n}\else \'{n}\fi{}ska}, \citenamefont {Piotrowska},
  \citenamefont {Springholz},\ and\ \citenamefont
  {Bauer}}]{grabecki2005disorder}%
  \BibitemOpen
  \bibfield  {author} {\bibinfo {author} {\bibfnamefont {G.}~\bibnamefont
  {Grabecki}}, \bibinfo {author} {\bibfnamefont {J.}~\bibnamefont {Wr\'obel}},
  \bibinfo {author} {\bibfnamefont {T.}~\bibnamefont {Dietl}}, \bibinfo
  {author} {\bibfnamefont {E.}~\bibnamefont {Janik}}, \bibinfo {author}
  {\bibfnamefont {M.}~\bibnamefont {Aleszkiewicz}}, \bibinfo {author}
  {\bibfnamefont {E.}~\bibnamefont {Papis}}, \bibinfo {author} {\bibfnamefont
  {E.}~\bibnamefont {Kami\ifmmode~\acute{n}\else \'{n}\fi{}ska}}, \bibinfo
  {author} {\bibfnamefont {A.}~\bibnamefont {Piotrowska}}, \bibinfo {author}
  {\bibfnamefont {G.}~\bibnamefont {Springholz}}, \ and\ \bibinfo {author}
  {\bibfnamefont {G.}~\bibnamefont {Bauer}},\ }\bibinfo {title} {Disorder
  suppression and precise conductance quantization in constrictions of {PbTe}
  quantum wells},\ \href {\doibase 10.1103/PhysRevB.72.125332} {\bibfield
  {journal} {\bibinfo  {journal} {Phys. Rev. B}\ }\textbf {\bibinfo {volume}
  {72}},\ \bibinfo {pages} {125332} (\bibinfo {year} {2005})}\BibitemShut
  {NoStop}%
\bibitem [{\citenamefont {Grabecki}\ \emph {et~al.}(2004)\citenamefont
  {Grabecki}, \citenamefont {Wrobel}, \citenamefont {Dietl}, \citenamefont
  {Papis}, \citenamefont {Kami{\'n}ska}, \citenamefont {Piotrowska},
  \citenamefont {Ratuszna}, \citenamefont {Springholz},\ and\ \citenamefont
  {Bauer}}]{grabecki2004ballistic}%
  \BibitemOpen
  \bibfield  {author} {\bibinfo {author} {\bibfnamefont {G.}~\bibnamefont
  {Grabecki}}, \bibinfo {author} {\bibfnamefont {J.}~\bibnamefont {Wrobel}},
  \bibinfo {author} {\bibfnamefont {T.}~\bibnamefont {Dietl}}, \bibinfo
  {author} {\bibfnamefont {E.}~\bibnamefont {Papis}}, \bibinfo {author}
  {\bibfnamefont {E.}~\bibnamefont {Kami{\'n}ska}}, \bibinfo {author}
  {\bibfnamefont {A.}~\bibnamefont {Piotrowska}}, \bibinfo {author}
  {\bibfnamefont {A.}~\bibnamefont {Ratuszna}}, \bibinfo {author}
  {\bibfnamefont {G.}~\bibnamefont {Springholz}}, \ and\ \bibinfo {author}
  {\bibfnamefont {G.}~\bibnamefont {Bauer}},\ }\bibinfo {title} {Ballistic
  transport in {PbTe}-based nanostructures},\ \href
  {https://doi.org/10.1016/j.physe.2003.08.010} {\bibfield  {journal} {\bibinfo
   {journal} {Phys. E}\ }\textbf {\bibinfo {volume} {20}},\ \bibinfo {pages}
  {236} (\bibinfo {year} {2004})}\BibitemShut {NoStop}%
\bibitem [{\citenamefont {Grabecki}\ \emph {et~al.}(2006)\citenamefont
  {Grabecki}, \citenamefont {Wr{\'o}bel}, \citenamefont {Dietl}, \citenamefont
  {Janik}, \citenamefont {Aleszkiewicz}, \citenamefont {Papis}, \citenamefont
  {Kami{\'n}ska}, \citenamefont {Piotrowska}, \citenamefont {Springholz},\ and\
  \citenamefont {Bauer}}]{grabecki2006pbte}%
  \BibitemOpen
  \bibfield  {author} {\bibinfo {author} {\bibfnamefont {G.}~\bibnamefont
  {Grabecki}}, \bibinfo {author} {\bibfnamefont {J.}~\bibnamefont
  {Wr{\'o}bel}}, \bibinfo {author} {\bibfnamefont {T.}~\bibnamefont {Dietl}},
  \bibinfo {author} {\bibfnamefont {E.}~\bibnamefont {Janik}}, \bibinfo
  {author} {\bibfnamefont {M.}~\bibnamefont {Aleszkiewicz}}, \bibinfo {author}
  {\bibfnamefont {E.}~\bibnamefont {Papis}}, \bibinfo {author} {\bibfnamefont
  {E.}~\bibnamefont {Kami{\'n}ska}}, \bibinfo {author} {\bibfnamefont
  {A.}~\bibnamefont {Piotrowska}}, \bibinfo {author} {\bibfnamefont
  {G.}~\bibnamefont {Springholz}}, \ and\ \bibinfo {author} {\bibfnamefont
  {G.}~\bibnamefont {Bauer}},\ }\bibinfo {title} {{PbTe}--{A} new medium for
  quantum ballistic devices},\ \href
  {https://doi.org/10.1016/j.physe.2006.03.100} {\bibfield  {journal} {\bibinfo
   {journal} {Phys. E}\ }\textbf {\bibinfo {volume} {34}},\ \bibinfo {pages}
  {560} (\bibinfo {year} {2006})}\BibitemShut {NoStop}%
\bibitem [{\citenamefont {Springholz}\ \emph {et~al.}(1993)\citenamefont
  {Springholz}, \citenamefont {Bauer},\ and\ \citenamefont
  {Ihninger}}]{springholz1993mbe}%
  \BibitemOpen
  \bibfield  {author} {\bibinfo {author} {\bibfnamefont {G.}~\bibnamefont
  {Springholz}}, \bibinfo {author} {\bibfnamefont {G.}~\bibnamefont {Bauer}}, \
  and\ \bibinfo {author} {\bibfnamefont {G.}~\bibnamefont {Ihninger}},\
  }\bibinfo {title} {{MBE} of high mobility {PbTe} films and
  {PbTe/Pb$_{1-x}$Eu$_x$Te} heterostructures},\ \href
  {https://doi.org/10.1016/0022-0248(93)90626-8} {\bibfield  {journal}
  {\bibinfo  {journal} {J. Cryst. Growth}\ }\textbf {\bibinfo {volume} {127}},\
  \bibinfo {pages} {302} (\bibinfo {year} {1993})}\BibitemShut {NoStop}%
\bibitem [{\citenamefont {Ueta}\ \emph {et~al.}(1997)\citenamefont {Ueta},
  \citenamefont {Springholz},\ and\ \citenamefont {Bauer}}]{ueta1997improved}%
  \BibitemOpen
  \bibfield  {author} {\bibinfo {author} {\bibfnamefont {A.}~\bibnamefont
  {Ueta}}, \bibinfo {author} {\bibfnamefont {G.}~\bibnamefont {Springholz}}, \
  and\ \bibinfo {author} {\bibfnamefont {G.}~\bibnamefont {Bauer}},\ }\bibinfo
  {title} {Improved nucleation and spiral growth of {PbTe} on {BaF$_2$}
  (111)},\ \href {https://doi.org/10.1016/S0022-0248(96)00985-2} {\bibfield
  {journal} {\bibinfo  {journal} {J. Cryst. Growth}\ }\textbf {\bibinfo
  {volume} {175}},\ \bibinfo {pages} {1022} (\bibinfo {year}
  {1997})}\BibitemShut {NoStop}%
\bibitem [{\citenamefont {Dziawa}\ \emph {et~al.}(2010)\citenamefont {Dziawa},
  \citenamefont {Sadowski}, \citenamefont {Dluzewski}, \citenamefont
  {Lusakowska}, \citenamefont {Domukhovski}, \citenamefont {Taliashvili},
  \citenamefont {Wojciechowski}, \citenamefont {Baczewski}, \citenamefont
  {Bukala}, \citenamefont {Galicka} \emph {et~al.}}]{dziawa2010defect}%
  \BibitemOpen
  \bibfield  {author} {\bibinfo {author} {\bibfnamefont {P.}~\bibnamefont
  {Dziawa}}, \bibinfo {author} {\bibfnamefont {J.}~\bibnamefont {Sadowski}},
  \bibinfo {author} {\bibfnamefont {P.}~\bibnamefont {Dluzewski}}, \bibinfo
  {author} {\bibfnamefont {E.}~\bibnamefont {Lusakowska}}, \bibinfo {author}
  {\bibfnamefont {V.}~\bibnamefont {Domukhovski}}, \bibinfo {author}
  {\bibfnamefont {B.}~\bibnamefont {Taliashvili}}, \bibinfo {author}
  {\bibfnamefont {T.}~\bibnamefont {Wojciechowski}}, \bibinfo {author}
  {\bibfnamefont {L.}~\bibnamefont {Baczewski}}, \bibinfo {author}
  {\bibfnamefont {M.}~\bibnamefont {Bukala}}, \bibinfo {author} {\bibfnamefont
  {M.}~\bibnamefont {Galicka}},  \emph {et~al.},\ }\bibinfo {title} {Defect
  free {PbTe} nanowires grown by molecular beam epitaxy on {GaAs}(111){B}
  substrates},\ \href {https://doi.org/10.1021/cg900575r} {\bibfield  {journal}
  {\bibinfo  {journal} {Cryst. Growth Des.}\ }\textbf {\bibinfo {volume}
  {10}},\ \bibinfo {pages} {109} (\bibinfo {year} {2010})}\BibitemShut
  {NoStop}%
\bibitem [{\citenamefont {Stanescu}\ \emph {et~al.}(2011)\citenamefont
  {Stanescu}, \citenamefont {Lutchyn},\ and\ \citenamefont {{Das
  Sarma}}}]{stanescu2011majorana}%
  \BibitemOpen
  \bibfield  {author} {\bibinfo {author} {\bibfnamefont {T.~D.}\ \bibnamefont
  {Stanescu}}, \bibinfo {author} {\bibfnamefont {R.~M.}\ \bibnamefont
  {Lutchyn}}, \ and\ \bibinfo {author} {\bibfnamefont {S.}~\bibnamefont {{Das
  Sarma}}},\ }\bibinfo {title} {{Majorana} fermions in semiconductor
  nanowires},\ \href {https://doi.org/10.1103/PhysRevB.84.144522} {\bibfield
  {journal} {\bibinfo  {journal} {Phys. Rev. B}\ }\textbf {\bibinfo {volume}
  {84}},\ \bibinfo {pages} {144522} (\bibinfo {year} {2011})}\BibitemShut
  {NoStop}%
\bibitem [{\citenamefont {Cole}\ \emph {et~al.}(2015)\citenamefont {Cole},
  \citenamefont {Das~Sarma},\ and\ \citenamefont {Stanescu}}]{cole2015effects}%
  \BibitemOpen
  \bibfield  {author} {\bibinfo {author} {\bibfnamefont {W.~S.}\ \bibnamefont
  {Cole}}, \bibinfo {author} {\bibfnamefont {S.}~\bibnamefont {Das~Sarma}}, \
  and\ \bibinfo {author} {\bibfnamefont {T.~D.}\ \bibnamefont {Stanescu}},\
  }\bibinfo {title} {Effects of large induced superconducting gap on
  semiconductor {Majorana} nanowires},\ \href {\doibase
  10.1103/PhysRevB.92.174511} {\bibfield  {journal} {\bibinfo  {journal} {Phys.
  Rev. B}\ }\textbf {\bibinfo {volume} {92}},\ \bibinfo {pages} {174511}
  (\bibinfo {year} {2015})}\BibitemShut {NoStop}%
\bibitem [{\citenamefont {Stanescu}\ and\ \citenamefont
  {Das~Sarma}(2017)}]{stanescu2017proximity}%
  \BibitemOpen
  \bibfield  {author} {\bibinfo {author} {\bibfnamefont {T.~D.}\ \bibnamefont
  {Stanescu}}\ and\ \bibinfo {author} {\bibfnamefont {S.}~\bibnamefont
  {Das~Sarma}},\ }\bibinfo {title} {Proximity-induced low-energy
  renormalization in hybrid semiconductor-superconductor {Majorana}
  structures},\ \href {\doibase 10.1103/PhysRevB.96.014510} {\bibfield
  {journal} {\bibinfo  {journal} {Phys. Rev. B}\ }\textbf {\bibinfo {volume}
  {96}},\ \bibinfo {pages} {014510} (\bibinfo {year} {2017})}\BibitemShut
  {NoStop}%
\bibitem [{\citenamefont {Reeg}\ \emph {et~al.}(2018)\citenamefont {Reeg},
  \citenamefont {Loss},\ and\ \citenamefont
  {Klinovaja}}]{reeg2018metallization}%
  \BibitemOpen
  \bibfield  {author} {\bibinfo {author} {\bibfnamefont {C.}~\bibnamefont
  {Reeg}}, \bibinfo {author} {\bibfnamefont {D.}~\bibnamefont {Loss}}, \ and\
  \bibinfo {author} {\bibfnamefont {J.}~\bibnamefont {Klinovaja}},\ }\bibinfo
  {title} {Metallization of a {Rashba} wire by a superconducting layer in the
  strong-proximity regime},\ \href {https://doi.org/10.1103/PhysRevB.97.165425}
  {\bibfield  {journal} {\bibinfo  {journal} {Phys. Rev. B}\ }\textbf {\bibinfo
  {volume} {97}},\ \bibinfo {pages} {165425} (\bibinfo {year}
  {2018})}\BibitemShut {NoStop}%
\bibitem [{\citenamefont {Nadj-Perge}\ \emph {et~al.}(2014)\citenamefont
  {Nadj-Perge}, \citenamefont {Drozdov}, \citenamefont {Li}, \citenamefont
  {Chen}, \citenamefont {Jeon}, \citenamefont {Seo}, \citenamefont {MacDonald},
  \citenamefont {Bernevig},\ and\ \citenamefont
  {Yazdani}}]{nadj2014observation}%
  \BibitemOpen
  \bibfield  {author} {\bibinfo {author} {\bibfnamefont {S.}~\bibnamefont
  {Nadj-Perge}}, \bibinfo {author} {\bibfnamefont {I.~K.}\ \bibnamefont
  {Drozdov}}, \bibinfo {author} {\bibfnamefont {J.}~\bibnamefont {Li}},
  \bibinfo {author} {\bibfnamefont {H.}~\bibnamefont {Chen}}, \bibinfo {author}
  {\bibfnamefont {S.}~\bibnamefont {Jeon}}, \bibinfo {author} {\bibfnamefont
  {J.}~\bibnamefont {Seo}}, \bibinfo {author} {\bibfnamefont {A.~H.}\
  \bibnamefont {MacDonald}}, \bibinfo {author} {\bibfnamefont {B.~A.}\
  \bibnamefont {Bernevig}}, \ and\ \bibinfo {author} {\bibfnamefont
  {A.}~\bibnamefont {Yazdani}},\ }\bibinfo {title} {Observation of {Majorana}
  fermions in ferromagnetic atomic chains on a superconductor},\ \href
  {https://link.aps.org/10.1126/science.1259327} {\bibfield  {journal}
  {\bibinfo  {journal} {Science}\ }\textbf {\bibinfo {volume} {346}},\ \bibinfo
  {pages} {602} (\bibinfo {year} {2014})}\BibitemShut {NoStop}%
\bibitem [{\citenamefont {Li}\ \emph {et~al.}(2014)\citenamefont {Li},
  \citenamefont {Chen}, \citenamefont {Drozdov}, \citenamefont {Yazdani},
  \citenamefont {Bernevig},\ and\ \citenamefont
  {MacDonald}}]{li2014topological}%
  \BibitemOpen
  \bibfield  {author} {\bibinfo {author} {\bibfnamefont {J.}~\bibnamefont
  {Li}}, \bibinfo {author} {\bibfnamefont {H.}~\bibnamefont {Chen}}, \bibinfo
  {author} {\bibfnamefont {I.~K.}\ \bibnamefont {Drozdov}}, \bibinfo {author}
  {\bibfnamefont {A.}~\bibnamefont {Yazdani}}, \bibinfo {author} {\bibfnamefont
  {B.~A.}\ \bibnamefont {Bernevig}}, \ and\ \bibinfo {author} {\bibfnamefont
  {A.~H.}\ \bibnamefont {MacDonald}},\ }\bibinfo {title} {Topological
  superconductivity induced by ferromagnetic metal chains},\ \href {\doibase
  10.1103/PhysRevB.90.235433} {\bibfield  {journal} {\bibinfo  {journal} {Phys.
  Rev. B}\ }\textbf {\bibinfo {volume} {90}},\ \bibinfo {pages} {235433}
  (\bibinfo {year} {2014})}\BibitemShut {NoStop}%
\bibitem [{\citenamefont {Hui}\ \emph {et~al.}(2015{\natexlab{b}})\citenamefont
  {Hui}, \citenamefont {Brydon}, \citenamefont {Sau}, \citenamefont {Tewari},\
  and\ \citenamefont {Das~Sarma}}]{hui2015majorana}%
  \BibitemOpen
  \bibfield  {author} {\bibinfo {author} {\bibfnamefont {H.-Y.}\ \bibnamefont
  {Hui}}, \bibinfo {author} {\bibfnamefont {P.}~\bibnamefont {Brydon}},
  \bibinfo {author} {\bibfnamefont {J.~D.}\ \bibnamefont {Sau}}, \bibinfo
  {author} {\bibfnamefont {S.}~\bibnamefont {Tewari}}, \ and\ \bibinfo {author}
  {\bibfnamefont {S.}~\bibnamefont {Das~Sarma}},\ }\bibinfo {title} {Majorana
  fermions in ferromagnetic chains on the surface of bulk spin-orbit coupled
  s-wave superconductors},\ \href {https://doi.org/10.1038/srep08880}
  {\bibfield  {journal} {\bibinfo  {journal} {Sci. Rep.}\ }\textbf {\bibinfo
  {volume} {5}},\ \bibinfo {pages} {8880} (\bibinfo {year}
  {2015}{\natexlab{b}})}\BibitemShut {NoStop}%
\bibitem [{\citenamefont {Winkler}\ \emph {et~al.}(2019)\citenamefont
  {Winkler}, \citenamefont {Antipov}, \citenamefont {van Heck}, \citenamefont
  {Soluyanov}, \citenamefont {Glazman}, \citenamefont {Wimmer},\ and\
  \citenamefont {Lutchyn}}]{winkler2019unified}%
  \BibitemOpen
  \bibfield  {author} {\bibinfo {author} {\bibfnamefont {G.~W.}\ \bibnamefont
  {Winkler}}, \bibinfo {author} {\bibfnamefont {A.~E.}\ \bibnamefont
  {Antipov}}, \bibinfo {author} {\bibfnamefont {B.}~\bibnamefont {van Heck}},
  \bibinfo {author} {\bibfnamefont {A.~A.}\ \bibnamefont {Soluyanov}}, \bibinfo
  {author} {\bibfnamefont {L.~I.}\ \bibnamefont {Glazman}}, \bibinfo {author}
  {\bibfnamefont {M.}~\bibnamefont {Wimmer}}, \ and\ \bibinfo {author}
  {\bibfnamefont {R.~M.}\ \bibnamefont {Lutchyn}},\ }\bibinfo {title} {Unified
  numerical approach to topological semiconductor-superconductor
  heterostructures},\ \href {\doibase 10.1103/PhysRevB.99.245408} {\bibfield
  {journal} {\bibinfo  {journal} {Phys. Rev. B}\ }\textbf {\bibinfo {volume}
  {99}},\ \bibinfo {pages} {245408} (\bibinfo {year} {2019})}\BibitemShut
  {NoStop}%
\bibitem [{\citenamefont {Carter}\ and\ \citenamefont
  {Bate}(1971)}]{carter1971physics}%
  \BibitemOpen
  \bibfield  {author} {\bibinfo {author} {\bibfnamefont {D.~L.}\ \bibnamefont
  {Carter}}\ and\ \bibinfo {author} {\bibfnamefont {R.~T.}\ \bibnamefont
  {Bate}},\ }\href@noop {} {\emph {\bibinfo {title} {The Physics of Semimetals
  and Narrow-gap Semiconductors: Proceedings}}},\ Vol.~\bibinfo {volume} {32}\
  (\bibinfo  {publisher} {Pergamon},\ \bibinfo {year} {1971})\BibitemShut
  {NoStop}%
\bibitem [{\citenamefont {de~Andrada~e Silva}(1999)}]{silva1999optical}%
  \BibitemOpen
  \bibfield  {author} {\bibinfo {author} {\bibfnamefont {E.~A.}\ \bibnamefont
  {de~Andrada~e Silva}},\ }\bibinfo {title} {Optical transition energies for
  lead-salt semiconductor quantum wells},\ \href {\doibase
  10.1103/PhysRevB.60.8859} {\bibfield  {journal} {\bibinfo  {journal} {Phys.
  Rev. B}\ }\textbf {\bibinfo {volume} {60}},\ \bibinfo {pages} {8859}
  (\bibinfo {year} {1999})}\BibitemShut {NoStop}%
\bibitem [{\citenamefont {Hasegawa}\ and\ \citenamefont {de~Andrada~e
  Silva}(2003)}]{hasegawa2003spin}%
  \BibitemOpen
  \bibfield  {author} {\bibinfo {author} {\bibfnamefont {M.~M.}\ \bibnamefont
  {Hasegawa}}\ and\ \bibinfo {author} {\bibfnamefont {E.~A.}\ \bibnamefont
  {de~Andrada~e Silva}},\ }\bibinfo {title} {Spin-orbit-split subbands in
  {IV-VI} asymmetric quantum wells},\ \href {\doibase
  10.1103/PhysRevB.68.205309} {\bibfield  {journal} {\bibinfo  {journal} {Phys.
  Rev. B}\ }\textbf {\bibinfo {volume} {68}},\ \bibinfo {pages} {205309}
  (\bibinfo {year} {2003})}\BibitemShut {NoStop}%
\bibitem [{\citenamefont {Ridolfi}\ \emph {et~al.}(2015)\citenamefont
  {Ridolfi}, \citenamefont {de~Andrada~e Silva},\ and\ \citenamefont
  {La~Rocca}}]{ridolfi2015effective}%
  \BibitemOpen
  \bibfield  {author} {\bibinfo {author} {\bibfnamefont {E.}~\bibnamefont
  {Ridolfi}}, \bibinfo {author} {\bibfnamefont {E.~A.}\ \bibnamefont
  {de~Andrada~e Silva}}, \ and\ \bibinfo {author} {\bibfnamefont {G.~C.}\
  \bibnamefont {La~Rocca}},\ }\bibinfo {title} {Effective $g$-factor tensor for
  carriers in {IV-VI} semiconductor quantum wells},\ \href {\doibase
  10.1103/PhysRevB.91.085313} {\bibfield  {journal} {\bibinfo  {journal} {Phys.
  Rev. B}\ }\textbf {\bibinfo {volume} {91}},\ \bibinfo {pages} {085313}
  (\bibinfo {year} {2015})}\BibitemShut {NoStop}%
\bibitem [{\citenamefont {Peres}\ \emph {et~al.}(2014)\citenamefont {Peres},
  \citenamefont {Monteiro}, \citenamefont {Chitta}, \citenamefont {de~Castro},
  \citenamefont {Mengui}, \citenamefont {Rappl}, \citenamefont {Oliveira~Jr.},
  \citenamefont {Abramof},\ and\ \citenamefont
  {Maude}}]{peres2014experimental}%
  \BibitemOpen
  \bibfield  {author} {\bibinfo {author} {\bibfnamefont {M.~L.}\ \bibnamefont
  {Peres}}, \bibinfo {author} {\bibfnamefont {H.~S.}\ \bibnamefont {Monteiro}},
  \bibinfo {author} {\bibfnamefont {V.~A.}\ \bibnamefont {Chitta}}, \bibinfo
  {author} {\bibfnamefont {S.}~\bibnamefont {de~Castro}}, \bibinfo {author}
  {\bibfnamefont {U.~A.}\ \bibnamefont {Mengui}}, \bibinfo {author}
  {\bibfnamefont {P.~H.~O.}\ \bibnamefont {Rappl}}, \bibinfo {author}
  {\bibfnamefont {N.~F.}\ \bibnamefont {Oliveira~Jr.}}, \bibinfo {author}
  {\bibfnamefont {E.}~\bibnamefont {Abramof}}, \ and\ \bibinfo {author}
  {\bibfnamefont {D.~K.}\ \bibnamefont {Maude}},\ }\bibinfo {title}
  {Experimental investigation of spin-orbit coupling in n-type {PbTe} quantum
  wells},\ \href {https://doi.org/10.1063/1.4867627} {\bibfield  {journal}
  {\bibinfo  {journal} {J. Appl. Phys.}\ }\textbf {\bibinfo {volume} {115}},\
  \bibinfo {pages} {093704} (\bibinfo {year} {2014})}\BibitemShut {NoStop}%
\bibitem [{\citenamefont {Rahman}\ \emph {et~al.}(2005)\citenamefont {Rahman},
  \citenamefont {Lundstrom},\ and\ \citenamefont
  {Ghosh}}]{rahman2005generalized}%
  \BibitemOpen
  \bibfield  {author} {\bibinfo {author} {\bibfnamefont {A.}~\bibnamefont
  {Rahman}}, \bibinfo {author} {\bibfnamefont {M.~S.}\ \bibnamefont
  {Lundstrom}}, \ and\ \bibinfo {author} {\bibfnamefont {A.~W.}\ \bibnamefont
  {Ghosh}},\ }\bibinfo {title} {Generalized effective-mass approach for n-type
  metal-oxide-semiconductor field-effect transistors on arbitrarily oriented
  wafers},\ \href {https://doi.org/10.1063/1.1845586} {\bibfield  {journal}
  {\bibinfo  {journal} {J. Appl. Phys.}\ }\textbf {\bibinfo {volume} {97}},\
  \bibinfo {pages} {053702} (\bibinfo {year} {2005})}\BibitemShut {NoStop}%
\bibitem [{\citenamefont {Aparecida~da Costa}\ and\ \citenamefont {de~Andrada~e
  Silva}(2010)}]{aparecida2010ballistic}%
  \BibitemOpen
  \bibfield  {author} {\bibinfo {author} {\bibfnamefont {V.}~\bibnamefont
  {Aparecida~da Costa}}\ and\ \bibinfo {author} {\bibfnamefont {E.~A.}\
  \bibnamefont {de~Andrada~e Silva}},\ }\bibinfo {title} {Ballistic conductance
  and thermoelectric power of lead-salt semiconductor nanowires},\ \href
  {\doibase 10.1103/PhysRevB.82.153302} {\bibfield  {journal} {\bibinfo
  {journal} {Phys. Rev. B}\ }\textbf {\bibinfo {volume} {82}},\ \bibinfo
  {pages} {153302} (\bibinfo {year} {2010})}\BibitemShut {NoStop}%
\bibitem [{\citenamefont {Boykin}\ \emph {et~al.}(2004)\citenamefont {Boykin},
  \citenamefont {Klimeck}, \citenamefont {Friesen}, \citenamefont
  {Coppersmith}, \citenamefont {von Allmen}, \citenamefont {Oyafuso},\ and\
  \citenamefont {Lee}}]{boykin2004valley}%
  \BibitemOpen
  \bibfield  {author} {\bibinfo {author} {\bibfnamefont {T.~B.}\ \bibnamefont
  {Boykin}}, \bibinfo {author} {\bibfnamefont {G.}~\bibnamefont {Klimeck}},
  \bibinfo {author} {\bibfnamefont {M.}~\bibnamefont {Friesen}}, \bibinfo
  {author} {\bibfnamefont {S.~N.}\ \bibnamefont {Coppersmith}}, \bibinfo
  {author} {\bibfnamefont {P.}~\bibnamefont {von Allmen}}, \bibinfo {author}
  {\bibfnamefont {F.}~\bibnamefont {Oyafuso}}, \ and\ \bibinfo {author}
  {\bibfnamefont {S.}~\bibnamefont {Lee}},\ }\bibinfo {title} {Valley splitting
  in low-density quantum-confined heterostructures studied using tight-binding
  models},\ \href {\doibase 10.1103/PhysRevB.70.165325} {\bibfield  {journal}
  {\bibinfo  {journal} {Phys. Rev. B}\ }\textbf {\bibinfo {volume} {70}},\
  \bibinfo {pages} {165325} (\bibinfo {year} {2004})}\BibitemShut {NoStop}%
\bibitem [{\citenamefont {Nestoklon}\ \emph {et~al.}(2006)\citenamefont
  {Nestoklon}, \citenamefont {Golub},\ and\ \citenamefont
  {Ivchenko}}]{nestoklon2006spin}%
  \BibitemOpen
  \bibfield  {author} {\bibinfo {author} {\bibfnamefont {M.~O.}\ \bibnamefont
  {Nestoklon}}, \bibinfo {author} {\bibfnamefont {L.~E.}\ \bibnamefont
  {Golub}}, \ and\ \bibinfo {author} {\bibfnamefont {E.~L.}\ \bibnamefont
  {Ivchenko}},\ }\bibinfo {title} {Spin and valley-orbit splittings in
  {SiGe}/{Si} heterostructures},\ \href {\doibase 10.1103/PhysRevB.73.235334}
  {\bibfield  {journal} {\bibinfo  {journal} {Phys. Rev. B}\ }\textbf {\bibinfo
  {volume} {73}},\ \bibinfo {pages} {235334} (\bibinfo {year}
  {2006})}\BibitemShut {NoStop}%
\bibitem [{\citenamefont {Avdeev}\ \emph {et~al.}(2017)\citenamefont {Avdeev},
  \citenamefont {Poddubny}, \citenamefont {Goupalov},\ and\ \citenamefont
  {Nestoklon}}]{avdeev2017valley}%
  \BibitemOpen
  \bibfield  {author} {\bibinfo {author} {\bibfnamefont {I.~D.}\ \bibnamefont
  {Avdeev}}, \bibinfo {author} {\bibfnamefont {A.~N.}\ \bibnamefont
  {Poddubny}}, \bibinfo {author} {\bibfnamefont {S.~V.}\ \bibnamefont
  {Goupalov}}, \ and\ \bibinfo {author} {\bibfnamefont {M.~O.}\ \bibnamefont
  {Nestoklon}},\ }\bibinfo {title} {Valley and spin splittings in {PbSe}
  nanowires},\ \href {\doibase 10.1103/PhysRevB.96.085310} {\bibfield
  {journal} {\bibinfo  {journal} {Phys. Rev. B}\ }\textbf {\bibinfo {volume}
  {96}},\ \bibinfo {pages} {085310} (\bibinfo {year} {2017})}\BibitemShut
  {NoStop}%
\bibitem [{\citenamefont {Potter}\ and\ \citenamefont
  {Lee}(2010)}]{potter2010multichannel}%
  \BibitemOpen
  \bibfield  {author} {\bibinfo {author} {\bibfnamefont {A.~C.}\ \bibnamefont
  {Potter}}\ and\ \bibinfo {author} {\bibfnamefont {P.~A.}\ \bibnamefont
  {Lee}},\ }\bibinfo {title} {Multichannel Generalization of {Kitaev's
  Majorana} End States and a Practical Route to Realize Them in Thin Films},\
  \href {\doibase 10.1103/PhysRevLett.105.227003} {\bibfield  {journal}
  {\bibinfo  {journal} {Phys. Rev. Lett.}\ }\textbf {\bibinfo {volume} {105}},\
  \bibinfo {pages} {227003} (\bibinfo {year} {2010})}\BibitemShut {NoStop}%
\bibitem [{\citenamefont {Lutchyn}\ \emph {et~al.}(2011)\citenamefont
  {Lutchyn}, \citenamefont {Stanescu},\ and\ \citenamefont
  {Das~Sarma}}]{lutchyn2011search}%
  \BibitemOpen
  \bibfield  {author} {\bibinfo {author} {\bibfnamefont {R.~M.}\ \bibnamefont
  {Lutchyn}}, \bibinfo {author} {\bibfnamefont {T.~D.}\ \bibnamefont
  {Stanescu}}, \ and\ \bibinfo {author} {\bibfnamefont {S.}~\bibnamefont
  {Das~Sarma}},\ }\bibinfo {title} {Search for {Majorana} Fermions in Multiband
  Semiconducting Nanowires},\ \href {\doibase 10.1103/PhysRevLett.106.127001}
  {\bibfield  {journal} {\bibinfo  {journal} {Phys. Rev. Lett.}\ }\textbf
  {\bibinfo {volume} {106}},\ \bibinfo {pages} {127001} (\bibinfo {year}
  {2011})}\BibitemShut {NoStop}%
\bibitem [{\citenamefont {Winkler}(2003)}]{winkler2003spin}%
  \BibitemOpen
  \bibfield  {author} {\bibinfo {author} {\bibfnamefont {R.}~\bibnamefont
  {Winkler}},\ }\href@noop {} {\emph {\bibinfo {title} {Spin-Orbit Coupling
  Effects in Two-Dimensional Electron and Hole Systems}}}\ (\bibinfo
  {publisher} {Springer, Berlin},\ \bibinfo {year} {2003})\BibitemShut
  {NoStop}%
\bibitem [{\citenamefont {Keyes}(1959)}]{keyes1959correlation}%
  \BibitemOpen
  \bibfield  {author} {\bibinfo {author} {\bibfnamefont {R.~W.}\ \bibnamefont
  {Keyes}},\ }\bibinfo {title} {Correlation between mobility and effective mass
  in semiconductors},\ \href {https://doi.org/10.1063/1.1735199} {\bibfield
  {journal} {\bibinfo  {journal} {J. Appl. Phys.}\ }\textbf {\bibinfo {volume}
  {30}},\ \bibinfo {pages} {454} (\bibinfo {year} {1959})}\BibitemShut
  {NoStop}%
\bibitem [{\citenamefont {Olsson}\ \emph {et~al.}(1996)\citenamefont {Olsson},
  \citenamefont {Andersson}, \citenamefont {H\aa{}kansson}, \citenamefont
  {Kanski}, \citenamefont {Ilver},\ and\ \citenamefont
  {Karlsson}}]{olsson1996charge}%
  \BibitemOpen
  \bibfield  {author} {\bibinfo {author} {\bibfnamefont {L.~O.}\ \bibnamefont
  {Olsson}}, \bibinfo {author} {\bibfnamefont {C.~B.~M.}\ \bibnamefont
  {Andersson}}, \bibinfo {author} {\bibfnamefont {M.~C.}\ \bibnamefont
  {H\aa{}kansson}}, \bibinfo {author} {\bibfnamefont {J.}~\bibnamefont
  {Kanski}}, \bibinfo {author} {\bibfnamefont {L.}~\bibnamefont {Ilver}}, \
  and\ \bibinfo {author} {\bibfnamefont {U.~O.}\ \bibnamefont {Karlsson}},\
  }\bibinfo {title} {Charge Accumulation at {InAs} Surfaces},\ \href {\doibase
  10.1103/PhysRevLett.76.3626} {\bibfield  {journal} {\bibinfo  {journal}
  {Phys. Rev. Lett.}\ }\textbf {\bibinfo {volume} {76}},\ \bibinfo {pages}
  {3626} (\bibinfo {year} {1996})}\BibitemShut {NoStop}%
\bibitem [{\citenamefont {Degtyarev}\ \emph {et~al.}(2017)\citenamefont
  {Degtyarev}, \citenamefont {Khazanova},\ and\ \citenamefont
  {Demarina}}]{degtyarev2017features}%
  \BibitemOpen
  \bibfield  {author} {\bibinfo {author} {\bibfnamefont {V.}~\bibnamefont
  {Degtyarev}}, \bibinfo {author} {\bibfnamefont {S.}~\bibnamefont
  {Khazanova}}, \ and\ \bibinfo {author} {\bibfnamefont {N.}~\bibnamefont
  {Demarina}},\ }\bibinfo {title} {Features of electron gas in {InAs} nanowires
  imposed by interplay between nanowire geometry, doping and surface states},\
  \href {https://doi.org/10.1038/s41598-017-03415-3} {\bibfield  {journal}
  {\bibinfo  {journal} {Sci. Rep.}\ }\textbf {\bibinfo {volume} {7}},\ \bibinfo
  {pages} {3411} (\bibinfo {year} {2017})}\BibitemShut {NoStop}%
\bibitem [{\citenamefont {W\'ojcik}\ \emph {et~al.}(2018)\citenamefont
  {W\'ojcik}, \citenamefont {Bertoni},\ and\ \citenamefont
  {Goldoni}}]{wojcik2018tuning}%
  \BibitemOpen
  \bibfield  {author} {\bibinfo {author} {\bibfnamefont {P.}~\bibnamefont
  {W\'ojcik}}, \bibinfo {author} {\bibfnamefont {A.}~\bibnamefont {Bertoni}}, \
  and\ \bibinfo {author} {\bibfnamefont {G.}~\bibnamefont {Goldoni}},\
  }\bibinfo {title} {Tuning {Rashba} spin-orbit coupling in homogeneous
  semiconductor nanowires},\ \href {\doibase 10.1103/PhysRevB.97.165401}
  {\bibfield  {journal} {\bibinfo  {journal} {Phys. Rev. B}\ }\textbf {\bibinfo
  {volume} {97}},\ \bibinfo {pages} {165401} (\bibinfo {year}
  {2018})}\BibitemShut {NoStop}%
\bibitem [{\citenamefont {Escribano}\ \emph {et~al.}(2019)\citenamefont
  {Escribano}, \citenamefont {Levy~Yeyati}, \citenamefont {Oreg},\ and\
  \citenamefont {Prada}}]{escribano2019effects}%
  \BibitemOpen
  \bibfield  {author} {\bibinfo {author} {\bibfnamefont {S.~D.}\ \bibnamefont
  {Escribano}}, \bibinfo {author} {\bibfnamefont {A.}~\bibnamefont
  {Levy~Yeyati}}, \bibinfo {author} {\bibfnamefont {Y.}~\bibnamefont {Oreg}}, \
  and\ \bibinfo {author} {\bibfnamefont {E.}~\bibnamefont {Prada}},\ }\bibinfo
  {title} {Effects of the electrostatic environment on superlattice {Majorana}
  nanowires},\ \href {\doibase 10.1103/PhysRevB.100.045301} {\bibfield
  {journal} {\bibinfo  {journal} {Phys. Rev. B}\ }\textbf {\bibinfo {volume}
  {100}},\ \bibinfo {pages} {045301} (\bibinfo {year} {2019})}\BibitemShut
  {NoStop}%
\bibitem [{\citenamefont {Escribano}\ \emph {et~al.}(2020)\citenamefont
  {Escribano}, \citenamefont {Yeyati},\ and\ \citenamefont
  {Prada}}]{escribano2020improved}%
  \BibitemOpen
  \bibfield  {author} {\bibinfo {author} {\bibfnamefont {S.~D.}\ \bibnamefont
  {Escribano}}, \bibinfo {author} {\bibfnamefont {A.~L.}\ \bibnamefont
  {Yeyati}}, \ and\ \bibinfo {author} {\bibfnamefont {E.}~\bibnamefont
  {Prada}},\ }\bibinfo {title} {Improved effective equation for the {Rashba}
  spin-orbit coupling in semiconductor nanowires},\ \href {\doibase
  10.1103/PhysRevResearch.2.033264} {\bibfield  {journal} {\bibinfo  {journal}
  {Phys. Rev. Research}\ }\textbf {\bibinfo {volume} {2}},\ \bibinfo {pages}
  {033264} (\bibinfo {year} {2020})}\BibitemShut {NoStop}%
\bibitem [{\citenamefont {Woods}\ \emph {et~al.}(2020)\citenamefont {Woods},
  \citenamefont {Das~Sarma},\ and\ \citenamefont
  {Stanescu}}]{woods2020subband}%
  \BibitemOpen
  \bibfield  {author} {\bibinfo {author} {\bibfnamefont {B.~D.}\ \bibnamefont
  {Woods}}, \bibinfo {author} {\bibfnamefont {S.}~\bibnamefont {Das~Sarma}}, \
  and\ \bibinfo {author} {\bibfnamefont {T.~D.}\ \bibnamefont {Stanescu}},\
  }\bibinfo {title} {Subband occupation in semiconductor-superconductor
  nanowires},\ \href {\doibase 10.1103/PhysRevB.101.045405} {\bibfield
  {journal} {\bibinfo  {journal} {Phys. Rev. B}\ }\textbf {\bibinfo {volume}
  {101}},\ \bibinfo {pages} {045405} (\bibinfo {year} {2020})}\BibitemShut
  {NoStop}%
\bibitem [{\citenamefont {Liu}\ \emph {et~al.}(2021)\citenamefont {Liu},
  \citenamefont {Schuwalow}, \citenamefont {Liu}, \citenamefont {Vilkelis},
  \citenamefont {Manesco}, \citenamefont {Krogstrup},\ and\ \citenamefont
  {Wimmer}}]{liu2021electronic}%
  \BibitemOpen
  \bibfield  {author} {\bibinfo {author} {\bibfnamefont {C.-X.}\ \bibnamefont
  {Liu}}, \bibinfo {author} {\bibfnamefont {S.}~\bibnamefont {Schuwalow}},
  \bibinfo {author} {\bibfnamefont {Y.}~\bibnamefont {Liu}}, \bibinfo {author}
  {\bibfnamefont {K.}~\bibnamefont {Vilkelis}}, \bibinfo {author}
  {\bibfnamefont {A.~L.~R.}\ \bibnamefont {Manesco}}, \bibinfo {author}
  {\bibfnamefont {P.}~\bibnamefont {Krogstrup}}, \ and\ \bibinfo {author}
  {\bibfnamefont {M.}~\bibnamefont {Wimmer}},\ }\bibinfo {title} {Electronic
  properties of {InAs/EuS/Al} hybrid nanowires},\ \href {\doibase
  10.1103/PhysRevB.104.014516} {\bibfield  {journal} {\bibinfo  {journal}
  {Phys. Rev. B}\ }\textbf {\bibinfo {volume} {104}},\ \bibinfo {pages}
  {014516} (\bibinfo {year} {2021})}\BibitemShut {NoStop}%
\bibitem [{\citenamefont {Mikkelsen}\ \emph {et~al.}(2018)\citenamefont
  {Mikkelsen}, \citenamefont {Kotetes}, \citenamefont {Krogstrup},\ and\
  \citenamefont {Flensberg}}]{mikkelsen2018hybridization}%
  \BibitemOpen
  \bibfield  {author} {\bibinfo {author} {\bibfnamefont {A.~E.~G.}\
  \bibnamefont {Mikkelsen}}, \bibinfo {author} {\bibfnamefont {P.}~\bibnamefont
  {Kotetes}}, \bibinfo {author} {\bibfnamefont {P.}~\bibnamefont {Krogstrup}},
  \ and\ \bibinfo {author} {\bibfnamefont {K.}~\bibnamefont {Flensberg}},\
  }\bibinfo {title} {Hybridization at Superconductor-Semiconductor
  Interfaces},\ \href {\doibase 10.1103/PhysRevX.8.031040} {\bibfield
  {journal} {\bibinfo  {journal} {Phys. Rev. X}\ }\textbf {\bibinfo {volume}
  {8}},\ \bibinfo {pages} {031040} (\bibinfo {year} {2018})}\BibitemShut
  {NoStop}%
\bibitem [{\citenamefont {Antipov}\ \emph {et~al.}(2018)\citenamefont
  {Antipov}, \citenamefont {Bargerbos}, \citenamefont {Winkler}, \citenamefont
  {Bauer}, \citenamefont {Rossi},\ and\ \citenamefont
  {Lutchyn}}]{antipov2018effects}%
  \BibitemOpen
  \bibfield  {author} {\bibinfo {author} {\bibfnamefont {A.~E.}\ \bibnamefont
  {Antipov}}, \bibinfo {author} {\bibfnamefont {A.}~\bibnamefont {Bargerbos}},
  \bibinfo {author} {\bibfnamefont {G.~W.}\ \bibnamefont {Winkler}}, \bibinfo
  {author} {\bibfnamefont {B.}~\bibnamefont {Bauer}}, \bibinfo {author}
  {\bibfnamefont {E.}~\bibnamefont {Rossi}}, \ and\ \bibinfo {author}
  {\bibfnamefont {R.~M.}\ \bibnamefont {Lutchyn}},\ }\bibinfo {title} {Effects
  of gate-induced electric fields on semiconductor {Majorana} nanowires},\
  \href {https://doi.org/10.1103/PhysRevX.8.031041} {\bibfield  {journal}
  {\bibinfo  {journal} {Phys. Rev. X}\ }\textbf {\bibinfo {volume} {8}},\
  \bibinfo {pages} {031041} (\bibinfo {year} {2018})}\BibitemShut {NoStop}%
\bibitem [{\citenamefont {Hoang}\ \emph {et~al.}(2007)\citenamefont {Hoang},
  \citenamefont {Mahanti},\ and\ \citenamefont {Jena}}]{hoang2007theoretical}%
  \BibitemOpen
  \bibfield  {author} {\bibinfo {author} {\bibfnamefont {K.}~\bibnamefont
  {Hoang}}, \bibinfo {author} {\bibfnamefont {S.~D.}\ \bibnamefont {Mahanti}},
  \ and\ \bibinfo {author} {\bibfnamefont {P.}~\bibnamefont {Jena}},\ }\bibinfo
  {title} {Theoretical study of deep-defect states in bulk {PbTe} and in thin
  films},\ \href {\doibase 10.1103/PhysRevB.76.115432} {\bibfield  {journal}
  {\bibinfo  {journal} {Phys. Rev. B}\ }\textbf {\bibinfo {volume} {76}},\
  \bibinfo {pages} {115432} (\bibinfo {year} {2007})}\BibitemShut {NoStop}%
\bibitem [{\citenamefont {Michaelson}(1977)}]{michaelson1977work}%
  \BibitemOpen
  \bibfield  {author} {\bibinfo {author} {\bibfnamefont {H.~B.}\ \bibnamefont
  {Michaelson}},\ }\bibinfo {title} {The work function of the elements and its
  periodicity},\ \href {https://doi.org/10.1063/1.323539} {\bibfield  {journal}
  {\bibinfo  {journal} {J. Appl. Phys.}\ }\textbf {\bibinfo {volume} {48}},\
  \bibinfo {pages} {4729} (\bibinfo {year} {1977})}\BibitemShut {NoStop}%
\bibitem [{\citenamefont {Lee}(2016)}]{lee2016thermoelectrics}%
  \BibitemOpen
  \bibfield  {author} {\bibinfo {author} {\bibfnamefont {H.}~\bibnamefont
  {Lee}},\ }\href@noop {} {\emph {\bibinfo {title} {Thermoelectrics: {Design}
  and {Materials}}}}\ (\bibinfo  {publisher} {John Wiley \& Sons},\ \bibinfo
  {year} {2016})\BibitemShut {NoStop}%
\bibitem [{\citenamefont {Grabecki}\ \emph {et~al.}(2010)\citenamefont
  {Grabecki}, \citenamefont {Kolwas}, \citenamefont {Wr{\'o}bel}, \citenamefont
  {Kapcia}, \citenamefont {Pu{\'z}niak}, \citenamefont {Jakie{\l}a},
  \citenamefont {Aleszkiewicz}, \citenamefont {Dietl}, \citenamefont
  {Springholz},\ and\ \citenamefont {Bauer}}]{grabecki2010contact}%
  \BibitemOpen
  \bibfield  {author} {\bibinfo {author} {\bibfnamefont {G.}~\bibnamefont
  {Grabecki}}, \bibinfo {author} {\bibfnamefont {K.}~\bibnamefont {Kolwas}},
  \bibinfo {author} {\bibfnamefont {J.}~\bibnamefont {Wr{\'o}bel}}, \bibinfo
  {author} {\bibfnamefont {K.}~\bibnamefont {Kapcia}}, \bibinfo {author}
  {\bibfnamefont {R.}~\bibnamefont {Pu{\'z}niak}}, \bibinfo {author}
  {\bibfnamefont {R.}~\bibnamefont {Jakie{\l}a}}, \bibinfo {author}
  {\bibfnamefont {M.}~\bibnamefont {Aleszkiewicz}}, \bibinfo {author}
  {\bibfnamefont {T.}~\bibnamefont {Dietl}}, \bibinfo {author} {\bibfnamefont
  {G.}~\bibnamefont {Springholz}}, \ and\ \bibinfo {author} {\bibfnamefont
  {G.}~\bibnamefont {Bauer}},\ }\bibinfo {title} {Contact superconductivity in
  {In--PbTe} junctions},\ \href {https://doi.org/10.1063/1.3475692} {\bibfield
  {journal} {\bibinfo  {journal} {J. Appl. Phys.}\ }\textbf {\bibinfo {volume}
  {108}},\ \bibinfo {pages} {053714} (\bibinfo {year} {2010})}\BibitemShut
  {NoStop}%
\bibitem [{\citenamefont {Kim}\ \emph {et~al.}(2017)\citenamefont {Kim},
  \citenamefont {Kim}, \citenamefont {Yang}, \citenamefont {Peng},
  \citenamefont {Yu},\ and\ \citenamefont {Doh}}]{kim2017strong}%
  \BibitemOpen
  \bibfield  {author} {\bibinfo {author} {\bibfnamefont {B.-K.}\ \bibnamefont
  {Kim}}, \bibinfo {author} {\bibfnamefont {H.-S.}\ \bibnamefont {Kim}},
  \bibinfo {author} {\bibfnamefont {Y.}~\bibnamefont {Yang}}, \bibinfo {author}
  {\bibfnamefont {X.}~\bibnamefont {Peng}}, \bibinfo {author} {\bibfnamefont
  {D.}~\bibnamefont {Yu}}, \ and\ \bibinfo {author} {\bibfnamefont {Y.-J.}\
  \bibnamefont {Doh}},\ }\bibinfo {title} {Strong superconducting proximity
  effects in $PbS$ semiconductor nanowires},\ \href
  {https://doi.org/10.1021/acsnano.6b04774} {\bibfield  {journal} {\bibinfo
  {journal} {ACS nano}\ }\textbf {\bibinfo {volume} {11}},\ \bibinfo {pages}
  {221} (\bibinfo {year} {2017})}\BibitemShut {NoStop}%
\bibitem [{\citenamefont {Nijholt}\ and\ \citenamefont
  {Akhmerov}(2016)}]{nijholt2016orbital}%
  \BibitemOpen
  \bibfield  {author} {\bibinfo {author} {\bibfnamefont {B.}~\bibnamefont
  {Nijholt}}\ and\ \bibinfo {author} {\bibfnamefont {A.~R.}\ \bibnamefont
  {Akhmerov}},\ }\bibinfo {title} {Orbital effect of magnetic field on the
  {Majorana} phase diagram},\ \href {\doibase 10.1103/PhysRevB.93.235434}
  {\bibfield  {journal} {\bibinfo  {journal} {Phys. Rev. B}\ }\textbf {\bibinfo
  {volume} {93}},\ \bibinfo {pages} {235434} (\bibinfo {year}
  {2016})}\BibitemShut {NoStop}%
\bibitem [{\citenamefont {Winkler}\ \emph {et~al.}(2017)\citenamefont
  {Winkler}, \citenamefont {Varjas}, \citenamefont {Skolasinski}, \citenamefont
  {Soluyanov}, \citenamefont {Troyer},\ and\ \citenamefont
  {Wimmer}}]{winkler2017orbital}%
  \BibitemOpen
  \bibfield  {author} {\bibinfo {author} {\bibfnamefont {G.~W.}\ \bibnamefont
  {Winkler}}, \bibinfo {author} {\bibfnamefont {D.}~\bibnamefont {Varjas}},
  \bibinfo {author} {\bibfnamefont {R.}~\bibnamefont {Skolasinski}}, \bibinfo
  {author} {\bibfnamefont {A.~A.}\ \bibnamefont {Soluyanov}}, \bibinfo {author}
  {\bibfnamefont {M.}~\bibnamefont {Troyer}}, \ and\ \bibinfo {author}
  {\bibfnamefont {M.}~\bibnamefont {Wimmer}},\ }\bibinfo {title} {Orbital
  Contributions to the Electron $g$ Factor in Semiconductor Nanowires},\ \href
  {\doibase 10.1103/PhysRevLett.119.037701} {\bibfield  {journal} {\bibinfo
  {journal} {Phys. Rev. Lett.}\ }\textbf {\bibinfo {volume} {119}},\ \bibinfo
  {pages} {037701} (\bibinfo {year} {2017})}\BibitemShut {NoStop}%
\bibitem [{\citenamefont {Stanescu}\ and\ \citenamefont
  {Das~Sarma}(2013)}]{stanescu2013superconducting}%
  \BibitemOpen
  \bibfield  {author} {\bibinfo {author} {\bibfnamefont {T.~D.}\ \bibnamefont
  {Stanescu}}\ and\ \bibinfo {author} {\bibfnamefont {S.}~\bibnamefont
  {Das~Sarma}},\ }\bibinfo {title} {Superconducting proximity effect in
  semiconductor nanowires},\ \href {\doibase 10.1103/PhysRevB.87.180504}
  {\bibfield  {journal} {\bibinfo  {journal} {Phys. Rev. B}\ }\textbf {\bibinfo
  {volume} {87}},\ \bibinfo {pages} {180504(R)} (\bibinfo {year}
  {2013})}\BibitemShut {NoStop}%
\bibitem [{\citenamefont {Rex}\ and\ \citenamefont
  {Sudb\o{}}(2014)}]{rex2014tilting}%
  \BibitemOpen
  \bibfield  {author} {\bibinfo {author} {\bibfnamefont {S.}~\bibnamefont
  {Rex}}\ and\ \bibinfo {author} {\bibfnamefont {A.}~\bibnamefont {Sudb\o{}}},\
  }\bibinfo {title} {Tilting of the magnetic field in {Majorana} nanowires:
  Critical angle and zero-energy differential conductance},\ \href {\doibase
  10.1103/PhysRevB.90.115429} {\bibfield  {journal} {\bibinfo  {journal} {Phys.
  Rev. B}\ }\textbf {\bibinfo {volume} {90}},\ \bibinfo {pages} {115429}
  (\bibinfo {year} {2014})}\BibitemShut {NoStop}%
\bibitem [{\citenamefont {Groth}\ \emph {et~al.}(2014)\citenamefont {Groth},
  \citenamefont {Wimmer}, \citenamefont {Akhmerov},\ and\ \citenamefont
  {Waintal}}]{groth2014kwant}%
  \BibitemOpen
  \bibfield  {author} {\bibinfo {author} {\bibfnamefont {C.~W.}\ \bibnamefont
  {Groth}}, \bibinfo {author} {\bibfnamefont {M.}~\bibnamefont {Wimmer}},
  \bibinfo {author} {\bibfnamefont {A.~R.}\ \bibnamefont {Akhmerov}}, \ and\
  \bibinfo {author} {\bibfnamefont {X.}~\bibnamefont {Waintal}},\ }\bibinfo
  {title} {Kwant: a software package for quantum transport},\ \href {\doibase
  10.1088/1367-2630/16/6/063065} {\bibfield  {journal} {\bibinfo  {journal}
  {New J. Phys.}\ }\textbf {\bibinfo {volume} {16}},\ \bibinfo {pages} {063065}
  (\bibinfo {year} {2014})}\BibitemShut {NoStop}%
\bibitem [{\citenamefont {Woods}\ \emph {et~al.}(2021)\citenamefont {Woods},
  \citenamefont {Das~Sarma},\ and\ \citenamefont {Stanescu}}]{woods2021charge}%
  \BibitemOpen
  \bibfield  {author} {\bibinfo {author} {\bibfnamefont {B.~D.}\ \bibnamefont
  {Woods}}, \bibinfo {author} {\bibfnamefont {S.}~\bibnamefont {Das~Sarma}}, \
  and\ \bibinfo {author} {\bibfnamefont {T.~D.}\ \bibnamefont {Stanescu}},\
  }\bibinfo {title} {Charge-Impurity Effects in Hybrid {Majorana} Nanowires},\
  \href {\doibase 10.1103/PhysRevApplied.16.054053} {\bibfield  {journal}
  {\bibinfo  {journal} {Phys. Rev. Applied}\ }\textbf {\bibinfo {volume}
  {16}},\ \bibinfo {pages} {054053} (\bibinfo {year} {2021})}\BibitemShut
  {NoStop}%
\bibitem [{\citenamefont {Osca}\ \emph {et~al.}(2014)\citenamefont {Osca},
  \citenamefont {Ruiz},\ and\ \citenamefont {Serra}}]{osca2014effects}%
  \BibitemOpen
  \bibfield  {author} {\bibinfo {author} {\bibfnamefont {J.}~\bibnamefont
  {Osca}}, \bibinfo {author} {\bibfnamefont {D.}~\bibnamefont {Ruiz}}, \ and\
  \bibinfo {author} {\bibfnamefont {L.}~\bibnamefont {Serra}},\ }\bibinfo
  {title} {Effects of tilting the magnetic field in one-dimensional {Majorana}
  nanowires},\ \href {\doibase 10.1103/PhysRevB.89.245405} {\bibfield
  {journal} {\bibinfo  {journal} {Phys. Rev. B}\ }\textbf {\bibinfo {volume}
  {89}},\ \bibinfo {pages} {245405} (\bibinfo {year} {2014})}\BibitemShut
  {NoStop}%
\bibitem [{\citenamefont {{Das Sarma}}\ \emph {et~al.}(2012)\citenamefont {{Das
  Sarma}}, \citenamefont {Sau},\ and\ \citenamefont
  {Stanescu}}]{sarma2012splitting}%
  \BibitemOpen
  \bibfield  {author} {\bibinfo {author} {\bibfnamefont {S.}~\bibnamefont {{Das
  Sarma}}}, \bibinfo {author} {\bibfnamefont {J.~D.}\ \bibnamefont {Sau}}, \
  and\ \bibinfo {author} {\bibfnamefont {T.~D.}\ \bibnamefont {Stanescu}},\
  }\bibinfo {title} {Splitting of the zero-bias conductance peak as smoking gun
  evidence for the existence of the {Majorana} mode in a
  superconductor-semiconductor nanowire},\ \href
  {https://doi.org/10.1103/PhysRevB.86.220506} {\bibfield  {journal} {\bibinfo
  {journal} {Phys. Rev. B}\ }\textbf {\bibinfo {volume} {86}},\ \bibinfo
  {pages} {220506(R)} (\bibinfo {year} {2012})}\BibitemShut {NoStop}%
\bibitem [{\citenamefont {Stanescu}\ \emph {et~al.}(2012)\citenamefont
  {Stanescu}, \citenamefont {Tewari}, \citenamefont {Sau},\ and\ \citenamefont
  {Das~Sarma}}]{stanescu2012to}%
  \BibitemOpen
  \bibfield  {author} {\bibinfo {author} {\bibfnamefont {T.~D.}\ \bibnamefont
  {Stanescu}}, \bibinfo {author} {\bibfnamefont {S.}~\bibnamefont {Tewari}},
  \bibinfo {author} {\bibfnamefont {J.~D.}\ \bibnamefont {Sau}}, \ and\
  \bibinfo {author} {\bibfnamefont {S.}~\bibnamefont {Das~Sarma}},\ }\bibinfo
  {title} {To Close or Not to Close: The Fate of the Superconducting Gap Across
  the Topological Quantum Phase Transition in Majorana-Carrying Semiconductor
  Nanowires},\ \href {\doibase 10.1103/PhysRevLett.109.266402} {\bibfield
  {journal} {\bibinfo  {journal} {Phys. Rev. Lett.}\ }\textbf {\bibinfo
  {volume} {109}},\ \bibinfo {pages} {266402} (\bibinfo {year}
  {2012})}\BibitemShut {NoStop}%
\bibitem [{\citenamefont {Ford}\ \emph {et~al.}(2012)\citenamefont {Ford},
  \citenamefont {Kumar}, \citenamefont {Kapadia}, \citenamefont {Guo},\ and\
  \citenamefont {Javey}}]{ford2012observation}%
  \BibitemOpen
  \bibfield  {author} {\bibinfo {author} {\bibfnamefont {A.~C.}\ \bibnamefont
  {Ford}}, \bibinfo {author} {\bibfnamefont {S.~B.}\ \bibnamefont {Kumar}},
  \bibinfo {author} {\bibfnamefont {R.}~\bibnamefont {Kapadia}}, \bibinfo
  {author} {\bibfnamefont {J.}~\bibnamefont {Guo}}, \ and\ \bibinfo {author}
  {\bibfnamefont {A.}~\bibnamefont {Javey}},\ }\bibinfo {title} {Observation of
  degenerate one-dimensional sub-bands in cylindrical {InAs} nanowires},\ \href
  {https://doi.org/10.1021/nl203895x} {\bibfield  {journal} {\bibinfo
  {journal} {Nano Lett.}\ }\textbf {\bibinfo {volume} {12}},\ \bibinfo {pages}
  {1340} (\bibinfo {year} {2012})}\BibitemShut {NoStop}%
\bibitem [{\citenamefont {van Weperen}\ \emph {et~al.}(2013)\citenamefont {van
  Weperen}, \citenamefont {Plissard}, \citenamefont {Bakkers}, \citenamefont
  {Frolov},\ and\ \citenamefont {Kouwenhoven}}]{van2013quantized}%
  \BibitemOpen
  \bibfield  {author} {\bibinfo {author} {\bibfnamefont {I.}~\bibnamefont {van
  Weperen}}, \bibinfo {author} {\bibfnamefont {S.~R.}\ \bibnamefont
  {Plissard}}, \bibinfo {author} {\bibfnamefont {E.~P.}\ \bibnamefont
  {Bakkers}}, \bibinfo {author} {\bibfnamefont {S.~M.}\ \bibnamefont {Frolov}},
  \ and\ \bibinfo {author} {\bibfnamefont {L.~P.}\ \bibnamefont
  {Kouwenhoven}},\ }\bibinfo {title} {Quantized conductance in an {InSb}
  nanowire},\ \href {https://doi.org/10.1021/nl3035256} {\bibfield  {journal}
  {\bibinfo  {journal} {Nano Lett.}\ }\textbf {\bibinfo {volume} {13}},\
  \bibinfo {pages} {387} (\bibinfo {year} {2013})}\BibitemShut {NoStop}%
\bibitem [{\citenamefont {van Weperen}\ \emph {et~al.}(2015)\citenamefont {van
  Weperen}, \citenamefont {Tarasinski}, \citenamefont {Eeltink}, \citenamefont
  {Pribiag}, \citenamefont {Plissard}, \citenamefont {Bakkers}, \citenamefont
  {Kouwenhoven},\ and\ \citenamefont {Wimmer}}]{weperen2015spin}%
  \BibitemOpen
  \bibfield  {author} {\bibinfo {author} {\bibfnamefont {I.}~\bibnamefont {van
  Weperen}}, \bibinfo {author} {\bibfnamefont {B.}~\bibnamefont {Tarasinski}},
  \bibinfo {author} {\bibfnamefont {D.}~\bibnamefont {Eeltink}}, \bibinfo
  {author} {\bibfnamefont {V.~S.}\ \bibnamefont {Pribiag}}, \bibinfo {author}
  {\bibfnamefont {S.~R.}\ \bibnamefont {Plissard}}, \bibinfo {author}
  {\bibfnamefont {E.~P. A.~M.}\ \bibnamefont {Bakkers}}, \bibinfo {author}
  {\bibfnamefont {L.~P.}\ \bibnamefont {Kouwenhoven}}, \ and\ \bibinfo {author}
  {\bibfnamefont {M.}~\bibnamefont {Wimmer}},\ }\bibinfo {title} {Spin-orbit
  interaction in {InSb} nanowires},\ \href {\doibase
  10.1103/PhysRevB.91.201413} {\bibfield  {journal} {\bibinfo  {journal} {Phys.
  Rev. B}\ }\textbf {\bibinfo {volume} {91}},\ \bibinfo {pages} {201413(R)}
  (\bibinfo {year} {2015})}\BibitemShut {NoStop}%
\bibitem [{\citenamefont {Kammhuber}\ \emph {et~al.}(2016)\citenamefont
  {Kammhuber}, \citenamefont {Cassidy}, \citenamefont {Zhang}, \citenamefont
  {G\"{u}l}, \citenamefont {Pei}, \citenamefont {De~Moor}, \citenamefont
  {Nijholt}, \citenamefont {Watanabe}, \citenamefont {Taniguchi}, \citenamefont
  {Car} \emph {et~al.}}]{kammhuber2016conductance}%
  \BibitemOpen
  \bibfield  {author} {\bibinfo {author} {\bibfnamefont {J.}~\bibnamefont
  {Kammhuber}}, \bibinfo {author} {\bibfnamefont {M.~C.}\ \bibnamefont
  {Cassidy}}, \bibinfo {author} {\bibfnamefont {H.}~\bibnamefont {Zhang}},
  \bibinfo {author} {\bibfnamefont {O.}~\bibnamefont {G\"{u}l}}, \bibinfo
  {author} {\bibfnamefont {F.}~\bibnamefont {Pei}}, \bibinfo {author}
  {\bibfnamefont {M.~W.}\ \bibnamefont {De~Moor}}, \bibinfo {author}
  {\bibfnamefont {B.}~\bibnamefont {Nijholt}}, \bibinfo {author} {\bibfnamefont
  {K.}~\bibnamefont {Watanabe}}, \bibinfo {author} {\bibfnamefont
  {T.}~\bibnamefont {Taniguchi}}, \bibinfo {author} {\bibfnamefont
  {D.}~\bibnamefont {Car}},  \emph {et~al.},\ }\bibinfo {title} {Conductance
  quantization at zero magnetic field in {InSb} nanowires},\ \href
  {https://doi.org/10.1021/acs.nanolett.6b00051} {\bibfield  {journal}
  {\bibinfo  {journal} {Nano Lett.}\ }\textbf {\bibinfo {volume} {16}},\
  \bibinfo {pages} {3482} (\bibinfo {year} {2016})}\BibitemShut {NoStop}%
\bibitem [{\citenamefont {Heedt}\ \emph {et~al.}(2016)\citenamefont {Heedt},
  \citenamefont {Prost}, \citenamefont {Schubert}, \citenamefont
  {Gr\"{u}tzmacher},\ and\ \citenamefont {Sch\"{a}pers}}]{heedt2016ballistic}%
  \BibitemOpen
  \bibfield  {author} {\bibinfo {author} {\bibfnamefont {S.}~\bibnamefont
  {Heedt}}, \bibinfo {author} {\bibfnamefont {W.}~\bibnamefont {Prost}},
  \bibinfo {author} {\bibfnamefont {J.}~\bibnamefont {Schubert}}, \bibinfo
  {author} {\bibfnamefont {D.}~\bibnamefont {Gr\"{u}tzmacher}}, \ and\ \bibinfo
  {author} {\bibfnamefont {T.}~\bibnamefont {Sch\"{a}pers}},\ }\bibinfo {title}
  {Ballistic transport and exchange interaction in {InAs} nanowire quantum
  point contacts},\ \href {https://doi.org/10.1021/acs.nanolett.6b00414}
  {\bibfield  {journal} {\bibinfo  {journal} {Nano Lett.}\ }\textbf {\bibinfo
  {volume} {16}},\ \bibinfo {pages} {3116} (\bibinfo {year}
  {2016})}\BibitemShut {NoStop}%
\bibitem [{\citenamefont {Kammhuber}\ \emph {et~al.}(2017)\citenamefont
  {Kammhuber}, \citenamefont {Cassidy}, \citenamefont {Pei}, \citenamefont
  {Nowak}, \citenamefont {Vuik}, \citenamefont {G{\"u}l}, \citenamefont {Car},
  \citenamefont {Plissard}, \citenamefont {Bakkers}, \citenamefont {Wimmer}
  \emph {et~al.}}]{kammhuber2017conductance}%
  \BibitemOpen
  \bibfield  {author} {\bibinfo {author} {\bibfnamefont {J.}~\bibnamefont
  {Kammhuber}}, \bibinfo {author} {\bibfnamefont {M.~C.}\ \bibnamefont
  {Cassidy}}, \bibinfo {author} {\bibfnamefont {F.}~\bibnamefont {Pei}},
  \bibinfo {author} {\bibfnamefont {M.~P.}\ \bibnamefont {Nowak}}, \bibinfo
  {author} {\bibfnamefont {A.}~\bibnamefont {Vuik}}, \bibinfo {author}
  {\bibfnamefont {{\"O}.}~\bibnamefont {G{\"u}l}}, \bibinfo {author}
  {\bibfnamefont {D.}~\bibnamefont {Car}}, \bibinfo {author} {\bibfnamefont
  {S.}~\bibnamefont {Plissard}}, \bibinfo {author} {\bibfnamefont
  {E.}~\bibnamefont {Bakkers}}, \bibinfo {author} {\bibfnamefont
  {M.}~\bibnamefont {Wimmer}},  \emph {et~al.},\ }\bibinfo {title} {Conductance
  through a helical state in an Indium antimonide nanowire},\ \href
  {https://doi.org/10.1038/s41467-017-00315-y} {\bibfield  {journal} {\bibinfo
  {journal} {Nat. Commun.}\ }\textbf {\bibinfo {volume} {8}},\ \bibinfo {pages}
  {478} (\bibinfo {year} {2017})}\BibitemShut {NoStop}%
\bibitem [{\citenamefont {Heedt}\ \emph {et~al.}(2017)\citenamefont {Heedt},
  \citenamefont {Ziani}, \citenamefont {Cr{\'e}pin}, \citenamefont {Prost},
  \citenamefont {Schubert}, \citenamefont {Gr{\"u}tzmacher}, \citenamefont
  {Trauzettel}, \citenamefont {Sch{\"a}pers} \emph
  {et~al.}}]{heedt2017signatures}%
  \BibitemOpen
  \bibfield  {author} {\bibinfo {author} {\bibfnamefont {S.}~\bibnamefont
  {Heedt}}, \bibinfo {author} {\bibfnamefont {N.~T.}\ \bibnamefont {Ziani}},
  \bibinfo {author} {\bibfnamefont {F.}~\bibnamefont {Cr{\'e}pin}}, \bibinfo
  {author} {\bibfnamefont {W.}~\bibnamefont {Prost}}, \bibinfo {author}
  {\bibfnamefont {J.}~\bibnamefont {Schubert}}, \bibinfo {author}
  {\bibfnamefont {D.}~\bibnamefont {Gr{\"u}tzmacher}}, \bibinfo {author}
  {\bibfnamefont {B.}~\bibnamefont {Trauzettel}}, \bibinfo {author}
  {\bibfnamefont {T.}~\bibnamefont {Sch{\"a}pers}},  \emph {et~al.},\ }\bibinfo
  {title} {Signatures of interaction-induced helical gaps in nanowire quantum
  point contacts},\ \href {https://doi.org/10.1038/nphys4070} {\bibfield
  {journal} {\bibinfo  {journal} {Nat. Phys.}\ }\textbf {\bibinfo {volume}
  {13}},\ \bibinfo {pages} {563} (\bibinfo {year} {2017})}\BibitemShut
  {NoStop}%
\bibitem [{\citenamefont {Estrada~Salda{\~n}a}\ \emph
  {et~al.}(2018)\citenamefont {Estrada~Salda{\~n}a}, \citenamefont {Niquet},
  \citenamefont {Cleuziou}, \citenamefont {Lee}, \citenamefont {Car},
  \citenamefont {Plissard}, \citenamefont {Bakkers},\ and\ \citenamefont
  {De~Franceschi}}]{estrada2018split}%
  \BibitemOpen
  \bibfield  {author} {\bibinfo {author} {\bibfnamefont {J.~C.}\ \bibnamefont
  {Estrada~Salda{\~n}a}}, \bibinfo {author} {\bibfnamefont {Y.-M.}\
  \bibnamefont {Niquet}}, \bibinfo {author} {\bibfnamefont {J.-P.}\
  \bibnamefont {Cleuziou}}, \bibinfo {author} {\bibfnamefont {E.~J.}\
  \bibnamefont {Lee}}, \bibinfo {author} {\bibfnamefont {D.}~\bibnamefont
  {Car}}, \bibinfo {author} {\bibfnamefont {S.~R.}\ \bibnamefont {Plissard}},
  \bibinfo {author} {\bibfnamefont {E.~P.}\ \bibnamefont {Bakkers}}, \ and\
  \bibinfo {author} {\bibfnamefont {S.}~\bibnamefont {De~Franceschi}},\
  }\bibinfo {title} {Split-channel ballistic transport in an {InSb} nanowire},\
  \href {https://doi.org/10.1021/acs.nanolett.7b03854} {\bibfield  {journal}
  {\bibinfo  {journal} {Nano Lett.}\ }\textbf {\bibinfo {volume} {18}},\
  \bibinfo {pages} {2282} (\bibinfo {year} {2018})}\BibitemShut {NoStop}%
\bibitem [{\citenamefont {Zhang}\ \emph {et~al.}(2019)\citenamefont {Zhang},
  \citenamefont {Liu}, \citenamefont {Wimmer},\ and\ \citenamefont
  {Kouwenhoven}}]{zhang2019next}%
  \BibitemOpen
  \bibfield  {author} {\bibinfo {author} {\bibfnamefont {H.}~\bibnamefont
  {Zhang}}, \bibinfo {author} {\bibfnamefont {D.~E.}\ \bibnamefont {Liu}},
  \bibinfo {author} {\bibfnamefont {M.}~\bibnamefont {Wimmer}}, \ and\ \bibinfo
  {author} {\bibfnamefont {L.~P.}\ \bibnamefont {Kouwenhoven}},\ }\bibinfo
  {title} {Next steps of quantum transport in {Majorana} nanowire devices},\
  \href {https://doi.org/10.1038/s41467-019-13133-1} {\bibfield  {journal}
  {\bibinfo  {journal} {Nat. Commun.}\ }\textbf {\bibinfo {volume} {10}},\
  \bibinfo {pages} {5128} (\bibinfo {year} {2019})}\BibitemShut {NoStop}%
\bibitem [{\citenamefont {Darnhofer}\ and\ \citenamefont
  {R\"ossler}(1993)}]{darnhofer1993effects}%
  \BibitemOpen
  \bibfield  {author} {\bibinfo {author} {\bibfnamefont {T.}~\bibnamefont
  {Darnhofer}}\ and\ \bibinfo {author} {\bibfnamefont {U.}~\bibnamefont
  {R\"ossler}},\ }\bibinfo {title} {Effects of band structure and spin in
  quantum dots},\ \href {\doibase 10.1103/PhysRevB.47.16020} {\bibfield
  {journal} {\bibinfo  {journal} {Phys. Rev. B}\ }\textbf {\bibinfo {volume}
  {47}},\ \bibinfo {pages} {16020} (\bibinfo {year} {1993})}\BibitemShut
  {NoStop}%
\end{thebibliography}

%

\end{document}